\documentclass[aps, prd, 11pt, a4paper, numbers,sort&compress]{revtex4-1}

\usepackage{amsmath}
\usepackage{comment}
\usepackage{graphicx}
\usepackage{dcolumn}
\usepackage{bm}
\usepackage{subfigure}
\usepackage{epsfig}
\usepackage{amsmath}
\usepackage{amsfonts}
\usepackage{graphicx}
\usepackage{amssymb}
\usepackage{nccmath}
\usepackage{amssymb,amsmath,amsfonts}
\usepackage{xfrac}
\usepackage{color}
\usepackage{tikz}
\usepackage[T1]{fontenc}
\usepackage[utf8]{inputenc}
\usepackage{ulem}

\usepackage{tcolorbox}
\usepackage{hyperref}
\usepackage{booktabs}
\hypersetup{colorlinks=true, citecolor=red, linkcolor=blue, urlcolor=blue}

\usepackage[T1]{fontenc}
\usepackage{lmodern}
\usepackage{amsmath,amssymb,mathtools,bm}
\usepackage{mathrsfs}
\usepackage{booktabs}
\usepackage{array}
\usepackage{microtype}
\usepackage[margin=1in]{geometry}

\newcommand{\M}{\mathcal{M}}
\newcommand{\F}{\mathcal{F}}

\newcommand{\Tr}{\operatorname{Tr}}
\newcommand{\supp}{\operatorname{supp}}
\newcommand{\dd}{\mathrm{d}}
\newcommand{\I}{\mathbb{I}}

\newcommand{\MM}{\mathbb{M}}
\newcommand{\POVM}{\mathsf{E}}
\newcommand{\Effect}{\mathsf{F}}
\newcommand{\Instr}{\mathcal{I}}
\newcommand{\e}{\mathrm{e}}

\begin{document}

\title{Integrating Quantum Fields and Gravitation}

\author{Vasilios Zarikas}
\email{vzarikas@uth.gr}
\affiliation{Department of Mathematics, University of Thessaly, 35100, Lamia, Greece}

\begin{abstract}
We propose a novel construction that couples quantum fields to classical
gravity using the tangent bundle of spacetime.
The integration has been acheived by formulating field local dynamics on the Minkowski tangent space associated with each event of the classical spacetime. Fields are quantized and renormalized in these tangent Minkowski
fibers, and a geometrical transport law relates their profiles at
different spacetime events. A localization positive-operator-valued
measure (POVM) resolves each field into localized modes, assigns them
positive shares that sum to the total renormalized energy, and
characterizes their spatial extent and quantum uncertainties.
Another central ingredient of the theoy is a geometric criterion for how much of each
mode's energy gravitates. A mode contributes the energy contained in
the region where the full classical metric, pulled back to the tangent
space, agrees with the Minkowski metric within a universal tolerance.
The contributions of all modes enter a single classical Einstein
equation, with a stress response that balances changes in gravitating
energy and momentum. The transport law recovers the proper-time phase
and gravitational frequency shift of quantum matter, in accord with
matter-wave experiments. We also construct an explicit unitary transport
for a complex scalar field in a radiation universe.
Two gravitational consequences follow. The tangent Minkowski reference
vacuum contributes no gravitational source at any epoch. In the
expanding cosmologies studied here, a progressively larger fraction of
the energy of a thermal gas of bosons and fermions gravitates, and we
calculate the quantum and thermal fluctuations of that contribution.
For a freely spreading complex-scalar packet, we identify the mechanism
that determines the outcome: its gravitating fraction grows when the
region selected by the metric expands faster than the packet's energy
profile spreads.
\end{abstract}

\maketitle


\section{Motivation and the physical construction}
\label{sec:introduction}

Several major research programmes seek to reconcile gravitation with quantum
theory, including string theory \cite{Polchinski1998StringTheory}, asymptotic
safety \cite{Reuter1998QuantumGravity}, loop quantum gravity and spin-foam
models \cite{AshtekarLewandowski2004,Perez2013SpinFoam}, holographic approaches
such as gauge/gravity duality \cite{Maldacena1998AdSCFT}, and discrete
approaches to geometry, including causal sets and causal dynamical
triangulations \cite{BombelliEtAl1987CausalSets,Loll2020CDT}. These programmes differ in their fundamental variables and
strategies for describing quantum geometry. Despite substantial theoretical
progress, it remains unsettled which, if any, describes gravity in nature;
experimental tests of the quantum character of gravity are themselves an
active area of research \cite{MarlettoVedral2025GravityTests}. This motivates
the investigation of other ways of coupling gravity to quantum fields,
including formulations in which the spacetime metric remains classical.

Published work in this broader alternative direction encompasses several
distinct proposals. Di\'osi's dynamical reduction models
\cite{Diosi1989Reduction} and Penrose's arguments for gravitationally induced
state reduction \cite{Penrose1996Reduction} explore modifications of quantum
dynamics motivated by gravity; these proposals do not by themselves provide
a complete dynamical theory of a classical metric coupled to quantum matter.
More explicit constructions include the classical measurement-and-feedback
model of gravitational interaction developed by Kafri, Taylor and Milburn
\cite{KafriTaylorMilburn2014}, the Newtonian semiclassical model of Tilloy and
Di\'osi in which stochastic signals associated with spontaneous localization
source a classical gravitational potential \cite{TilloyDiosi2016}, and
Oppenheim's formulation of stochastic classical--quantum dynamics for a
classical metric coupled to quantum fields
\cite{Oppenheim2023ClassicalGravity}. These approaches differ in their
treatment of state reduction, stochasticity and gravitational backreaction.
The present work belongs to this second, broader line of inquiry: it retains
a classical spacetime metric, describes quantum fields in tangent Minkowski
fibers, and prescribes their gravitational contribution through a geometrical
comparison of the full metrics.

A quantum field can carry inertial energy without a proposed classical--quantum theory having already specified how much of that energy enters Einstein's equation. 
Conventional semiclassical gravity takes the renormalized expectation
value $\langle\widehat T_{\mu\nu}\rangle_{\rm ren}$ as the source of a
classical metric. Whether this prescription describes the gravitational
field in an individual experiment depends on which quantum state is
used to evaluate that expectation value. Page and Geilker tested this
distinction by using radioactive decays to select between two arrangements
of laboratory masses. The measured gravitational field followed the
arrangement actually realized, rather than the field predicted from a
source averaged over both possible arrangements
\cite{PageGeilker1981}. This result constrains that specific use of an
averaged source; it does not exclude every theory in which quantum
matter sources a classical metric.

At the same time, gravitational phases and free fall of quantum matter are well tested \cite{ColellaOverhauserWerner1975,Asenbaum2017,Asenbaum2020}. These experiments test the propagation of matter in a prescribed field. They provide a benchmark for the transport law, distinct from a test of the gravitational field that the matter itself produces.

The present proposal retains one classical Lorentzian spacetime $(\M,g)$ and quantum matter described by renormalized Minkowski theories in tangent fibers. Here $\M$ is the spacetime manifold and $g$ its metric, with signature $(-,+,+,+)$. The classical connection compares local frames; a prescribed finite-profile transport determines the comparison of field observables and states. There is no quantized metric or added graviton field. The first part of the construction specifies the tangent field theory, its admissible transport, physical localization, positive energy shares, and uncertainty profiles.

The new gravitational postulate compares the finite tangent description with the surrounding classical geometry. A tangent space is an abstract vector space, so there is no intrinsic ambient distance between it and the spacetime. Instead, the exponential map associates tangent displacement vectors with spacetime events. Pulling the full metric back by the differential of that map makes an exact tensor comparison possible. At the contact point the two metrics agree. Away from it they generally differ. Standard normal-coordinate and geodesic-deviation results quantify that difference \cite{PoissonPoundVega2011,Nesterov1999}. The further rule that only energy within a specified tolerance gravitates is the new physical postulate.

A universal dimensionless constant $\zeta>0$ specifies the tolerance. A physical timelike observer supplies a positive norm, needed because the Lorentzian metric is not a positive norm on vectors. In a slowly varying, nearly isotropic region the rule reduces to an energy window of radius approximately $\zeta L_{\rm curv}$, where $L_{\rm curv}$ is the local curvature length derived below. The fundamental definition uses the full pulled-back metric and can be directional. For matter today, the predicted gravitational mass must agree with the inertial mass to the precision of measurements that compare them. A particle horizon does not enter the criterion.

The word ``mode'' refers to a localized outcome cell of a specified positive-operator-valued measure (POVM), not to a freely selected Fourier oscillator. Completeness of the effects gives an exact positive energy partition. In a detector-free universe the initial state, observer, window shape and localization widths remain initial physical data. Homogeneous isotropic maximum-entropy states and spherical windows provide a concrete cosmological realization. Infinitely many outcomes or widths require a normalized outcome measure; they do not provide infinitely many copies of the field energy.

The difference between three operations is essential. Quantum transport evolves the field and its moments in the prescribed bundle. The metric comparison determines which part of its assigned energy gravitates. A classical reconstruction uses the selected energy, momentum and stress data to source the common metric. A geometrical weight can change without creating or destroying a quantum excitation. The resulting gravitating matter stress is generally not separately conserved; a determined classical response must satisfy the total Bianchi balance.

The paper is organized as follows. Section~\ref{sec:geometry} introduces the tangent Minkowski fibers and their renormalized quantum fields. Section~\ref{sec:propagation} distinguishes the relevant covariant derivatives and describes the transport of profiles, observables and quantum states. Section~\ref{sec:povm} defines the physical localization POVM and the positive, additive partition of quantum energy. Section~\ref{sec:worldtube} explains how spatial, momentum and energy uncertainties determine the localized profiles, their spreading and their worldtubes. Section~\ref{sec:metric-comparison} introduces the full metric-comparison criterion that selects the gravitating portion of each mode's energy and derives its local curvature-radius approximation. Section~\ref{sec:source} constructs the common classical source, explains its energy--momentum balance, and introduces a candidate response for a detector interacting with separated scalar packets. Section~\ref{sec:vacuum-gravity} states the assumptions under which the reference vacuum contributes zero gravitational source. Section~\ref{sec:cosmology} applies the construction to expanding FRW universes containing a thermal boson--fermion fluid or a complex scalar field, including quantum uncertainties and the conditions for increasing gravitating fractions. Subsection~\ref{sec:collapse} examines the corresponding decrease that can occur during contraction. Section~\ref{sec:algorithm} summarizes the complete construction as a step-by-step algorithm and identifies its theoretical and observational questions. The final main-text section presents the conclusions. Eight appendices retain the supporting domain and implementation conditions, scalar generators, energy-form construction, uncertainty and metric derivations, the detector-response test and cosmological estimates.

\section{Quantum fields in tangent Minkowski fibers}
\label{sec:geometry}\label{sec:qft}

The first step specifies what is classical and what is quantum. There is one classical spacetime $(\M,g)$, with signature $(-,+,+,+)$, and at every event $x$ an auxiliary Minkowski space $\MM_x$ and a Hilbert space $\F_x$. We use $c=\hbar=1$. A local orthonormal tetrad $e_{\hat a}{}^\mu$ converts tangent components, labelled by hatted indices, into spacetime components, labelled by Greek indices. The Minkowski space is a tangent model, not a finite flat region cut out of the universe. Its origin corresponds to $x$; placing a finite packet in spacetime requires a separate geometrical map.

The fibers carry representations of one fixed renormalized Minkowski matter theory, with observable algebra $\mathfrak A_R^{\rm M}$. A bundle is the organization of these Hilbert spaces over the classical events; it does not, by itself, tell us how to compare states at different events. That comparison is the next step.

More explicitly, a local Hilbert bundle is a projection $p_{\mathscr F}:\mathscr F\to\M$ with fiber $p_{\mathscr F}^{-1}(x)=\F_x$. A representation $\pi_x$ assigns each abstract observable $O\in\mathfrak A_R^{\rm M}$ an operator $\pi_x(O)$ on $\F_x$, on its stated domain when it is unbounded. The bundle projection $p_{\mathscr F}$ and the observable representation $\pi_x$ are different maps. A section $\Psi:x\mapsto\Psi(x)\in\F_x$ assigns a state vector to each event; it is not a classical scalar field. The tangent metric is $\eta_{\hat a\hat b}=\operatorname{diag}(-1,1,1,1)$, and $g_{\mu\nu}e_{\hat a}{}^\mu e_{\hat b}{}^\nu=\eta_{\hat a\hat b}$ fixes the tetrad convention.

The construction is formulated for both free and interacting renormalized Minkowski quantum fields through their observable algebra $\mathfrak A_R^{\rm M}$, its representations on the Hilbert fibers $\F_x$, and their quantum states. Matter interactions belong to the renormalized tangent-Minkowski Hamiltonian; the complex scalar with quartic self-interaction, introduced below, provides an example. A Fock-space description supplies explicit particle and antiparticle examples in free theories and, when suitable scattering states exist, describes the incoming and outgoing particles of an interacting theory. These are called asymptotic states because they describe particles in the large-time limits before or after scattering; they do not replace the interacting dynamics at finite times \cite{Haag1996}. The general formulation therefore does not require a particle-number decomposition or assume that every interacting theory has the Fock representation of a free field. Its application to an interacting theory remains subject to the domain and quantum-transport conditions stated below. The later POVM decomposition labels localization outcomes of the quantum state, including an interacting state, rather than sectors of fixed particle number.

For a running example take a complex scalar field of renormalized mass $m_R$. The operator $\widehat\Phi_{R,x}(X)$ lives in the tangent theory at $x$, where $X=(X^0,\bm X)$ is a Minkowski coordinate. In its free sector a smooth particle packet has the state and profile
\begin{equation}
 |\psi_A\rangle=\widehat a_x^\dagger(\varphi_A)|0\rangle,
 \qquad
 \Phi_A(X)=\langle0|\widehat\Phi_{R,x}(X)|\psi_A\rangle.
 \label{eq:main-scalar-state-profile}
\end{equation}
Here $|0\rangle$ is the tangent vacuum, $\widehat a_x^\dagger(\varphi_A)$ creates a normalized particle with momentum amplitude $\varphi_A$, and $A$ provisionally labels a prepared packet. Its physical localization-cell meaning will be fixed in Sec.~\ref{sec:povm}. The complex-valued function $\Phi_A$ is a matrix element, often called a c-number profile. It describes the envelope and phase of this packet, while the state and operators retain its quantum content. The profile is an initial-fiber matrix element. At another event $x_2$, the corresponding matrix element uses the transported bra as well as the ket: $\langle 0|U^\dagger\widehat\Phi_{R,x_2}(X)U|\psi_A\rangle$, where $U$ is the field propagator between the two fibers and $X$ is the endpoint tangent coordinate. Replacing the bra by the endpoint reference vacuum generally changes the profile; see Eq.~\eqref{eq:vacuum-transport-preservation}. The invariant momentum measure, commutators, antiparticle expansion and packet normalization are given in Appendix~\ref{app:bundle-foundations}.

\subsection{Local quantum dynamics and the role of curvature}

The tangent Hamiltonian $\widehat H_R^{\rm M}$ contains the mass, quantum dispersion and renormalized matter interactions; we also write it as $\widehat H_R$, with the superscript M emphasizing its Minkowski meaning. For the complex scalar the Hermitian quartic potential is $V_R(|\Phi|^2)=m_R^2|\Phi|^2+\lambda_R|\Phi|^4/2$, with fixed renormalized coupling $\lambda_R$; its operator version is defined by Minkowski composite renormalization. Particles and antiparticles carry positive energy under the chosen vacuum convention $\widehat H_R^{\rm M}\ge0$ and $\widehat H_R^{\rm M}|0\rangle=0$.

The postulate excludes independently adjustable curvature vertices from this quantum theory: in particular there is no $\xi R\widehat\Phi_R^\dagger\widehat\Phi_R$ interaction, where $R$ is the Ricci scalar and $\xi$ would be an independent nonminimal coupling. Quantum loops and subtractions remain Minkowski ones. Curvature instead acts on finite profiles and their comparison between fibers. This separation must be demonstrated by the transport construction; it cannot be established merely by renaming a curved-space Hamiltonian. Appendix~\ref{app:bundle-foundations} states the renormalization restriction and finite-energy/stress domains, and Appendix~\ref{app:transport-tests} explains the nontrivial short-distance tests it requires.

To state the required finiteness quantitatively, let $\rho$ be a positive unit-trace density operator on the fiber, let $E_*>0$ be a reference energy, and let $r_*\ge1$ be an integer fixing the energy-moment regularity. For a real smooth test tensor $f^{\hat a\hat b}(X)$, the smeared stress and its admitted moments satisfy
\begin{align}
 \widehat T_R[f]&=\int d^4X\,f^{\hat a\hat b}(X)\widehat T^R_{\hat a\hat b}(X),
 \qquad \partial^{\hat a}\widehat T^R_{\hat a\hat b}=0,
 \label{eq:main-stress-Ward}\\
 \Tr\!\left[\rho(1+\widehat H_R/E_*)^{2r_*}\right]&<\infty,
 \qquad \Tr\bigl(\rho\widehat T_R[f]^2\bigr)<\infty
 \quad(f\in\mathcal T_A).
 \label{eq:main-state-moments}
\end{align}
Here $\mathcal T_A$ is the complete space of resolved bulk and boundary test tensors needed for the packet, with support in the prescribed finite resolution regions. The Ward identity in Eq.~\eqref{eq:main-stress-Ward} is a distributional conservation law, meaning that it holds after smearing. The domain $\mathfrak D_{\rm adm}(A)$ consists of density operators satisfying Eq.~\eqref{eq:main-state-moments}, together with the further propagation requirements below. For $r_*=1$ the mean energy and its variance are finite; a larger value is used only if the stress-energy bounds require it. 

Equation~\eqref{eq:main-stress-Ward} expresses local energy--momentum conservation in the tangent quantum theory and defines stress observables through smearing, as required because the stress tensor is an operator-valued distribution. The finiteness conditions in Eq.~\eqref{eq:main-state-moments} are necessary because normalization of a quantum state alone does not guarantee finite energy and stress fluctuations, which are required to determine the localized profiles, their uncertainties and the resulting classical gravitational source.
The smooth test functions specify which spacetime-resolved stress observables are evaluated. Requiring finite second moments for these observables is compatible with conventional treatments of quantum stress fluctuations. It does not eliminate fluctuations or require them to be small.

Appendix~\ref{app:bundle-foundations} gives the continuous bounds on all test tensors, rather than checking a few selected measurements. These requirements control the quantum inputs; conservation of the reconstructed spacetime source requires a further step.

For example, a Gaussian scalar momentum packet has finite energy moments, whereas a normalized packet with a sufficiently slow high-momentum tail may not. Similarly, the stress tensor is a distribution: resolved energy density, pressure and flux are obtained by smearing it with finite spatial and temporal responses. A response may be specified by a detector or by the localization prescription and initial data of a cosmological model. Finiteness of these quantities is required before constructing the classical source. This is the physical meaning of admissibility here, not an additional field species or a freely fitted suppression factor.

\section{Propagation: from finite profiles to quantum states}
\label{sec:propagation}

The second step answers how the same quantum preparation is compared at successive classical events. The metric, field representation, reference trajectory, cross-sections and endpoint field identifications specify one candidate transport. We first describe that prescription for a prepared packet; Secs.~\ref{sec:povm}--\ref{sec:worldtube} then supply its precise localization and extent. On a trial metric these data are recomputed together by their fixed laws. Thus the order of explanation does not assume that a completed worldtube is known before propagation. ``One prescription'' means a fixed evolution after these physical identifications and the required well-posedness and quantum-implementation conditions have been supplied.

At the initial and final events of each propagation interval, the prescription must specify how the profile data are identified with the field data in the corresponding tangent-Minkowski fibers. For a complex scalar, this identification includes the field amplitude and its canonical momentum, expressed in the chosen local frame; a detector, when present, additionally supplies its measurement resolution and response functions. These identifications are part of the transport prescription and must be specified before testing its short-distance consistency or calculating the gravitating energy fraction.

The same field profile may be described using different canonical
variables. For example, one may replace the complex scalar amplitude
$\Phi$ by $\chi=a\Phi$, where $a$ is the cosmological scale factor,
and transform its conjugate momenta accordingly. To compare the same
physical field data, both the evolution on the classical geometry and
the tangent-Minkowski reference evolution must be expressed in these
same variables at each endpoint, as in
Eq.~\eqref{eq:endpoint-identifications}. Rescaling the data in only one
of the two evolutions would change the physical comparison.
These identifications must be fixed by the physical clock, rulers and field responses independently of the desired result. A general prescription for them and an interacting all-field lift remain open.

\subsection{Which covariant derivative acts on which object?}

The covariant derivatives used in the construction are distinguished by the objects on which they act. First, spacetime covariant differentiation describes how tensor or spinor profiles vary over the classical geometry. For a complex scalar profile $\Phi_A$, the gravitational covariant derivative satisfies $\nabla_\mu\Phi_A=\partial_\mu\Phi_A$, whereas its second derivative is $\nabla_\mu\nabla_\nu\Phi_A=\partial_\mu\partial_\nu\Phi_A-\Gamma^\lambda_{\mu\nu}\partial_\lambda\Phi_A$, where $\Gamma^\lambda_{\mu\nu}$ are the Levi--Civita connection coefficients of $g_{\mu\nu}$. Consequently, the wave operator $\Box_g\Phi_A=g^{\mu\nu}\nabla_\mu\nabla_\nu\Phi_A$ depends on the classical metric. For spinor profiles, the corresponding covariant derivative contains the spin connection induced by the same Levi--Civita geometry; for profiles carrying an internal gauge charge, the full derivative also includes the appropriate gauge connection. These derivatives act on the spacetime dependence and component indices of the profiles. The Hilbert-bundle derivative introduced next acts instead on quantum-state sections, comparing state vectors belonging to fibers at different spacetime events.

The Hilbert-bundle derivative $\mathcal D^{\mathscr F,\rm Lor}_\mu$ instead compares state sections when the local Lorentz frame changes. Its connection is $\mathcal A^{\rm Lor}_\mu=-i\omega_\mu{}^{\hat a\hat b}\widehat J_{R,\hat a\hat b}/2$. 
Thus, the Hilbert-bundle covariant derivative $\mathcal D^{\mathscr F,\rm Lor}_\mu$ acts on quantum-state sections $\Psi(x)\in\F_x$. It is obtained by letting the Levi--Civita Lorentz connection act through the unitary Lorentz representation of the tangent quantum theory, using the spin covering group when required. This is the connection induced on an associated Hilbert bundle; an explicit treatment of the underlying bundle construction and induced parallel transport is given in Ref.~\cite{DrechslerTuckey1996}. In our conventions its local expression is Eq.~\eqref{eq:main-Lorentz-connection}, where $\omega_\mu{}^{\hat a\hat b}$ are the Lorentz-connection coefficients of the chosen tetrad and $\widehat J_{R,\hat a\hat b}$ are the self-adjoint total Lorentz generators of the renormalized tangent theory, acting on a common invariant domain. The factor $-i$ converts these generators into the anti-Hermitian connection appropriate for unitary frame transport, while $1/2$ compensates for summing over both orders of the antisymmetric index pair. For a complex scalar, intrinsic spin vanishes, but the packet's momentum and spatial profile still transform under rotations and boosts. This derivative describes the Lorentz-frame contribution; the complete finite-profile propagation is specified separately below.

With these definitions the frame derivative is
\begin{equation}
 \mathcal D^{\mathscr F,\rm Lor}_\mu\Psi
 =\partial_\mu\Psi+\mathcal A^{\rm Lor}_\mu\Psi,
 \qquad
 \mathcal A^{\rm Lor}_\mu=-\frac{i}{2}
 \omega_\mu{}^{\hat a\hat b}\widehat J_{R,\hat a\hat b}.
 \label{eq:main-Lorentz-connection}
\end{equation}
The antisymmetric operators $\widehat J_R^{\hat a\hat b}$ are the renormalized total Lorentz generators: their spatial components generate rotations and their time--space components generate boosts (generators of the already renormalized Minkowski quantum field theory). Lowering their indices uses $\eta_{\hat a\hat b}$. They include orbital and, for fields with spin, intrinsic-spin contributions. They are distinct from a gauge charge current. Abstractly the derivative maps a section of $\mathscr F$ to a section of $T^*\M\otimes\mathscr F$, where $T^*\M$ is the cotangent bundle of spacetime; contraction with a curve tangent gives a derivative taking values in the Hilbert fiber itself. Thus the additional covector index records the direction of differentiation, not a change of quantum state space.

Finally, to describe the effect of geometry on an extended quantum packet, the construction introduces the geometrical connection $\mathcal A_{{\rm Sch},A,s}^{\rm geo}$. Here $A$ labels the packet, $s$ parametrizes its propagation along the chosen reference trajectory, ``geo'' denotes the geometrical contribution, and ``Sch'' specifies its Schr\"odinger-picture representation. This operator generates the infinitesimal change of the quantum state associated with the geometrical deformation of its finite profile, after the ordinary tangent-Minkowski evolution has been accounted for. For example, a complex-scalar packet in an expanding universe can undergo changes of physical width and momentum beyond those produced by its Minkowski evolution. Subsection~\ref{subsec:profile-comparison} first isolates this additional geometrical change through the residual profile map in Eq.~\eqref{eq:P-flat-limit}. Subsection~\ref{subsec:observable-transport} then asks whether that map can be implemented by a unitary operator on quantum states and, when this implementation exists, defines the corresponding connection explicitly in Eq.~\eqref{eq:main-connection-from-lift}. The connection enters the state-evolution law together with the tangent-Minkowski quantum generator in Eq.~\eqref{eq:bundle-evolution}. To write these operators, the prescribed fiber identifications express states from different fibers in a common reference Hilbert space; the Schr\"odinger picture then specifies how the evolution is assigned to states and observables. Because this construction concerns a finite packet, the connection may depend on its specified history and cross-sections, rather than only on the spacetime event. Appendix~\ref{app:unitary-foundations} supplies the detailed implementation and domain conditions.

For a freely propagating massive packet, let $z_A(s)$ denote its timelike reference trajectory, whose successive points identify the tangent fibers used in the propagation. The parameter $s$ labels the successive evolution sections, whereas $\tau_A$ is proper time along that trajectory. Their relation and the free reference motion are
\begin{equation}
 \begin{aligned}
 N_A:=\frac{d\tau_A}{ds}
 &=\sqrt{-g_{\mu\nu}\dot z_A^\mu\dot z_A^\nu},
 &\qquad u_A^\mu&=\frac{\dot z_A^\mu}{N_A},\\
 u_A^\nu\nabla_\nu u_A^\mu&=0,
 &u_A^\nu\nabla_\nu e_{\hat a}{}^\mu&=0.
 \end{aligned}
 \label{eq:main-reference-trajectory-clock}
\end{equation}
Here a dot denotes $d/ds$, $N_A>0$ is the lapse along the reference trajectory, $u_A^\mu$ is its unit four-velocity, and the tetrad is parallel transported with $e_{\hat0}{}^\mu=u_A^\mu$. These equations prescribe the reference motion and frame; the evolution of the extended profile and quantum state is constructed in Subsections~\ref{subsec:profile-comparison} and~\ref{subsec:observable-transport}. The measured centroid of a finite packet need not coincide exactly with this reference geodesic; their agreement requires the appropriate narrow-packet, short-wavelength approximation.

The lapse converts local proper-time evolution into evolution with respect to $s$. In the pure time-translation case described after Eq.~\eqref{eq:bundle-evolution}, this gives $\widehat K_{A,R}^{\rm M}=N_A\widehat H_R^{\rm M}$. The corresponding phase of a stationary energy component is proportional to $\int E N_A\,ds=\int E\,d\tau_A$, where $E$ is its local energy. Thus changing the label $s$ does not change the physical phase. Choosing $s=\tau_A$ sets $N_A=1$ along that particular trajectory; it does not identify the proper times of different observers or remove gravitational redshift. This path lapse is not automatically identical to the lapse of an arbitrary spacetime foliation.

To make the redshift statement explicit, consider a scalar packet in the geometrical-optics regime. If $\theta_A$ is its rapidly varying phase, its wave covector is $k^A_\mu=\partial_\mu\theta_A$, and an observer with unit timelike four-velocity $u_{\rm obs}^\mu$ measures
\begin{equation}
 \omega_{\rm obs}=-u_{\rm obs}^\mu k^A_\mu .
 \label{eq:main-observed-packet-frequency}
\end{equation}
This frequency depends on the propagated phase and the observer, independently of how the trajectory is parametrized. For light exchanged between static observers, define $N_{\rm stat}(\bm x)=\sqrt{-g_{tt}(\bm x)}$ relative to a common static time coordinate $t$. Conservation of the ray's Killing energy gives $\omega_{\rm rec}/\omega_{\rm em}=N_{\rm stat}(\bm x_{\rm em})/N_{\rm stat}(\bm x_{\rm rec})$, where ``em'' and ``rec'' denote emission and reception. These static observers need not follow geodesics. In an FRW universe, comoving observers instead have $d\tau_{\rm com}/dt=1$, but a freely propagating massless component has $\omega(t)=|\bm k|/a(t)$ in the geometrical-optics limit, with $\bm k$ its conserved comoving wave vector and $a(t)$ the scale factor. Cosmological redshift therefore arises from profile propagation through the expanding geometry, even though the comoving clock lapse is unity \cite{HarveySchuckingSurowitz2006}. A null ray itself has no proper-time clock: massless localization uses spacelike sections with a specified timelike observer normal, supplied, for example, by the cosmological fluid without requiring a detector. For a split packet, recombination probes the phase of $\langle\psi|U_2^\dagger U_1|\psi\rangle$ when the overlap is nonzero; $U_1,U_2$ are the arm propagators in a common preparation and readout convention, including applied phases. The observed phase and free fall constrain forward propagation \cite{ColellaOverhauserWerner1975,Asenbaum2017,Asenbaum2020}; they do not measure what fraction of the packet's energy produces a gravitational field.

\subsection{The profile and its geometrical residual}
\label{subsec:profile-comparison}

Start with the scalar matrix element in Eq.~\eqref{eq:main-scalar-state-profile}. Inside a free fiber its operator satisfies the Minkowski Klein--Gordon equation. Across spacetime the prescribed \emph{profile} evolves by
\begin{equation}
 (\Box_g-m_R^2)\Phi_A=0.
 \label{eq:effective-KG-new}
\end{equation}
This equation propagates the initial finite profile with the classical metric; it does not put curvature vertices into the tangent quantum action. Both the scalar value and its normal derivative are needed as initial data. For a complex field their complete canonical description retains the conjugate field and conjugate momentum as well. Appendix~\ref{app:profile-foundations} gives the canonical transport test, the explicit scalar generators and the corresponding representation-connection requirements. The nonlinear quartic profile equation is an effective interacting scalar example; it does not determine the full hierarchy of interacting correlations.

For clarity, on an oriented spacelike cross-section $\Sigma_A$ with future unit normal $n$ and induced volume $d\Sigma$, use densitized scalar momenta $P_\varphi=\sqrt h\,n^\mu\nabla_\mu\Phi_A^*$ and $P_{\varphi^*}=\sqrt h\,n^\mu\nabla_\mu\Phi_A$, where $h$ is the determinant of the induced spatial metric, so $d\Sigma=\sqrt h\,d^3x$. This is also the canonical convention in the cosmological case. The full complex data and their canonical pairing are
\begin{align}
 \mathbf Z&=(\varphi,\varphi^*,P_\varphi,P_{\varphi^*})^{\mathsf T},
 \qquad \varphi=\Phi_A|_{\Sigma_A},
 \label{eq:main-complex-data}\\
 \Omega_{\mathbb C}(\delta_1\mathbf Z,\delta_2\mathbf Z)
 &=\int_{\Sigma_A}d^3x\,
 \bigl[\delta_1\varphi\,\delta_2P_\varphi
       -\delta_2\varphi\,\delta_1P_\varphi
       +\delta_1\varphi^*\,\delta_2P_{\varphi^*}
       -\delta_2\varphi^*\,\delta_1P_{\varphi^*}\bigr].
 \label{eq:main-complex-symplectic}
\end{align}
Here $\Omega_{\mathbb C}$ is the canonical symplectic form on the space of complex-scalar Cauchy data. It is a bilinear, antisymmetric pairing of two infinitesimal variations $\delta_1\mathbf Z$ and $\delta_2\mathbf Z$ of the field and its conjugate momenta. For a single mechanical coordinate $q$ with conjugate momentum $p$, the corresponding expression is $\Omega(\delta_1,\delta_2)=\delta_1q\,\delta_2p-\delta_2q\,\delta_1p$; Eq.~\eqref{eq:main-complex-symplectic} extends this pairing to a field by integrating over the spatial section and including both complex-conjugate canonical pairs. The subscript $\mathbb C$ identifies the complex-field description, while the form itself is real when the variations respect the complex-conjugacy conditions. Physically, it specifies which field and momentum variables are canonically conjugate and determines their Poisson brackets. Requiring profile transport to preserve this form ensures that it preserves the canonical structure underlying the quantum commutation relations.
The superscript $\mathsf T$ forms a column, the star denotes complex conjugation, and $\delta_1\mathbf Z,\delta_2\mathbf Z$ are two independent infinitesimal variations of the initial data, satisfying the conjugacy conditions. A symplectic form measures the oriented canonical pairing of two variations; hence two arguments are required. It is real on the physical complex-field phase space. The admitted space $\mathscr C_{A,s}$ contains data with the required regularity, finite moments and spatial support or sufficiently rapid decay. Decay means that tails decrease fast enough for the specified spatial integrals to converge; it does not require exactly compact support. In the free theory, preservation of Eq.~\eqref{eq:main-complex-symplectic} is the classical counterpart of preserving the canonical commutation relations.
In practice, we use $\Omega_{\mathbb C}$ to test the proposed profile propagators: both the geometrical evolution and its Minkowski reference must preserve this canonical pairing, as expressed in Eq.~\eqref{eq:scalar-symplectic-preservation}. Their residual comparison then preserves the same structure, supplying the canonical-commutation requirement for the free-field observable transport constructed in Subsection~\ref{subsec:observable-transport}; unitary implementation requires the additional conditions stated there.

We now compare two evolutions of the same initial profile: propagation determined by the classical geometry and the reference propagation supplied by the tangent-Minkowski theory. Both must be expressed in the same canonical variables and with the same endpoint identifications. Denoting their propagators by $\mathscr T_A^{\rm min}$ and $\mathscr T_A^{\rm M}$, respectively, we define their residual comparison by
\begin{equation}
 \mathfrak P_{I,A}[g;s_2,s_1]
 :=[\mathscr T_A^{\rm M}(s_2,s_1)]^{-1}
 \mathscr T_A^{\rm min}[g;s_2,s_1],
 \qquad
 \mathscr T_A^{\rm min}
 =\mathscr T_A^{\rm M}\mathfrak P_{I,A}.
 \label{eq:P-flat-limit}
\end{equation}
The superscript ``min'' specifies minimal geometrical propagation: the metric enters through the geometrical structures required by the field representation, without independently adjustable nonminimal curvature couplings. For the free complex scalar, this is the profile propagation prescribed by Eq.~\eqref{eq:effective-KG-new}. The superscript ``M'' identifies the corresponding Minkowski reference evolution, including the ordinary oscillation and spreading already present in flat spacetime.
The arguments specify the metric and the initial and final cross-sections at labels $s_1<s_2$. The inverse reference map brings the result back to the initial data space, so $\mathfrak P_{I,A}$ acts in that reference space. The subscript $I$ denotes interaction picture.

Physically it is the extra deformation left after accounting for the propagation that the Minkowski theory already supplies. In an inertial flat geometry it is the identity, while ordinary field oscillation, spreading and interactions continue.

For the scalar example, a localized massive packet spreads even in flat space. In an expanding universe it also redshifts and changes its physical width. Adding a full curved profile evolution to an already complete Minkowski evolution would count the flat spreading twice. Equation~\eqref{eq:P-flat-limit} prevents that duplication. At the generator level the subtraction must be conjugated by the reference propagation before being integrated. Its ordered exponential accumulates successive, generally noncommuting deformations. Appendix~\ref{app:profile-foundations} gives the composition rule and explicit complex-scalar matrices showing why the conjugation matters.

In a fixed common data chart, let $\mathscr L^{\rm min}_{A,s}$ and $\mathscr L^{\rm M}_{A,s}$ be the respective linear profile generators, defined by $\partial_s\mathbf Z=\mathscr L_{A,s}\mathbf Z$. Their difference and its exact interaction-picture form are
\begin{align}
 \Delta\mathscr L^{\rm geo}_{A,s}
 &=\mathscr L^{\rm min}_{A,s}-\mathscr L^{\rm M}_{A,s},
 \qquad
 \mathscr G^{\rm geo}_{I,A,s}
 =[\mathscr T_A^{\rm M}(s,s_1)]^{-1}
 \Delta\mathscr L^{\rm geo}_{A,s}\mathscr T_A^{\rm M}(s,s_1),
 \label{eq:main-residual-generator}\\
 \mathfrak P_{I,A}(s_2,s_1)
 &=\mathcal P\exp\!\left[\int_{s_1}^{s_2}
 \mathscr G^{\rm geo}_{I,A,s}\,ds\right].
 \label{eq:main-residual-exponential}
\end{align}
The ordering symbol $\mathcal P$ puts later factors to the left. The rightmost reference map advances the data to the current section, the difference acts there, and the inverse returns the change to the initial section. The exponential composes these infinitesimal changes. It has no $-i$ because its generator acts on classical canonical data, rather than on a quantum state. Without the conjugation, the result is generally wrong when the generators do not commute. If endpoint coordinates are changed in time, both generators must be transformed with the same time-dependent identification. Appendix~\ref{app:scalar-generators} displays the complete scalar example, including the fixed comoving chart used for the subtraction.

\subsection{Observables, unitary evolution and the admission tests}
\label{subsec:observable-transport}

The profile evolution constructed in Subsection~\ref{subsec:profile-comparison} describes how the packet's envelope and phase change. To describe quantum propagation, we must also specify how the field operators and the quantum state are transported, so that energy, fluctuations and correlations can be calculated consistently. The construction therefore proceeds from the profile map to a transformation of field operators and then, when possible, to a unitary transformation of states. The mathematical distinction between these steps is standard in the algebraic formulation of quantum field theory \cite{Haag1996,BaerGinouxPfaeffle2007}.

To understand the first step physically, recall that a quantum field is sampled with a smooth weighting function rather than evaluated as an ordinary operator at one mathematical point. The notation $\widehat\Phi_R[f]$ denotes such a smeared field, with $f$ specifying its spatial and temporal weighting. This weighting can describe a localization prescription without requiring a detector. In the canonical description, the test object may contain several weighting functions that sample the field, its conjugate and their momenta. Write $\langle f,Z\rangle$ for this weighted linear sampling of the Cauchy data $Z$. If the profile is transported by $\mathfrak P_{I,A}$, the corresponding response must be transported consistently. Its transformed version, denoted by $\mathfrak P_{I,A}^{\vee}f$, is defined by
\begin{equation}
 \langle\mathfrak P_{I,A}^{\vee}f,\mathfrak P_{I,A}Z\rangle
 =\langle f,Z\rangle,
 \qquad
 \mathfrak P_{I,A}^{\vee}
 =(\mathfrak P_{I,A}^{-1})'.
 \label{eq:main-dual-transport}
\end{equation}
The prime denotes the dual action on test objects: the transformed response first undoes the profile transformation and then applies the original weighting. Thus Eq.~\eqref{eq:main-dual-transport} compares a transported profile with its correspondingly transported response. It does not assert that a response held fixed would register no change in the propagating packet. Test functions that give the same smeared field by virtue of the field equation are identified.

For a free scalar field, preservation of the canonical form $\Omega_{\mathbb C}$ allows this rule to extend consistently to the field algebra \cite{BaerGinouxPfaeffle2007}. An algebra here means the collection of field quantities together with the rules for adding and multiplying them and taking their adjoints. Denote the induced transformation by $\alpha^{\rm geo}_{I,A}$. It is called a $*$-isomorphism because it is invertible and preserves these operations, including the adjoint denoted by $*$. Consequently, it preserves the commutation relations and transforms products of fields consistently, as required when calculating correlations.

The next question is whether this operator transformation can be represented by a unitary operator $\mathcal V_{I,A}^{\rm geo}$ on the chosen reference Hilbert space. This is the meaning of a \emph{unitary lift}: it realizes the already specified transformation of field quantities through an operator acting on quantum states. Let $O$ be an abstract observable and let $\pi_{z_A(s_1)}(O)$ be its operator representation in the initial reference fiber. The field transformation and its unitary implementation then satisfy
\begin{equation}
 \begin{aligned}
 \alpha^{\rm geo}_{I,A}\bigl(\widehat\Phi_R[f]\bigr)
 &=\widehat\Phi_R[\mathfrak P_{I,A}^{\vee}f],\\
 \pi_{z_A(s_1)}\bigl(\alpha^{\rm geo}_{I,A}(O)\bigr)
 &=\mathcal V_{I,A}^{\rm geo}
   \pi_{z_A(s_1)}(O)
   (\mathcal V_{I,A}^{\rm geo})^\dagger .
 \end{aligned}
 \label{eq:main-algebra-lift}
\end{equation}
Here $\dagger$ denotes the Hilbert-space adjoint, and unitarity means $(\mathcal V_{I,A}^{\rm geo})^\dagger=(\mathcal V_{I,A}^{\rm geo})^{-1}$. For a complex scalar, the conjugate field must be transported as well: the complex field operator alone is not self-adjoint, while physical observables are constructed from the fields with the required Hermiticity and symmetry properties. Preservation of the field algebra does not automatically guarantee unitary implementation in a chosen Hilbert representation; this is a separate condition, particularly for systems with infinitely many degrees of freedom \cite{Haag1996,AgulloAshtekar2015}. For interacting fields, the transport must additionally respect the renormalized composite operators, field equations and correlations; transporting one scalar profile does not establish these properties.

If that lift exists, denote it by $\mathcal V_{I,A}^{\rm geo}$. The complete quantum propagator is
\begin{equation}
 \mathcal U_A[g;s_2,s_1]
 =\mathcal U_A^{\rm M}(s_2,s_1)\mathcal V_{I,A}^{\rm geo}[g;s_2,s_1],
 \qquad
 \rho_A(s_2)=\mathcal U_A\rho_A(s_1)\mathcal U_A^\dagger.
 \label{eq:main-quantum-transport}
\end{equation}
Here $\rho_A$ is the packet density operator and $\mathcal U_A^{\rm M}$ supplies the full renormalized tangent quantum evolution. The residual lift acts in the initial reference fiber; the product maps to the final fiber. 
With the initial endpoint $s_1$ held fixed, the two connection representations are determined by the lift:
\begin{equation}
 \mathcal A_{I,A,s}^{\rm geo}
 =-(\partial_s\mathcal V_{I,A}^{\rm geo})(\mathcal V_{I,A}^{\rm geo})^{-1},
 \qquad
 \mathcal A_{{\rm Sch},A,s}^{\rm geo}
 =\mathcal U_A^{\rm M}\mathcal A_{I,A,s}^{\rm geo}(\mathcal U_A^{\rm M})^{-1}.
 \label{eq:main-connection-from-lift}
\end{equation}
Every factor on the right uses the same interval $(s,s_1)$. Thus the Schr\"odinger-picture connection is the interaction-picture connection carried forward by the Minkowski quantum evolution. It is fixed by the admitted geometrical lift, including its specified phase convention; it is not a separately selectable Hamiltonian.

Equivalently the state obeys
\begin{equation}
 i\left(\frac{d}{ds}+\mathcal A_{{\rm Sch},A,s}^{\rm geo}\right)
 |\Psi_A(s)\rangle
 =\widehat K_{A,R}^{\rm M}(s)|\Psi_A(s)\rangle,
 \qquad
 (\mathcal A_{{\rm Sch},A,s}^{\rm geo})^\dagger
 =-\mathcal A_{{\rm Sch},A,s}^{\rm geo}.
 \label{eq:bundle-evolution}
\end{equation}
Here $|\Psi_A(s)\rangle$ is a state vector in $\F_{z_A(s)}$, and the dagger denotes the Hilbert-space adjoint. The generator $\widehat K_{A,R}^{\rm M}$ is determined by the physical cross-section deformation. It equals $N_A\widehat H_R^{\rm M}$ only for pure translation of parallel rest hyperplanes, without shift, rotation or a more general deformation. In that case a scalar energy component acquires the familiar phase $-\int E\,d\tau_A$, where $E$ is the energy measured in that local frame. The connection accounts for geometrical comparison and finite deformation; the Minkowski generator supplies the quantum dynamics once.

An explicit example of both terms in
Eq.~\eqref{eq:bundle-evolution} is supplied by the free complex
scalar, with self-interaction $\lambda_R=0$ for simplicity, in a spatially flat FRW background.
For this illustration, use the fixed comoving canonical comparison
of Appendix~\ref{app:scalar-generators}, take $s=t$, and let the
Minkowski reference evolution translate parallel tangent rest
hyperplanes in proper time, so that $N_A=1$. To display the
unitary lift explicitly, first retain finitely many normalized
Fourier components in a periodic comoving box. The Fourier label
$\bm k$ is a computational label and is distinct from the
localization-outcome label $A$.

For one retained component, let $\widehat q_{\bm k}$ be its
complex field-amplitude operator and $\widehat p_{\bm k}$ its
canonically conjugate momentum. Their classical counterparts
satisfy $p_{\bm k}=a^3\dot q_{\bm k}^{\,*}$, and their canonical
commutators are
$[\widehat q_{\bm k},\widehat p_{\bm k}]
=[\widehat q_{\bm k}^\dagger,\widehat p_{\bm k}^\dagger]=i$,
with the other independent commutators vanishing.
Writing $\omega_{\bm k}^R=\sqrt{m_R^2+|\bm k|^2}$, the two
generators are
\begin{align}
 \widehat K_{\bm k,R}^{\rm M}
 &=
 :\!\left[
 \widehat p_{\bm k}^\dagger\widehat p_{\bm k}
 +(\omega_{\bm k}^R)^2
  \widehat q_{\bm k}^\dagger\widehat q_{\bm k}
 \right]\!:\!_{\rm M},
 \label{eq:scalar-finite-mode-K}
 \\
 \mathcal A_{{\rm Sch},\bm k,t}^{\rm geo}
 &=
 i:\!\left[
 (a^{-3}-1)\widehat p_{\bm k}^\dagger\widehat p_{\bm k}
 +\bigl(m_R^2(a^3-1)+|\bm k|^2(a-1)\bigr)
  \widehat q_{\bm k}^\dagger\widehat q_{\bm k}
 \right]\!:\!_{\rm M}.
 \label{eq:scalar-finite-mode-A}
\end{align}
Here $a=a(t)$ is dimensionless and normalized by $a(t_0)=1$,
and $:\cdots:_{\rm M}$ denotes normal ordering with respect to
the same fixed Minkowski reference vacuum. This convention fixes
the scalar phase of the finite-mode implementation consistently.
The operators act on the Minkowski oscillator representation;
the geometrical coefficients enter through the prescribed
canonical transport.

Equation~\eqref{eq:scalar-finite-mode-A} is the unitary lift of
the quadratic canonical-generator difference calculated in
Appendix~\ref{app:scalar-generators}. Indeed, the full generator
in Eq.~\eqref{eq:bundle-evolution} becomes
\begin{equation}
 \widehat K_{\bm k,R}^{\rm M}
 -i\mathcal A_{{\rm Sch},\bm k,t}^{\rm geo}
 =
 :\!\left[
 a^{-3}\widehat p_{\bm k}^\dagger\widehat p_{\bm k}
 +(a^3m_R^2+a|\bm k|^2)
  \widehat q_{\bm k}^\dagger\widehat q_{\bm k}
 \right]\!:\!_{\rm M},
 \label{eq:scalar-finite-mode-total}
\end{equation}
whose commutator equations reproduce the free complex canonical
profile evolution in
Eq.~\eqref{eq:FLRW-curved-canonical-generator}.
The connection is anti-Hermitian because the quadratic expression
multiplying $i$ is self-adjoint. Its interaction-picture form is
obtained by the conjugation in
Eq.~\eqref{eq:main-connection-from-lift}.

Physically, $\widehat K_{\bm k,R}^{\rm M}$ supplies the ordinary
Minkowski oscillation, while
$\mathcal A_{{\rm Sch},\bm k,t}^{\rm geo}$ supplies the additional
canonical deformation associated with the expanding geometry.
In the inertial reference geometry $a(t)\equiv1$, the connection
vanishes and the Minkowski evolution remains. For smooth positive
$a(t)$, the finite quadratic system has a well-defined unitary
evolution \cite{CombescureRobert2006}. This realizes both terms of Eq.~\eqref{eq:bundle-evolution} with a finite cutoff. Appendix~\ref{app:transport-tests} shows why this particular calibration fails as the cutoff is removed in a massless radiation example, and exhibits an alternative with an exact continuum lift.

The differential equation can be integrated as
\begin{equation}
 \mathcal U_A[g;s_2,s_1]
 =\mathcal P\exp\!\left[-\int_{s_1}^{s_2}
 \bigl(\mathcal A_{{\rm Sch},A,s}^{\rm geo}
       +i\widehat K_{A,R}^{\rm M}(s)\bigr)ds\right].
 \label{eq:main-full-ordered-propagator}
\end{equation}
On an admitted common domain the total generator is
$\widehat G_{A,s}=\widehat K_{A,R}^{\rm M}(s)-i\mathcal A_{{\rm Sch},A,s}^{\rm geo}$.
Its self-adjointness, suitable time dependence and a stable domain yield a unitary evolution; anti-Hermiticity of the connection then gives preservation of the norm. Self-adjoint summands alone do not establish self-adjointness of their unbounded sum. These are the conditions made explicit in Appendix~\ref{app:unitary-foundations}, including composition and the phase convention for the lift. For a scalar packet, the same $\mathcal U_A$ propagates its state and all observables used for energy, width and stress, rather than assigning a different propagation law to each quantity.

The algebra map and its implementation have separate existence requirements. For each admitted interval the representations must admit a unitary implementing the specified map, the implementers must form a strongly continuous evolution, and the differentiable generators must share a suitable stable dense domain. 

Preservation of the canonical pairing establishes the classical canonical consistency of the profile transport. Its realization as quantum evolution additionally requires the unitary-implementation, continuity and common-domain conditions stated in Appendix~\ref{app:unitary-foundations}. When these conditions are satisfied, the ordered exponential represents the resulting well-defined unitary propagator.
Appendix~\ref{app:transport-tests} adds preservation of the finite energy and stress moments and of the short-distance correlations. An infinite homogeneous system requires a specified local algebraic formulation or a justified finite-volume limit; it need not have a trace-class density matrix or one global Fock representation.

For the complex scalar, a smooth low-energy packet illustrates the distinction. Its classical profile can solve Eq.~\eqref{eq:effective-KG-new}, while a proposed transformation of its full quantum state can still fail to preserve the domain of the stress tensor. The scalar matrix element alone is therefore insufficient. The transported field, its conjugate, and the composite observables entering the source must satisfy the same quantum admission requirements. These requirements concern whether the proposed transport exists; the metric-comparison rule of Sec.~\ref{sec:metric-comparison} subsequently determines which part of its assigned energy gravitates.

Quantum transport must preserve the short-distance structure needed
by the tangent-Minkowski renormalization prescription. The proposed
two-point condition is
\begin{equation}
 W_{2,x}^{\rm tr}(X,Y)-W_2^{\rm M}(X,Y)\in C^\infty,
 \label{eq:main-tangent-smoothness}
\end{equation}
where $W_{2,x}^{\rm tr}$ and $W_2^{\rm M}$ are respectively the
transported and reference two-point distributions, expressed in the
same tangent-Minkowski fiber at the endpoint $x$. The arguments $X,Y$
are points in that tangent spacetime, and $C^\infty$ means that their
difference is smooth in both arguments.

Physically, this condition permits finite changes of quantum
correlations while preserving the reference short-distance
singularity. For a free complex scalar, it gives a finite change in
the renormalized mean stress tensor, because the relevant derivatives
of the smooth difference have finite coincidence limits. Finite
stress fluctuations and interacting composite observables require
the additional conditions discussed in
Appendix~\ref{app:transport-tests}. Equation~\eqref{eq:main-tangent-smoothness}
is a proposed compatibility requirement whose preservation must be
established for the admitted transport. The comparison with
conventional curved-spacetime Hadamard singularities is explained in
Appendix~\ref{subsec:endpoint-short-distance}. This quantum consistency
requirement is separate from the subsequent metric-comparison rule
that determines the gravitating energy.

\section{A physical POVM decomposition and positive energy shares}
\label{sec:povm}

The third step specifies the localization observable. A positive-operator-valued measure (POVM) assigns positive operators to measurable spatial outcomes and permits smooth, overlapping responses. It supplies probabilities without requiring sharp position eigenstates. A detector can realize such an observable, but a detector is not necessary for its mathematical definition or for the proposed gravitational rule in this study. In a cosmological application the initial state, physical rest frame and localization prescription are part of the model's initial data. Homogeneity and isotropy restrict the prescription to an appropriate covariant, isotropic family; they do not determine its width uniquely.
For example, a Gaussian width $\ell_{b,i}$ for bosons or $\ell_{f,i}$ for fermions can be specified at an initial epoch. Their subsequent evolution must follow the stated prescription and field dynamics. 
Appendix~\ref{app:localization} adds more details.

\subsection{Localization and conditional quantum data}

We describe localization using a normalized positive-operator-valued
measure (POVM), the standard generalization of a projective quantum
observable that permits effects which are not orthogonal projections
\cite{Busch2016QuantumMeasurement}. Let $w_A(X)$ be nonnegative
smooth responses on a tangent spacelike surface, normalized by
$\sum_A w_A(X)=1$, and let $\POVM$ denote the normalized spatial
POVM. The cell effects, probabilities, and conditional states
associated with the chosen square-root instrument are
\begin{equation}
 \Effect_A=\int w_A(X)\POVM(dX),\qquad
 p_A=\Tr(\rho\Effect_A),\qquad
 \rho_A=\frac{\Effect_A^{1/2}\rho\Effect_A^{1/2}}{p_A}
 \quad(p_A>0).
 \label{eq:main-localization}
\end{equation}
Here $\rho$ is the unconditioned tangent state and $A$ labels a localization outcome, not a Fourier oscillator or a fixed particle-number sector. A continuous outcome family replaces the sum by an integral with its specified measure. The effects resolve the identity on the localization sector; a separate ``no excitation'' vacuum outcome, when needed, completes it and carries zero energy.

The conditional state in Eq.~\eqref{eq:main-localization} is obtained
from the generalized L\"uders, or square-root, instrument
\cite{BuschSingh1998,Busch2016QuantumMeasurement}.
When $\Effect_A$ is an orthogonal projector, its positive square
root equals itself and the expression reduces to the original
L\"uders projection rule \cite{Luders2006}. For general POVM
effects, the square-root instrument is an additional prescription:
the POVM determines outcome probabilities but does not uniquely
determine the conditional state. Here this prescription defines
the conditional quantum data used in the construction; its use
does not assert that an irreversible measurement has occurred.

Covariance makes this prescription independent of its inertial description. For a Poincar\'e transformation $\mathsf g=(a,\Lambda)$, with translation $a$ and Lorentz transformation $\Lambda$, let $U_x(\mathsf g)$ be its unitary tangent representation. For a measurable region $\Delta$ of a tangent spacelike surface $\Sigma$,
\begin{equation}
 U_x(\mathsf g)\POVM_\Sigma(\Delta)U_x(\mathsf g)^\dagger
 =\POVM_{\mathsf g\Sigma}(\mathsf g\Delta),
 \qquad
 \Instr_A(\rho)=\Effect_A^{1/2}\rho\Effect_A^{1/2}.
 \label{eq:main-POVM-covariance}
\end{equation}
Here $\mathsf g$ denotes the group element, distinct from the metric $g$. The unnormalized output $\Instr_A(\rho)$ has trace $p_A$. Under the same change of inertial description the response transforms as $w_{\mathsf g A}(\mathsf g X)=w_A(X)$, so the transformed effect and its positive square root are the unitary conjugates of the original ones. Transforming the state, hyperplane and observer together therefore gives the same outcome probabilities and the corresponding transformed conditional states. The instrument used for widths and stress must preserve the domain in Eq.~\eqref{eq:main-state-moments}.

Let $J$ be a group of outcome labels of the fixed physical instrument. If only membership in $J$ is reported, its instrument operation is $\Instr_J(\rho):=\sum_{A\in J}\Instr_A(\rho)$. A new square-root update using the combined effect $\sum_{A\in J}\Effect_A$ can produce a different conditional state. Reporting several labels together therefore does not prescribe a second localization of the field.

A complex-scalar packet can overlap two Gaussian responses. Their effects assign two positive shares of its energy, while their conditional states have separately calculable centroids, energy distributions and momentum covariances. In a many-particle state, an energy-weighted effect can allocate energy near one cell while its square-root update leaves distant excitations in the conditional state. Thus a cell's energy profile and the position distribution of its entire conditional state must not be identified without a calculation.

If one needs an operational statement that the conditional state is concentrated in a specified region $\mathcal R_A^{\rm loc}$, it can be tested by
\begin{equation}
 \Tr\!\left[\rho_A\POVM(\mathcal R_A^{\rm loc})\right]
 \ge 1-\epsilon_A^{\rm loc},\qquad 0<\epsilon_A^{\rm loc}<\tfrac12.
 \label{eq:localization-concentration-admission}
\end{equation}
The region and allowed leakage $\epsilon_A^{\rm loc}$ are prescribed before the test. This is a diagnostic of the chosen conditional localization, not a gravitational eligibility condition. The metric-comparison rule acts directly on the positive assigned energy measure defined in Sec.~\ref{sec:metric-comparison}, including its tails, and remains meaningful when a particular conditional state is not concentrated in one cell.

At each endpoint, first specify the unconditioned field state from the admitted field evolution and the complete localization POVM in one representation and energy frame. Compute every $p_A$ and $\rho_A$ there by Eq.~\eqref{eq:main-localization}. If a packet lift is used to propagate an earlier conditional state separately, it describes the same outcome only when its result agrees with this endpoint conditioning, wherever both are defined. The probabilities and energy shares are evaluated anew. Independently transported cell effects need not sum to the identity, and the mixture $\sum_A p_A\rho_A$ is the state after the unrecorded instrument, which need not equal $\rho$. Neither procedure constructs the unconditioned state. Appendix~\ref{app:localization} gives the energy-form and multiparticle details.

\subsection{Why the conditional mean energy is not the source share}

The form domain $D(\widehat H_R^{1/2})$ consists of vectors with finite mean energy. For a normalized vector $\psi$ in this domain, define
\begin{equation}
 q_A[\psi]=\|\Effect_A^{1/2}\widehat H_R^{1/2}\psi\|^2,
 \qquad
 \sum_Aq_A[\psi]=\|\widehat H_R^{1/2}\psi\|^2.
 \label{eq:main-energy-form}
\end{equation}
This is a quadratic form: $q_A[c\psi]=|c|^2q_A[\psi]$ for a complex number $c$. Its value is assigned energy, not probability. Since $0\le\Effect_A\le\I$, it is positive and no larger than the total energy. If it is closable, its specified closure defines a positive self-adjoint operator $\widehat{\mathcal H}_A$. The form definition avoids multiplying unbounded operators on an unspecified domain.

For a finite-energy density operator $\rho=\sum_n r_n|\psi_n\rangle\langle\psi_n|$, with $r_n\ge0$ and $\sum_nr_n=1$, every eigenvector of positive weight lies in the form domain. Averaging the pure-state forms gives $\mathcal E_A=\sum_nr_nq_A[\psi_n]$, or equivalently
\begin{equation}
 \mathcal E_A=\Tr\!\left[
 \widehat H_R^{1/2}\rho\widehat H_R^{1/2}\Effect_A\right]\ge0,
 \qquad \sum_A\mathcal E_A=\Tr(\rho\widehat H_R).
 \label{eq:main-energy-partition}
\end{equation}
The trace is defined by the positive energy form as detailed in Appendix~\ref{app:localization}. This is the density-operator version of Eq.~\eqref{eq:main-energy-form}; no second energy prescription is introduced. The statistical expectation within one quantum state is distinct from combining sources: contributions from different cells add, with no division by their number.

The conditional mean $\bar E_A=\Tr(\rho_A\widehat H_R)$ describes the energy distribution after the chosen conditioning. Generally $\mathcal E_A\ne p_A\bar E_A$ because the Hamiltonian and localization effect need not commute. For $\Effect_1=\Effect_2=\I/2$, both conditional states equal $\rho$, but each assigned share equals half the total energy. Adding the conditional means would double count it. Consequently $\mathcal E_A$ must not be multiplied by $p_A$ a second time. It is possible to have $\mathcal E_A>0$ but $p_A=0$: the effect can annihilate $\psi$ without annihilating $\widehat H_R^{1/2}\psi$. The positive share remains defined. Any reconstruction that needs $\rho_A$ applies only where every such energy-bearing cell has $p_A>0$; otherwise that reconstruction is unavailable, and the cell's energy cannot be set to zero.

A transparent free-scalar realization uses the one-particle profile $\Phi_f$ and its positive classical energy form $\varepsilon_f=|\partial_{X^0}\Phi_f|^2+|\bm\nabla\Phi_f|^2+m_R^2|\Phi_f|^2$. With the corresponding energy-based localization POVM, the assigned share is $\mathcal E_A=\int w_A\varepsilon_f\,d^3X$. If two responses are $0.4$ and $0.6$ in their overlap, they allocate those fractions of its energy exactly once. This free-field example does not assert that a renormalized QFT stress density is pointwise positive in arbitrary states.

For precision, let $\Omega=\sqrt{m_R^2+\bm p^2}$ be the positive one-particle energy and define the form $\mathsf K_A$ by $\langle f,\mathsf K_Af\rangle=\int w_A\varepsilon_f\,d^3X$, using the consistent packet normalization of Appendix~\ref{app:bundle-foundations}. In the massive sector,
\begin{equation}
 \Effect_A^{(1)}=\Omega^{-1/2}\mathsf K_A\Omega^{-1/2},
 \qquad \sum_A\mathsf K_A=\Omega,
 \qquad \sum_A\Effect_A^{(1)}=\I.
 \label{eq:main-scalar-energy-POVM}
\end{equation}
The superscript $(1)$ denotes the one-particle sector. The energy sandwich of Eq.~\eqref{eq:main-energy-partition} then returns $\mathsf K_A$. Inverse powers in a massless sector need a form-domain construction or a specified spectral restriction; they cannot simply be treated as bounded operators. Appendix~\ref{app:povm-fock} treats the bosonic and fermionic sums, the complex scalar's antiparticles and the vacuum. The interacting quartic energy must likewise be allocated once by an admitted interacting prescription. The free formula illustrates this requirement without proving the interacting extension.

This energy-density localization is one physical choice. Another legitimate fixed-frame prescription, used for the explicit thermal calculation below, starts from a position observable $Q$ and the one-particle Hamiltonian $h$ and assigns the positive form $h^{1/2}w_A(Q)h^{1/2}$. Here $h$ represents the same one-particle energy in the chosen position representation, and $w_A(Q)$ is the response function applied to that position observable. The two forms need not coincide: local scalar stress and an energy-weighted position observable define different localization measurements. Their respective positive POVMs obey the same exact energy-partition principle, but a numerical gravitating fraction must use one specified prescription consistently. The thermal example states that choice explicitly; it is not obtained by silently replacing the local stress-density observable.

\section{Uncertainties, finite profiles and their worldtubes}
\label{sec:worldtube}

The next step describes the size and evolution of each mode. A finite mode remains a state of a tangent Minkowski theory; its worldtube records the transported profile and its uncertainties. It does not carry a separate curved classical metric. The gravitating part will be obtained from the full metric comparison of Sec.~\ref{sec:metric-comparison}. The worldtube and its covariance are useful summaries of the profile, but an exact energy fraction requires the full assigned energy measure, including tails.

\subsection{Spatial shape and quantum spreading}
The purpose of this subsection is to determine the spatial extent
and evolution of each localized quantum mode from its conditional
state and the prescribed POVM. This information connects the
localization construction of Sec.~\ref{sec:povm} with the finite
profiles used in the subsequent metric comparison. The central
trajectory specifies the reference motion, while position and
momentum uncertainties, together with their correlations, determine
how the packet spreads or focuses around that trajectory. For
example, two nonrelativistic complex-scalar packets with the same
initial position width can develop different later widths if their
momentum spreads or position--momentum correlations differ.
We therefore calculate the centroid and covariance, distinguish
quantum fluctuations from the resolution of the localization
observable, and require their evolution to respect the quantum
uncertainty relations. These quantities define the spatial contours
used to describe the mode's worldtube. Their physical relevance
to gravitation is that spreading or focusing can change how the
assigned energy overlaps the region selected by the full metric
criterion of Sec.~\ref{sec:metric-comparison}, thereby changing
the gravitating fraction. The exact fraction must nevertheless
be calculated from the full assigned energy measure, including
its tails; the position covariance and its associated contour
provide a spatial characterization rather than a complete energy
distribution.

Let $d\nu_A(X)=\Tr[\rho_A\POVM(dX)]$ be the position-outcome probability measure of a conditional state. This probability measure is distinct from the assigned energy measure $d\mu_A$ introduced below. Define its centroid and covariance by
\begin{equation}
 \bar X_A^i=\int X^i\,d\nu_A(X),\qquad
 (C_{X,A})^{ij}=\int(X^i-\bar X_A^i)(X^j-\bar X_A^j)\,d\nu_A(X).
 \label{eq:main-position-moments}
\end{equation}
The outcome measure includes the intrinsic resolution of the selected POVM; an additional apparatus resolution is included only when there is an apparatus. A strictly positive covariance in the resolved directions defines the contour
\begin{equation}
 \mathcal B_A(s)=\{X:\xi_A^i(C_{X,A}^{-1})_{ij}\xi_A^j\le q_X^2\},
 \qquad \xi_A^i=X^i-\bar X_A^i,
 \label{eq:ellipsoid-new}
\end{equation}
where $q_X>0$ is a fixed contour number. Its semiaxes are $q_X\sqrt{\lambda_i(C_{X,A})}$, with $\lambda_i$ the covariance eigenvalues. For a Gaussian it is a familiar probability contour. For a general state its enclosed probability must be calculated; its tails are never discarded merely because they lie outside this contour.

Position width alone does not fix its future value. For a nonrelativistic, initially uncorrelated minimum-uncertainty complex-scalar packet in flat space,
\begin{equation}
 \sigma_{\rm M}^2(t)=\sigma_0^2+
 \frac{(t-t_1)^2}{4m_R^2\sigma_0^2},
 \label{eq:Minkowski-Gaussian-spreading}
\end{equation}
where $\sigma_0$ is the initial intrinsic position standard deviation and $t_1$ the preparation time. The second term is quantum spreading caused by the momentum uncertainty. A position-readout variance is included separately and only once. In curved geometry the full covariance also carries focusing and correlations; a central geodesic alone cannot determine these quantities.

To state the uncertainty condition, let $\mathbf R_A=(X_A^1,X_A^2,X_A^3,P_{A,1},P_{A,2},P_{A,3})^{\mathsf T}$ collect the self-adjoint first-moment position and momentum operators in the chosen sector, and set $\Delta\mathbf R_{A,I}=\mathbf R_{A,I}-\Tr(\rho_A\mathbf R_{A,I})\I$. With $I,J$ indexing these six components, define
\begin{equation}
 (\mathbf C_A^{\rm op})_{IJ}
 =\tfrac12\Tr[\rho_A\{\Delta\mathbf R_{A,I},\Delta\mathbf R_{A,J}\}],
 \qquad
 (\boldsymbol\Omega_A)_{IJ}
 =-i\Tr[\rho_A[\mathbf R_{A,I},\mathbf R_{A,J}]].
 \label{eq:main-intrinsic-covariance}
\end{equation}
Braces denote the anticommutator. If the centered Heisenberg observables evolve exactly linearly through a real matrix $\mathcal M_A=\mathcal M_A(s,s_1)$, then
\begin{align}
 \mathbf C_A^{\rm op}(s)&=\mathcal M_A\mathbf C_A^{\rm op}(s_1)\mathcal M_A^{\mathsf T},
 \qquad C_{X,A}(s)=[\mathbf C_A^{\rm op}(s)]_{XX}+N_{X,A}^{\rm res}(s),
 \label{eq:main-covariance-evolution}\\
 \mathbf C_A^{\rm op}+\frac{i}{2}\boldsymbol\Omega_A&\ge0,
 \qquad \boldsymbol\Omega_A(s)=\mathcal M_A\boldsymbol\Omega_A(s_1)\mathcal M_A^{\mathsf T}.
 \label{eq:main-covariance-uncertainty}
\end{align}
The block $[\cdot]_{XX}$ is the position covariance. The matrix $N_{X,A}^{\rm res}$ is the difference between the actual position-outcome covariance and the first-moment operator covariance. It includes intrinsic POVM unsharpness and any specified detector noise, each once. A state-independent additive-noise form is justified only for a corresponding unbiased realization. The Robertson--Schr\"odinger inequality follows from positivity of $\langle B^\dagger B\rangle$ for every complex linear combination $B$ of the centered observables \cite{Robertson1929,Schrodinger1930}.

For canonical position and momentum,
$[X_A^i,P_{A,j}]=i\delta^i{}_j\I$, while the position--position
and momentum--momentum commutators vanish. Consequently, the
$XX$, $XP$, $PX$, and $PP$ blocks of $\boldsymbol\Omega_A$
are respectively $0$, $\I_3$, $-\I_3$, and $0$, where
$\I_3$ is the three-dimensional identity matrix.
Nonlinear evolution requires the actual higher-moment hierarchy rather than an assumed linear covariance law. Appendix~\ref{app:worldtube-details} supplies the block derivation and domain qualifications.

\subsection{Duration, tails and resolved flux}
The spatial contour defined above tells us where a mode is concentrated at
a given instant. In this subsection we follow that contour in time to
construct the mode's worldtube and evaluate the energy and momentum that
flow across its moving boundary. In cosmology we can do this over
successive intervals of fluid time without introducing a detector. A
finite time window specifies the resolution at which stress fluctuations
are evaluated. A separate question concerns the evolution of the quantum
phases: two complex scalar packets can have the same spatial width but
different energy distributions, so their phases can disperse on different
time scales.

Let $\widehat H^{\rm loc}_{A,R}$ be the renormalized Hamiltonian that
generates evolution in the local observer's time $\varsigma$, and assume
that it is time independent over the interval considered. During an
elapsed time $\delta\varsigma$, a component of energy $E$ acquires the
phase factor $e^{-iE\delta\varsigma}$. The coherence diagnostic identifies
the first time, if any, at which the magnitude of the probability-weighted
sum of these phase factors falls to a prescribed fraction of its initial
value. This time characterizes phase dispersion; it does not set the
mode's lifetime or determine when its energy gravitates. Writing
$\bar E_A^{\rm loc}=\Tr(\rho_A\widehat H^{\rm loc}_{A,R})$ for the mean
energy in the localized state $\rho_A$, we remove the common phase
associated with this mean and define:
\begin{equation}
 \mathcal C_A^{\rm coh}(\delta\varsigma)
 =e^{i\bar E_A^{\rm loc}\delta\varsigma}
 \Tr[\rho_Ae^{-i\widehat H^{\rm loc}_{A,R}\delta\varsigma}],\qquad
 T_A=\inf\{\delta\varsigma>0:|\mathcal C_A^{\rm coh}(\delta\varsigma)|\le\gamma_\star\}.
 \label{eq:main-time-characteristic}
\end{equation}
Here $\varsigma$ is the physical observer time, $0<\gamma_\star<1$ is a fixed diagnostic threshold, and the infimum of an empty set is infinity. An approximately Gaussian energy distribution gives $T_A=\sqrt{2\ln(\gamma_\star^{-1})}/\Delta E_A^{\rm loc}$, with $\Delta E_A^{\rm loc}$ its standard deviation. This diagnostic describes energy-phase dephasing, not irreversible decoherence or a universal time operator. In particular, it is not a delayed permission for the mode to gravitate.

At each value of the curve parameter $s$, the position uncertainties
define a region $\mathcal B_A(s)$ in the tangent Minkowski space of
the mode. To describe its worldtube, we must identify which events
of the classical spacetime correspond to the points of this region.
Start at the reference event $z_A(s)$ and use the three spatial axes
$e_{\hat i}(s)$ of the observer's orthonormal frame. A point with
spatial coordinates $\bm X=(X^1,X^2,X^3)$ specifies the displacement
vector $X^i e_{\hat i}(s)$ at that event. In flat spacetime one would simply add this
displacement to the reference point. In curved spacetime we instead
follow the spacelike geodesic of the metric $g$ that leaves $z_A(s)$
in the specified direction, for a proper length $|\bm X|$. Its
endpoint is the spacetime exponential map applied to the displacement.
The endpoint map $\mathfrak E_A^g$ and the resulting worldtube $W_A^g$
are therefore
\begin{equation}
\begin{aligned}
 \mathfrak E_A^g(s,\bm X)
 &:=\exp_{z_A(s)}\!\left(X^i e_{\hat i}(s)\right),\\
 W_A^g
 &:=\bigcup_{s\in I_A}
 \left\{\mathfrak E_A^g(s,\bm X)\;\middle|\;
                  \bm X\in\mathcal B_A(s)\right\}.
\end{aligned}
\label{eq:worldtube-map-new}
\end{equation}
The first line locates one point of the profile in spacetime; in
particular, $\bm X=0$ gives the reference event itself. The spacelike
geodesic labels a displacement across the packet, not the trajectory
of a particle. The second line takes the entire spatial region at each
$s$ and collects its images over the chosen interval $I_A$.

For example, consider a complex scalar packet centered on $z_A(s)$
with an isotropic Gaussian position distribution. If $\sigma_A(s)$
is its position standard deviation along each spatial axis, the contour
in Eq.~\eqref{eq:ellipsoid-new} is a ball of radius $q_X\sigma_A(s)$
in the tangent space, where $q_X$ is the prescribed number of standard
deviations. If the packet spreads, these balls grow wider as their
centers follow the reference curve. Their images form the worldtube.
Its boundary marks the chosen uncertainty contour: the Gaussian tails
outside it remain part of the energy calculation. The labels
$(s,\bm X)$ give a single coordinate chart only where the map is
one-to-one and its differential is nonsingular; a larger profile may
require additional charts or an explicitly specified extension.
Section~\ref{sec:metric-comparison} uses the exponential map at a
fixed reference event to compare the full classical metric with the
tangent Minkowski metric and determine how much of the assigned
energy gravitates.

A moving contour has energy and momentum flux even when it follows the evolving uncertainty width. To resolve that flux, let $b_A(X)$ increase outward through the contour, let $\kappa_A(b)$ be a smooth normalized normal response, let $r_A(X^0)$ be a smooth temporal response, and let $v^{\hat a}(X)$ select the measured stress component. The corresponding flux operator and expectation are
\begin{equation}
 \widehat J_A[v]
 =\int d^4X\,r_A(X^0)\kappa_A(b_A(X))v^{\hat a}(X)
 \widehat T^R_{\hat a\hat b}(X)\partial^{\hat b}b_A(X),\qquad
 \mathcal J_A[v]=\Tr(\rho_A\widehat J_A[v]).
 \label{eq:main-resolved-boundary-flux}
\end{equation}
This $\widehat J_A[v]$ is a smeared stress flux, distinct from the Lorentz generators. Its connected symmetrized second moment gives the flux covariance. Both spatial and temporal widths are retained in stress-fluctuation calculations. In local components the outward energy flux across a boundary of speed $v_b$ is $T^{i0}s_i-v_bT^{00}$, where $s_i$ is its spatial unit normal. The second term accounts for the motion of the boundary itself.

The required inputs are consequently the unconditioned state, the complete localization prescription, its instrument when conditional data are used, assigned energy measures, momentum and position moments with their correlations, resolved stress and flux, the reference observer and trajectory, and field constraints. These data are propagated by their prescribed laws. They are not separately fitted to make a cell gravitate. The metric window changes the gravitating share of these data; Sec.~\ref{sec:source} and Appendix~\ref{app:source-details} explain how changing shares enter the classical stress balance.

\section{Gravitating energy from a comparison of the full metrics}
\label{sec:metric-comparison}

The quantum construction above specifies what is transported, how it is localized, and how its energy is partitioned. A further physical postulate is required to say how much of this energy gravitates. We now formulate that postulate by comparing the classical metric with its tangent Minkowski metric over the localization region. A mode retains its quantum state in a tangent Minkowski fiber; no separate classical interior metric is assigned to it. Its gravitating contribution is the positive energy contained where the prescribed comparison regards the tangent and classical geometries as sufficiently close.

The familiar picture of a curved surface departing from its tangent plane motivates this construction, but it does not supply an invariant distance between the two spaces. The spacetime and its tangent space are different manifolds, and the theory does not introduce an ambient embedding space. The comparison must therefore specify how points and infinitesimal vectors in the tangent space are related to those in spacetime. The exponential map provides this relation using the classical metric itself. The geometry of this map is standard; selecting gravitational energy by its metric distortion is the new postulate.

\subsection{A point-based comparison with the tangent metric}

The physical question is whether clocks and rulers in the classical
spacetime assign the same intervals to neighboring points of a mode's
profile as they would in its tangent Minkowski space. Fix the event
$x\in\M$ at which the comparison is made, and let $u\in T_x\M$ be the
four-velocity of the observer who specifies the mode's energy and spatial
profile, with $g_x(u,u)=-1$. In a homogeneous cosmology, $u$ may be the
cosmic-fluid four-velocity. For a massless mode, $u$ is still the
observer's timelike four-velocity; the mode's null direction describes
its propagation. At $x$ itself the tangent metric is exactly
$\eta_x:=g_x$, which has the Minkowski form in the observer's
orthonormal frame. To extend the comparison across the profile, let
$X\in T_x\M$ be a displacement and let $\mathcal I_x(X)$ be the event
reached by following the geodesic from $x$ with initial tangent $X$ to
unit affine parameter. We use a \emph{star-shaped normal domain}
$\mathcal V_x$: if it contains $X$, it contains every shorter displacement
$\alpha X$ with $0\leq\alpha\leq1$, and each event in its image has a
unique displacement label that varies smoothly with the event. The
endpoint map and the classical metric expressed in these labels are
\begin{equation}
 \mathcal I_x(X):=\exp_x(X),\qquad
 \widetilde g_{x,X}(U,V)
 :=g_{\mathcal I_x(X)}
 \bigl((d\mathcal I_x)_X U,(d\mathcal I_x)_X V\bigr).
 \label{eq:metric-pullback}
\end{equation}
The distinction between $X$ and $V$ is useful: $X$ selects the point
of the profile, whereas $V$ specifies how we vary that point. If the
initial displacement changes from $X$ to $X+\epsilon V$, then
$(d\mathcal I_x)_X V$ is the change of the spacetime endpoint per unit
$\epsilon$, evaluated at $\epsilon=0$. The same interpretation applies
to a second variation $U$. Thus $\widetilde g_{x,X}(V,V)$ gives the
classical squared interval for that infinitesimal variation, while
$\eta_x(V,V)$ gives the tangent-Minkowski prediction for the same
variation. This is the physical meaning of the \emph{pullback} metric
$\widetilde g_x=\mathcal I_x^*g$. The two predictions agree at $X=0$;
away from the origin, curvature can change the separation of neighboring
geodesics. Appendix~\ref{app:metric-comparison} gives the corresponding
normal-coordinate construction \cite{PoissonPoundVega2011,Nesterov1999}.

To express this difference as one number at each $X$, we ask: among
displacements of the same size in the observer's frame, which gives the
largest change in squared interval? The Lorentzian interval itself
cannot serve as the measure of size, since it vanishes for a nonzero
null displacement. We therefore use the positive form
$q_x(U,V):=\eta_x(U,V)+2\eta_x(u,U)\eta_x(u,V)$. In the observer's
orthonormal frame, $q_x(V,V)$ is simply the sum of the squares of the
time and three spatial components of $V$. It sets the scale for
comparing directions; the intervals being compared are still those of
the two Lorentzian metrics. The dimensionless mismatch is
\begin{equation}
 \begin{split}
 D_{x,u}(X)
 &:=\sup_{V\ne0}
 \frac{\left|\widetilde g_{x,X}(V,V)-\eta_x(V,V)\right|}
      {q_x(V,V)}\\
 &=\left\|q_x^{-1}\bigl(\widetilde g_{x,X}-\eta_x\bigr)
   \right\|_{\mathrm{op},q_x}.
 \end{split}
 \label{eq:metric-mismatch}
\end{equation}
The first line gives the physical prescription: normalize $V$ to unit
size using $q_x$, examine every direction, and retain the largest
absolute difference. The second line gives a practical way to calculate
it: in the observer's orthonormal frame, form the symmetric matrix
$\widetilde g_{x,X}-\eta_x$ and take the largest absolute value of its
eigenvalues. Thus $D_{x,u}(X)=0$ means that the metrics agree in every
direction at that displacement. The result is independent of the
coordinates used to calculate it. Its dependence on $u$ records the
physical choice of observer, as does the energy assigned to the mode.

Consider a complex scalar packet centered at the event $x$. To evaluate
the metric mismatch across the packet, choose a spatial displacement
$\bm X$ in the local rest frame of the observer with four-velocity $u$.
In that frame, $X=(0,\bm X)$ belongs to $u^\perp$, the three-dimensional
space orthogonal to $u$. It is useful to distinguish the roles of $X$
and $V$ in Eq.~\eqref{eq:metric-mismatch}: $X$ tells us \emph{where in
the packet} to compare the metrics, whereas $V$ tells us \emph{which
interval} to compare there. Although $X$ is spatial, $V$ can point in
any spacetime direction. A purely temporal $V$ tests clock intervals,
a purely spatial $V$ tests lengths, and a $V$ with both components also
tests the time--space terms of the metric. This distinction matters in
a spatially flat expanding FRW universe: the geometry within each
constant-cosmic-time slice is Euclidean, although the four-dimensional
spacetime is curved. For a comoving observer, the spacelike geodesics
used to locate the packet start tangent to that slice at $x$, but they
need not remain on it. Their endpoints specify where the metrics are
compared. Expressing the selected energy and stress as a source on the
hypersurface used to evolve the classical geometry is a separate step,
described in Sec.~\ref{sec:source}.

\subsection{A positive gravitational portion of each energy share}

Let $\mu_A$ denote the positive energy measure assigned to mode $A$ on its tangent spatial section $\Sigma_A$. Its total mass is the additive energy share, $\mu_A(\Sigma_A)=\mathcal E_A$. More explicitly, a spatial localization POVM $\POVM(dX)$ and nonnegative partition functions $w_A$ give $\Effect_A(B):=\int_Bw_A(X)\POVM(dX)$ for a measurable region $B$. With $\sum_Aw_A=1$, the corresponding measure is
\begin{equation}
 \mu_A(B)
 :=\Tr\!\left[\widehat H_R^{1/2}\rho\widehat H_R^{1/2}
                   \Effect_A(B)\right],
 \qquad
 \sum_A\mu_A(\Sigma_A)=\Tr(\rho\widehat H_R).
 \label{eq:metric-energy-measure}
\end{equation}
All effects, the state, and the Hamiltonian in this partition are compared in the same endpoint representation and energy frame; this is not an assertion of a conserved global energy in a general spacetime. The trace is understood in the finite-energy form sense already used for the energy partition. 

Here a positive energy measure means that every spatial region $B$ is
assigned a nonnegative energy share $\mu_A(B)$, and that the shares of
disjoint regions add. This positivity follows from the localization
effect and the Hamiltonian in Eq.~\eqref{eq:metric-energy-measure}.
In general, $\mu_A(B)$ is not the integral over $B$ of the renormalized
local energy density $\langle\widehat T^R_{\hat0\hat0}(X)\rangle$,
which can be negative in some quantum states. The metric criterion
therefore selects a fraction of the mode's assigned energy without
requiring the local stress tensor to have positive energy density
at every point.

Let $x_A$ be the central event of mode $A$ and $u_A$ the four-velocity
of its reference observer. We now state the \emph{criterion for mode
$A$ to gravitate}. We postulate that only the energy assigned to the
part of the mode's profile where the metric mismatch satisfies
$D_{x_A,u_A}(X)\leq\zeta^2/3$ contributes to the gravitational source
of the classical spacetime. Here $\zeta>0$ is a universal dimensionless
constant setting the tolerance. Energy assigned to the rest of the
profile remains part of the quantum field's energy but does not enter
the gravitational source. Denoting the region that satisfies this
criterion by $\mathcal G_A[g]$, we define
\begin{equation}
 \begin{split}
 \mathcal G_A[g]&:=
 \left\{X\in\Sigma_A:\ D_{x_A,u_A}(X)\le\frac{\zeta^2}{3}\right\},\\
 \mathcal E_A^{\rm grav}[g,\rho]
 &:=\int_{\Sigma_A}
 \mathbf 1_{\mathcal G_A[g]}(X)\,d\mu_A(X),
 \qquad
 f_A:=\frac{\mathcal E_A^{\rm grav}}{\mathcal E_A}
 \quad(\mathcal E_A>0).
 \end{split}
 \label{eq:metric-gravitating-energy}
\end{equation}

Here $\mathbf{1}_{\mathcal G_A[g]}:\Sigma_A\to\{0,1\}$ is the
\emph{indicator function} of the region $\mathcal G_A[g]$ selected by
the gravitational criterion. At each position $X$ in the mode's tangent
spatial section $\Sigma_A$, its value is
$\mathbf{1}_{\mathcal G_A[g]}(X)=1$ if
$D_{x_A,u_A}(X)\leq\zeta^2/3$, and
$\mathbf{1}_{\mathcal G_A[g]}(X)=0$ if
$D_{x_A,u_A}(X)>\zeta^2/3$. The bold $1$ thus denotes a scalar function
of position, not the identity operator on quantum states. Multiplying
$d\mu_A(X)$ by this function restricts the energy integral to the
accepted region, giving
$\mathcal E_A^{\rm grav}=\mu_A(\mathcal G_A[g])$.
The zero-or-one decision applies at each position separately: if the
region contains only part of the mode's assigned energy, only that
part gravitates.
For a mode with $\mathcal E_A>0$, its entire assigned energy gravitates
when $f_A=1$, only part gravitates when $0<f_A<1$, and none gravitates
when $f_A=0$. For example, if the accepted region contains one third of
the assigned energy of a complex scalar packet, that packet contributes
$\mathcal E_A/3$ to the classical gravitational source. What matters is
the energy contained in the accepted region, not the fraction of the
packet's spatial volume that it occupies. A mode with zero assigned
energy contributes zero, and no fraction need be defined for it.
The factor $1/3$ fixes the convention for $\zeta$ so that the leading isotropic curvature approximation below has radius $\zeta L_{\rm curv}$. It is not a separate fitted coefficient. The same $\zeta$ applies to every mode and epoch. For a source described by finite quantum profiles, a present-day comparison of gravitational and inertial mass can restrict $\zeta$ only if the observable depends on the fraction of energy that gravitates. Subsection~\ref{subsec:results-scope} distinguishes such a comparison from a test of free fall.

Because the mask lies between zero and one, $0\le\mathcal E_A^{\rm grav}\le\mathcal E_A$. Its complement accounts for the rest of the same tangent quantum energy; it does not remove that energy from the quantum Hamiltonian or change its spectrum. The sources from different outcomes are added with their already assigned energy shares. No second factor equal to the probability of an outcome is inserted. Overlapping windows therefore retain the original positive, non-double-counting partition. For a complex scalar state, changing the mask changes the part of its spatial energy distribution used in the classical source; it does not project the state, perform a measurement, or delete high-energy field states.

For fixed geometry, observer and normalized energy profile, the fraction is independent of the overall energy amplitude. Energy dependence enters through differences of profile, localization and state, and through their feedback on the common metric. In global Minkowski spacetime with its standard exponential map, $D_{x,u}=0$ everywhere and the complete assigned energy gravitates.

Equation~\eqref{eq:metric-gravitating-energy} gives the ideal, sharp
criterion for gravitation: the assigned energy at a point of the profile
counts in full if the metric mismatch is within tolerance, and counts
for nothing if it exceeds the tolerance. This rule gives a well-defined
mean energy because $\mu_A$ assigns an energy to every measurable
region. A difficulty arises when we also want to calculate fluctuations
of the gravitational source. At a sharp boundary the weight jumps from
one to zero over an arbitrarily short distance. The renormalized stress
tensor is an operator-valued distribution, and its variance need not be
finite when sampled with such a discontinuous weight. To describe the
criterion at finite resolution, we therefore replace the sharp step by
a smooth response. Let $y=3D_{x_A,u_A}(X)/\zeta^2$ be the metric
mismatch at $X$ in units of the prescribed tolerance, so that $y=1$
is the threshold. We choose a single smooth, nonincreasing function
$W(y)$, with values between zero and one, equal to one near $y=0$ and
zero for $y\geq1$. The mean energy of mode $A$ that gravitates at this
resolution is
\begin{equation}
 \mathcal E_{A,W}^{\rm grav}
 :=\int_{\Sigma_A}
 W\!\left(\frac{3D_{x_A,u_A}(X)}{\zeta^2}\right)\,d\mu_A(X).
 \label{eq:metric-smooth-energy}
\end{equation}
The physical meaning of $W$ is the fraction of the assigned energy
that gravitates at that part of the profile. For example, if $W=1/2$
over a portion of a complex scalar packet, half of the energy assigned
to that portion enters the classical source. This is an energy fraction,
not a probability that the whole packet gravitates. Thus mode $A$
gravitates precisely when a region with $W>0$ contains nonzero assigned
energy. Compared with the sharp criterion, the finite-resolution rule
can reduce the contribution from within the tolerance as the threshold
is approached; it admits no energy from beyond the threshold. The same
$W$ is fixed for all modes and all epochs.

Equation~\eqref{eq:metric-smooth-energy} still gives a mean, not a
variance. For a fluctuation calculation, the resulting weight must be
smooth as a function of position $X$ and the stress must also be sampled
over a finite time interval. If the eigenvalue that determines the
mismatch changes and produces a kink in $D_{x_A,u_A}(X)$, the spatial
weight itself must be smoothed there. The variance is then calculated
from the correlations of the sampled observables, including correlations
between modes when their contributions are added.
The sharp rule gives the ideal mean. A relative fluctuation must use the mean and variance of the same observable at the same resolution. For the spatially weighted energy share this is the mean in Eq.~\eqref{eq:metric-smooth-energy}, which generally differs from the sharp mean. For stress sampled in space and time, both moments use that spacetime sampling.

The quantum uncertainty of a mode enters through $\rho$, its energy measure and its resolved stress correlations. Thus the rule does not compare a classical point particle with a sphere while ignoring the spatial and momentum spreads of the underlying field. Nor does it impose a particle-horizon condition. The normal domain is a mathematical domain of the chosen comparison map. For profiles extending beyond a single normal chart, a continuation prescription is needed; energy outside that chart is not automatically declared nongravitating. Finite-support examples lie within a specified comparison domain, while infinite-tail calculations require either a defined extension or an explicit tail-error estimate.

\subsection{What the curvature-radius approximation actually says}

At the central event, the two metrics agree and their first normal-coordinate derivatives agree. With the curvature convention of Appendix~\ref{app:metric-comparison}, their first difference is
\begin{equation}
 \widetilde g_{ab}(X)-\eta_{ab}
 =-\frac13R_{acbd}(x)X^cX^d+O(|X|_q^3).
 \label{eq:metric-local-expansion}
\end{equation}
Here $R_{acbd}$ are orthonormal-frame components of the classical Riemann tensor and $|X|_q^2=q_x(X,X)$; the remainder includes derivatives of curvature. Equation~\eqref{eq:metric-local-expansion} gives a controlled approximation only where those higher terms are small. The exact rule remains Eq.~\eqref{eq:metric-mismatch}.

For an FRW spacetime, let $a(t)$ be dimensionless with $a(t_0)=1$ today, $H:=\dot a/a$, and $k/a^2$ the spatial curvature in inverse-length-squared units. At a comoving event, put $X=(0,r\bm n)$ with $\bm n$ a unit spatial vector. The leading mismatch is isotropic:
\begin{equation}
 D_{x,u}(0,r\bm n)
 =\frac{r^2}{3}\max\!\left\{
       \left|\frac{\ddot a}{a}\right|,
       \left|H^2+\frac{k}{a^2}\right|\right\}
   +O(r^3),
 \qquad
 L_{\rm curv}^{-2}:=
 \max\!\left\{\left|\frac{\ddot a}{a}\right|,
                \left|H^2+\frac{k}{a^2}\right|\right\}.
 \label{eq:metric-frw-leading}
\end{equation}
Consequently, the sharp gravitating region is approximately the ball $r\le\zeta L_{\rm curv}$ when the quadratic expansion is valid. The exact exponential-map comparison need not be a ball in a general spacetime, and even in FRW its radial function contains higher-order terms. The cosmological calculations below distinguish this approximation from the fundamental full-metric rule.

A Gaussian scalar energy profile illustrates the physical content. If its width remains much smaller than the permitted region, nearly all its assigned energy gravitates. If the permitted region is much smaller than its width, only a small central portion gravitates. A growing comparison region gives a growing fraction only relative to the evolving energy profile: a sufficiently rapidly spreading packet can offset that growth. The proposed rule therefore permits increasing gravitation during expansion, but the geometry alone does not prove this for every quantum state. The following examples state the packet or ensemble assumptions needed for that result.

\section{The common classical source and changes in gravitating energy}
\label{sec:source}\label{sec:bianchi}

The metric comparison selects an amount and a spatial distribution of positive assigned energy. Einstein's equation needs more: momentum density, pressure, shear, their transport to the classical spacetime, and a conservation law for the complete source. This step supplies that distinction. There is only one classical metric. No separate interior metric is solved for each mode.

\subsection{Positive energy selection and resolved stress}

Let $\mu_A$ be the positive assigned-energy measure of mode $A$, and let $\Gamma_A(X;g)$ denote the indicator $\mathbf1_{\mathcal G_A[g]}(X)$ of Sec.~\ref{sec:metric-comparison}, or its specified smooth finite-resolution realization. Its gravitating energy and fraction are
\begin{equation}
 \mathcal E_A^{\rm grav}[g]=\int\Gamma_A(X;g)\,d\mu_A(X),
 \qquad f_A[g]=\frac{\mathcal E_A^{\rm grav}[g]}{\mathcal E_A},
 \qquad 0\leq f_A\leq1.
 \label{eq:source-energy-fraction}
\end{equation}
The fraction is defined only when $\mathcal E_A>0$; a zero-energy cell has zero gravitational contribution. The integral uses the tangent localization measure, not an extra classical volume factor. A classical volume factor enters later, when that measure is represented as a density on a spacetime slice. Neither $f_A$ nor the already normalized share $\mathcal E_A$ is multiplied again by the localization probability $p_A$.

For a complex scalar, the resolved stress data are obtained by smearing the Minkowski-renormalized composite tensor
\begin{equation}
 \begin{split}
 \widehat T^R_{\hat a\hat b}=\bigl[&
 \partial_{\hat a}\widehat\Phi_R^\dagger\partial_{\hat b}\widehat\Phi_R+
 \partial_{\hat b}\widehat\Phi_R^\dagger\partial_{\hat a}\widehat\Phi_R\\
 &-\eta_{\hat a\hat b}\bigl(
 \partial^{\hat c}\widehat\Phi_R^\dagger\partial_{\hat c}\widehat\Phi_R+
 m_R^2\widehat\Phi_R^\dagger\widehat\Phi_R+
 \tfrac{\lambda_R}{2}(\widehat\Phi_R^\dagger\widehat\Phi_R)^2\bigr)
 \bigr]_R^{\rm M}.
 \end{split}
 \label{eq:main-complex-scalar-stress}
\end{equation}
Here $[\cdot]_R^{\rm M}$ denotes the fixed renormalized Minkowski composite prescription, including the Ward identity. The mixed components give momentum flow and the spatial components give pressure and shear. The complete interacting tensor includes every interaction once. A positive Hamiltonian does not make the renormalized pointwise energy density positive. Equation~\eqref{eq:source-energy-fraction} therefore uses the positive POVM energy measure; it is not an integral of an assumed positive local $\langle\widehat T_{00}\rangle$.

Denote by $\mathfrak D_A$ the endpoint energy measure and the resolved stress, covariance and flux data required by the reconstruction. When conditional moments supply these data, they are evaluated from the common endpoint state and the specified instrument as in Sec.~\ref{sec:povm}. A specified deterministic reconstruction $\mathfrak R_{A,g}$ produces a resolved classical tensor
\begin{equation}
 T^{{\rm grav},A}_{\mu\nu}[g]
 =\mathfrak R_{A,g}[\mathfrak D_A,\Gamma_A]_{\mu\nu},
 \qquad
 \int_{\Sigma_s}n^\mu n^\nu T^{{\rm grav},A}_{\mu\nu}\,d\Sigma_g
 =\mathcal E_A^{\rm grav}(s).
 \label{eq:source-reconstruction}
\end{equation}
The slice $\Sigma_s$ is the specified classical spacelike evolution slice, $n^\mu$ its future unit normal, and $d\Sigma_g$ its proper volume element. A spatial reconstruction map $\iota_{A,s}:\Sigma_A\to\Sigma_s$, together with an observer-frame identification, must be supplied to place the selected tangent energy on this slice. It is distinct from the exponential map used only to compute the metric mismatch: the geodesic comparison section $\exp_{x_A}(u_A^\perp)$ generally is not $\Sigma_s$. Both sides of Eq.~\eqref{eq:source-reconstruction} use the same calibrated observer energy; any change of observer must include the corresponding transformation of momentum and energy. The reconstruction includes $\iota_{A,s}$, its spatial Jacobian, the prescribed finite resolution, and the transformation of flux and stress. Its energy normalization is imposed on the complete retained profile, including tails. When this normalization changes in time, its derivative belongs to the source balance below.

The scalar energy measure alone does not determine $\mathfrak R_{A,g}$. Stress moments and a constitutive reconstruction are additional data. For a slowly moving massive scalar packet, a controlled leading approximation is a smooth mass density with the packet's measured velocity, pressure and dispersion corrections. For an isotropic thermal massless complex scalar, the ensemble source has radiation pressure. These examples specify such a reconstruction explicitly; a general interacting construction remains conditional on its existence and compatibility with the quantum data. Appendix~\ref{app:source-details} states the energy normalization and conservation problem without treating this missing tensor information as already derived.

\subsection{Addition of contributions and Bianchi balance}

At each stage of the evolution, first apply the metric criterion and reconstruct the stress tensor for each outcome $A$ of the prescribed physical localization. Once these tensors have been placed on the same classical slice, they may be reported individually or in groups. Let $\mathscr P$ be a partition of the outcome labels into disjoint groups $J$. Disjointness refers to the labels: the spatial profiles of two outcomes may overlap. At the same spacetime event $x$, the group contribution, total source and Einstein equation are
\begin{equation}
 \begin{aligned}
 T^{{\rm grav},J}_{\mu\nu}(x)
   &:=\sum_{A\in J}T^{{\rm grav},A}_{\mu\nu}(x),\\
 T^{\rm grav}_{\mu\nu}(x)
   &=T^{\rm base}_{\mu\nu}(x)
     +\sum_{J\in\mathscr P}T^{{\rm grav},J}_{\mu\nu}(x)
    =T^{\rm base}_{\mu\nu}(x)
     +\sum_A T^{{\rm grav},A}_{\mu\nu}(x),\\
 G_{\mu\nu}[g]+\Lambda g_{\mu\nu}
   &=8\pi G\bigl(T^{\rm grav}_{\mu\nu}+C_{\mu\nu}\bigr).
 \end{aligned}
 \label{eq:source-einstein}
\end{equation}
Here $T^{\rm base}$ is any separately specified classical source and $C_{\mu\nu}$ is a classical stress response that balances changes in gravitating energy--momentum and other reconstruction fluxes. For continuous labels, the contribution of a measurable group is the integral of the reconstructed tensor density over those labels, with the same outcome measure used for the original family and no extra factor of the outcome probability. Overlapping cells do not require a division by the number of modes: completeness already partitions the energy. Interaction stress must be assigned once in a common reconstruction, not copied into independently normalized packets.

For example, suppose two overlapping complex-scalar packets are reported together. Each retains its own center, observer, metric mask and reconstructed tensor. In the same observer frame, adding those tensors adds their energy densities, momentum fluxes, pressures and shears. No new center or mask is assigned to the group, and Eq.~\eqref{eq:source-reconstruction} is not applied again to a grouped conditional state. This order matters because the reconstruction of an individual cell need not be linear in its conditional data. For the same physical history, initial data and response prescription, regrouping leaves the total source in Einstein's equation unchanged throughout the evolution. It therefore leaves the predicted curvature unchanged. Changing the physical localization widths, centers or instrument is a different operation and need not give the same source.

\noindent\textbf{Why a classical body gravitates in full.}
A classical continuum can extend over a large region even though each material element has a definite position. Let $B$ be such a body on the spacelike slice $\Sigma_s$, with classical stress tensor $T^B_{\mu\nu}$, nonnegative energy density in the normal frame and finite positive total matter energy $E_B$. Its energy element in the slice's normal frame is $dE_B(x)=T^B_{\mu\nu}(x)n^\mu n^\nu\,d\Sigma_g$. If the body is supplied as classical matter, its tensor enters $T^{\rm base}_{\mu\nu}$ in Eq.~\eqref{eq:source-einstein} with coefficient one for energy, momentum and stress alike. Its matter is not also counted among the quantum modes.

The metric rule gives the same result in the ideal local classical description. Each material element at $x$ has its energy at $X=0$ in its own tangent space; the body's radius is not the uncertainty of that element. Denote the sharp or resolved weight of this element by $\Gamma_x(X;g)$. Since $(d\exp_x)_0$ is the identity, the pulled-back metric at $X=0$ is $g_x$ itself and $D_{x,u}(0)=0$ for each specified observer $u$. Both weights therefore equal one at the element: the sharp criterion accepts zero mismatch, and $W(0)=1$ for the resolved criterion. Adding the energies of all the elements gives the fraction $F_B$ of the body's matter energy that enters the source:
\begin{equation*}
 F_B=\frac{\int_B\Gamma_x(0;g)\,dE_B(x)}{\int_B dE_B(x)}
     =\frac{E_B}{E_B}=1 .
 \tag{\theequation a}\label{eq:classical-full-fraction}
\end{equation*}
The result holds at every regular contact event, without a weak-curvature assumption. Full inclusion of local matter energy is distinct from identifying its integral with the total gravitational mass of a strongly bound system.

A finite quantum complex-scalar packet illustrates how the classical result is approached. By continuity of $D$ and because $W$ equals one near zero, there is a neighborhood of the contact point with unit weight. If at most a fraction $\epsilon$ of the packet's assigned energy lies outside that neighborhood, then $0\leq1-f_A\leq\epsilon$, since the weight elsewhere lies between zero and one. Concentration of energy at the contact point thus gives $f_A\to1$; at finite width exact equality requires unit weight almost everywhere with respect to the packet's energy measure.

Environmental interactions can both suppress spatial coherence and help
localize a packet. If the interaction concentrates the mode's assigned
energy in the region where the gravitational weight $\Gamma_A$ is one,
or keeps the energy localized there as the packet evolves, it can bring
the gravitating fraction $f_A$ close to one or keep it there. Loss of
coherence alone does not ensure this: two separated peaks of a scalar
packet can cease to interfere while the energy carried by each peak
remains unchanged. If the assigned energy profile and the metric weight
are unchanged, so is $f_A$.
The latter property could be used in principle to design an experiment to verify the proposed theory.

Define the gain/loss covector $\mathcal J^{\rm grav}_\nu$ by
\begin{equation}
 \mathcal J^{\rm grav}_\nu:=\nabla^\mu T^{\rm grav}_{\mu\nu},
 \qquad \nabla^\mu C_{\mu\nu}=-\mathcal J^{\rm grav}_\nu,
 \qquad \nabla^\mu(T^{\rm grav}_{\mu\nu}+C_{\mu\nu})=0.
 \label{eq:source-balance}
\end{equation}
This $\mathcal J^{\rm grav}_\nu$ is neither the scalar charge current nor a quantum operator. It measures the balance of the gravitationally contributing matter sector. That sector can gain or lose energy while the underlying quantum field continues its prescribed Minkowski dynamics and transport. The tensor $C$ need not cancel the gain by an equal negative energy density; a pressure or momentum flux can supply the required balance. It is not identified with all non-gravitating quantum energy, nor with a material shell.

Equation~\eqref{eq:source-balance} is a balance requirement, not a unique response law. Subsection~\ref{sec:measurement} proposes an explicit detector-response completion in linearized gravity; Appendix~\ref{app:source-details} gives its conservation and early-readout proof. The homogeneous cosmological example instead uses the zero-energy-density pressure closure in Eq.~\eqref{eq:cosmo-response-closure}. These are specified applications of the same balance requirement, not an established universal constitutive law.

The geometrical weight is evaluated on the supplied slice and metric without a particle-horizon or communication-certificate cutoff. Retarded evolution of a chosen classical response is a separate dynamical specification. The resulting source functional is spatially nonlocal through the finite profile and the comparison metric. The candidate below establishes an early no-signalling result only under its stated weak-field and local-instrument assumptions; a general theorem for the coupled model remains open.

\subsection{Separated scalar packets, a local detector, and the classical response}
\label{sec:measurement}

This subsection asks how a definite detector record affects the classical metric when a quantum excitation is shared between two distant packets. Three operations must be distinguished: the detector transfers quantum energy locally; conditioning on its record changes the state used to describe the complete correlated system; and a further response law determines which change of the classical metric a distant observer can measure. The geometrical criterion in Eq.~\eqref{eq:metric-gravitating-energy} specifies the gravitating portion of a given energy profile, but does not by itself complete the third operation. We develop an explicit candidate response, first in the weak-field regime and then on a prescribed spacetime of arbitrary curvature, and identify the additional assumptions needed for a nonlinear metric theorem.

\paragraph{The experiment and the gravitational regime.}
Consider a massive complex scalar field with two nearly orthogonal, spatially separated wave packets L and R. A quantum detector D is placed close to R. Its coupling to the field is confined to a bounded spacetime region $\mathcal O_R$ containing the right packet during the interaction and the formation of the detector record. The detector does not couple to L. A separate gravitational probe near L measures, for example, a relative acceleration of freely falling test bodies, or a phase sensitive to the local tidal field. Its complete preparation and readout interaction occupies a region $\mathcal O_L$ spacelike separated from $\mathcal O_R$. In a nearly inertial laboratory, separation $d$ and experiment duration $\Delta t<d$ illustrate this condition, since $c=1$; the precise condition is spacelike separation of the full interaction regions. The gravitational probe is a second device, not the detector absorbing R.

The same energy criterion applies in highly curved regions. One or both packets can then have very small gravitating fractions, or zero contribution if their positive energy measures assign no energy to the accepted region. High curvature alone does not imply exact zero: the metrics agree at the contact point, and a smooth centered packet can retain a small central gravitating share. For such a packet ``does not gravitate'' means effectively negligible contribution when $f_A\ll1$, unless $f_A=0$ has actually been established. Localization widths and the energy profile matter as well as curvature. Suppressing the original field packets also does not guarantee that energy subsequently concentrated in a much narrower detector profile is suppressed. We study the more challenging case in which L, R and the detector excitation have $f_L\simeq f_R\simeq f_D\simeq1$ throughout the experiment. Exact statements below use full acceptance; controlled tail errors give the corresponding approximation.

One spatial localization POVM of the field can have left, right and remaining outcomes. Two packets therefore do not require two ``entangled POVMs'': entanglement is a property of the quantum state. A second POVM describes the detector pointer. Their joint effects are $\Effect_A\otimes P_r$, where $\Effect_A$ is a field localization effect and $P_r$ is the pointer effect for record $r$. The field--detector interaction $U_{\rm int}$ maps these to input effects $U_{\rm int}^\dagger(\Effect_A\otimes P_r)U_{\rm int}$. The joint state and its correlations change; the interaction does not require an arbitrary new decomposition. Field and detector observables after the interaction are marginals of the same joint description. An instrument, rather than a POVM alone, specifies the state associated with a recorded outcome \cite{DaviesLewis1970,Busch2016QuantumMeasurement,FewsterVerch2020}.

To define that state, let $\rho_{\Phi D}^{\rm out}$ be the density operator of the complete outgoing field--detector system. Let $\Instr_r^D$ be its completely positive pointer operation, acting as the identity on the field factor. Define the positive operator $\sigma_r$, its trace $p_r$, and the unit-trace conditional state by
\begin{equation}
 \sigma_r:=\Instr_r^D(\rho_{\Phi D}^{\rm out}),\qquad
 p_r:=\Tr\sigma_r,\qquad
 \rho_{\Phi D,r}:=\frac{\sigma_r}{p_r},\qquad
 \Tr\rho_{\Phi D,r}=1\quad(p_r>0).
 \label{eq:detector-conditional-state}
\end{equation}
Thus $\sigma_r=p_r\rho_{\Phi D,r}$ stores both the probability of the record and the state conditioned on it. It is called a \emph{subnormalized positive state operator} because its trace is $p_r$, generally smaller than one. Dividing by $p_r$ normalizes conditional probabilities; it has nothing to do with ultraviolet renormalization or removing divergences. The sum of the pointer operations is trace preserving. Their physical realization follows from the detector coupling and readout and need not equal the localization L\"uders instrument.

\paragraph{Absorption and the energy that gravitates.}
For a transparent example, prepare one excitation coherently shared between the two packets: $|L\rangle=|1_L,0_R\rangle$ and $|R\rangle=|0_L,1_R\rangle$. This is one particle in a superposition of locations, not one particle in each packet. Assume negligible packet overlap and equal narrow energy distributions with mean $E$. The detector begins in its ground state $|g\rangle_D$, adjacent to R. An ideal absorption transfers the right excitation into a detector state $|e\rangle_D$ while leaving the left branch unchanged. This motivates the bookkeeping map
\begin{equation}
 |\Psi_{\rm in}\rangle=
 \frac{|L\rangle+|R\rangle}{\sqrt2}|g\rangle_D
 \ \longmapsto\ 
 |\Psi_{\rm out}\rangle=
 \frac{|L\rangle|g\rangle_D+|0\rangle|e\rangle_D}{\sqrt2}.
 \label{eq:detector-packet-absorption}
\end{equation}
Here $|0\rangle$ is the field vacuum. The excited detector carries the absorbed energy and, for a charged scalar excitation, the corresponding charge; the complete coupling conserves charge. A phase acquired during absorption can be included in the definition of $|e\rangle_D$. The first term means that an excitation in L does not reach or excite the right detector. The second means that the excitation in R is absorbed, leaving the field vacuum and a click. This finite-mode idealization explains the energy transfer; exact spacelike-locality statements below require the underlying local field--detector interaction, not an arbitrary single-mode truncation \cite{FewsterVerch2020,FunaiMartinMartinez2019}.

For an ideal efficient detector, click and no click have probabilities $1/2$. Omitting the unchanged detector rest energy, the additive energy shares $(\mathcal E_L,\mathcal E_R,\mathcal E_D)$ are $(E/2,E/2,0)$ before interaction, $(0,0,E)$ conditioned on a click, and $(E,0,0)$ conditioned on no click. Each sum is $E$. Quantum energy is transferred to the right detector; no energy quantum travels from L to R during the click. The change assigned to L is a change of the state conditioned on the record. If the right record is ignored, the left occupation statistics remain one half occupied and one half empty, just as before the interaction. In the finite-mode notation its reduced state is $\rho_L=\tfrac12|1_L\rangle\langle1_L|+\tfrac12|0_L\rangle\langle0_L|$ in both cases. For the different initial state $|1_L,1_R\rangle$, absorption on R leaves the left particle and its energy. A click does not remove an independently occupied left packet.

Applying the gravitational rule to these states gives, when the packet shapes and their fractions are retained,
\begin{equation}
 \mathcal E^{\rm grav}_{\rm before}
 =\tfrac E2(f_L+f_R),\qquad
 \mathcal E^{\rm grav}_{\rm click}=Ef_D,\qquad
 \mathcal E^{\rm grav}_{\rm no\ click}=Ef_L.
 \label{eq:detector-gravitating-budgets}
\end{equation}
Here $f_D$ belongs to the actual absorbed-energy profile; it is not automatically the original $f_R$. In the full-acceptance case each total equals $E$, but its spatial assignment changes. The fractions are recalculated from the appropriate states and profiles if the shapes change. Detector, control and interaction stress must all be counted once. The click-conditioned source is not multiplied again by $1/2$: that probability enters comparisons of many runs, not the energy budget of a specified run.

\paragraph{What would constitute superluminal signalling?}
The controllable choice is whether to operate the right detector, not whether it clicks. A superluminal channel exists if this choice changes a distribution of left readouts before ordinary causal contact, without the left observer receiving or sorting by the right record. Conditional correlations alone do not establish such a channel. Equally, equality of mean gravitational forces is insufficient.

For example, suppose a proposed classical readout gives a fixed half-strength tidal signal when the detector is off, but zero or full strength with equal probabilities when it is on. Let $Y$ be the measured signal, $y_*$ its full-strength value, and $s_{\mathrm{pr}}^2$ the same independent probe-noise variance in both settings. Then
\begin{equation}
 \begin{aligned}
 \langle Y\rangle_{\rm off}&=\langle Y\rangle_{\rm on}=y_*/2,\\
 \operatorname{Var}(Y)_{\rm off}&=s_{\mathrm{pr}}^2,
 &\operatorname{Var}(Y)_{\rm on}&=s_{\mathrm{pr}}^2+y_*^2/4.
 \end{aligned}
 \label{eq:detector-signalling-variance}
\end{equation}
These are distinguishable unconditioned distributions, even though their means agree. With repeated spacelike separated trials, their difference would transmit the on/off choice. This example assumes that these are actual classical gravitational readout laws. It is not the distribution of an ordinary noisy particle-counting POVM, for which the same local reduced quantum state gives the same complete distribution. Quantum backreaction of the probe must be included in an actual experiment.

Three alternatives therefore have different physical consequences:
\begin{enumerate}
\item \emph{Immediate conditional matter update with a compensating classical response.} The actual record supplies the matter state on the specified slice, including its distant conditional changes. The tensor $C_{\mu\nu}$ makes the early measurable gravitational field insensitive to the right choice. The advantage is that the record-conditioned source can be retained while operational signalling is avoided. The consequence is a temporary response energy, pressure or flux, possibly of either sign relative to the reference source. Its later relaxation and gravitational transients become physical predictions that require a response law. We develop this candidate below.
\item \emph{Condition the source only on causally available records.} At an event, the sourcing state incorporates detector records in its causal past. Near L it is initially unchanged and incorporates the right click only after causal arrival. This makes the treatment of distant records explicit but replaces a single simultaneous global conditional sourcing state by an event-dependent prescription. It must still be combined with conservation and a consistent dynamical metric. Related causal-state and conditional semiclassical proposals exist, but do not establish a general covariant completion of the present model \cite{Kent2021Readout,LiuEtAl2023CausalConditional}. Delayed tidal changes and their record correlations could distinguish a specified realization from alternative~1.
\item \emph{Allow an immediately readable distant metric change.} If a right measurement really changes the left unconditioned gravitational readout distribution at spacelike separation, the theory admits superluminal signalling. Random click outcomes do not prevent the controllable on/off channel in Eq.~\eqref{eq:detector-signalling-variance}. This alternative requires a physical prescription for the ordering of such updates, for example a preferred slicing, and cannot simultaneously retain ordinary relativistic no-signalling. A null result in a sufficiently sensitive spacelike control experiment constrains it; noise or a small signal may hide the effect in practice but does not remove the channel in principle.
\end{enumerate}

\paragraph{A concrete response function for alternative~1.}
The response must distinguish an immediate assignment of conditional matter energy from a causal change of the total classical source. Define $\delta T^{(r)}_{\mu\nu}=T^{{\rm grav},r}_{\mu\nu}-T^{\rm ref}_{\mu\nu}$, where $T^{\rm ref}$ describes the unconditioned field--detector evolution with its actual controls. Together with a specified $C^{\rm ref}$ it gives a conserved reference source $\Theta^{\rm ref}=T^{\rm ref}+C^{\rm ref}$. This reference is a baseline solution used to define the response; it is not an average of the metrics in different recorded runs. When comparing detector on and off, their reference histories must themselves have identical local dynamics before a causal influence can reach the left probe.

Introduce a dimensionless scalar $0\leq\chi_R\leq1$. It describes how the record-dependent change is admitted into the \emph{total classical source}: zero before causal arrival, rising during a response interval, and one after that interval in a simple relaxation model. It does not alter the POVM, the quantum state, or the fraction selected by Eq.~\eqref{eq:metric-gravitating-energy}.

An explicit flat-background example uses the detector rest frame. Let $t_0$ be the start of the record response, $\tau_D>0$ its calibrated relaxation time, and $\varrho_D(\bm z)\geq0$ a smooth compact spatial response density of the right detector, normalized by $\int\varrho_Dd^3z=1$. Define
\begin{equation}
 \begin{aligned}
 \chi_R(t,\bm x)&=\int d^3z\,\varrho_D(\bm z)
 F_{\rm resp}\!\left(\frac{t-t_0-|\bm x-\bm z|}{\tau_D}\right),\\
 F_{\rm resp}(v)&=\frac{\vartheta(v)}{\vartheta(v)+\vartheta(1-v)},
 &\vartheta(v)&=
 \begin{cases}e^{-1/v},&v>0,\\0,&v\leq0.\end{cases}
 \end{aligned}
 \label{eq:detector-response-function}
\end{equation}
The dimensionless variable $v$ measures time since a signal emitted at the detector could arrive, in units of $\tau_D$. The smooth step is zero for $v\leq0$, one for $v\geq1$, and increasing in between. The spatial integral represents the detector's finite extent, not an averaging of branch energies. For a point well outside that extent the transition starts after the light-travel time and takes approximately $\tau_D$. A finite record-production time can similarly be included by convolution with a specified temporal response. The function is bounded and has support in the causal future of the source region.

This provides a concrete response family rather than a bare instruction to choose an unspecified function. The causal delay follows from the proposed response dynamics; $\tau_D$ and the shape of the transition require microscopic detector/response physics or experimental calibration. They cannot be chosen after inspecting a distant result. The geometrical tolerance $\zeta$ does not determine them. The absence of an early signal proved below is independent of the detailed rise shape, but the later metric is not. In curved spacetime the time argument may be defined by retarded emission proper time along the detector worldline, on a region with a unique smooth null connection; caustics and multiple arrivals require a smooth extended response prescription. A globally covariant alternative is given below.

\paragraph{Conservation and the weak-field theorem.}
Multiplying the source change by $\chi_R$ alone would generally violate conservation: both $\chi_R\partial^\mu\delta T^{(r)}_{\mu\nu}$ and $(\partial^\mu\chi_R)\delta T^{(r)}_{\mu\nu}$ can be nonzero. We therefore add a determined compensating flux. In a fixed weak-field background set
\begin{equation}
 \begin{aligned}
 Q^{(r)}_{\mu\nu}&=\chi_R\delta T^{(r)}_{\mu\nu},
 &\Box b^{(r)}_\nu&=-\partial^\mu Q^{(r)}_{\mu\nu},\\
 (\mathcal Lb)_{\mu\nu}&=\partial_\mu b_\nu+\partial_\nu b_\mu
       -\eta_{\mu\nu}\partial^\alpha b_\alpha,\\
 \delta C^{(r)}_{\mu\nu}&=(\chi_R-1)\delta T^{(r)}_{\mu\nu}
       +(\mathcal Lb^{(r)})_{\mu\nu},\\
 \delta\Theta^{(r)}_{\mu\nu}&=Q^{(r)}_{\mu\nu}
       +(\mathcal Lb^{(r)})_{\mu\nu}.
 \end{aligned}
 \label{eq:record-candidate-response}
\end{equation}
Here $\Box=\eta^{\alpha\beta}\partial_\alpha\partial_\beta$, $b_\nu$ is an auxiliary classical covector, $\mathcal L$ is the displayed differential operator, $\delta C=C^{(r)}-C^{\rm ref}$, and $\delta\Theta=\delta T+\delta C$ is the total Einstein-source change. Choose the retarded solution for $b_\nu$, with zero perturbation data before the experiment. Its purpose is to supply the pressure and flux required by the divergence of $Q$; it is not another quantum field or a freely adjustable random force.

Direct differentiation gives $\partial^\mu(\mathcal Lb)_{\mu\nu}=\Box b_\nu$, so $\partial^\mu\delta\Theta^{(r)}_{\mu\nu}=0$. Before the response reaches L, both $Q$ and the retarded $b$ vanish there. Hence $\delta C^{(r)}=-\delta T^{(r)}$ and the total local source is unchanged, despite the immediate conditional matter assignment. \textbf{In the equal-packet example the response temporarily restores the missing left $E/2$ in click runs and cancels the extra left $E/2$ in no-click runs.} These are source-energy differences relative to the baseline, not a claim that a new scalar particle appears or disappears. Once the response arrives, the metric can begin to distinguish the records. The remaining $\mathcal Lb$ term need not vanish even when $\chi_R=1$; a final branch metric cannot be inferred by simply removing the response.

\emph{Weak-field statement.} Assume smooth resolved sources, identical initial metric perturbations, retarded response and Einstein evolution, and a left probe whose coupling is local and depends on the response only through the metric. Assume also that the reference histories agree throughout the causal past of the probe and that the actual right instrument is local. Then the right on/off choice cannot change the early unconditioned left readout distribution.

To see the geometrical part, the retarded wave equation implies that $b^{(r)}$ and $\delta\Theta^{(r)}$ are supported in $J^+(\mathcal O_R)$. In Lorenz gauge the trace-reversed metric perturbation obeys $\Box\delta\overline h^{(r)}_{\mu\nu}=-16\pi G\delta\Theta^{(r)}_{\mu\nu}$, where $\delta\overline h_{\mu\nu}=\delta h_{\mu\nu}-\eta_{\mu\nu}\eta^{\alpha\beta}\delta h_{\alpha\beta}/2$. Its retarded solution has the same support property. The locally observable linearized curvature, and thus the early probe dynamics, are unchanged outside that future \cite{Carroll1997GR,BaerGinouxPfaeffle2007}.

The cancellation changes $C_{\mu\nu}$ immediately wherever the conditional matter assignment changes. It is the total source $\Theta_{\mu\nu}$, not each term separately, whose change has causal support. This is a substantive response postulate. The proposed no-signalling result assumes that the compensating contribution has no independently readable local coupling beyond its contribution to the metric; such a coupling would require a separate signalling analysis.

For the statistical part let $\mathsf d$ denote detector on or off, $M_{\mathsf d r\alpha}$ its local instrument operators, and $F_y$ the effect describing the left readout $y$, evolved with the common early metric. The label $\alpha$ denotes unreported detector details. Local commutativity gives $[F_y,M_{\mathsf d r\alpha}]=0$, and normalization gives $\sum_{r,\alpha}M_{\mathsf d r\alpha}^\dagger M_{\mathsf d r\alpha}=\I$. For the complete initial state $\rho$, including the probe, one obtains
\begin{equation}
 \begin{aligned}
 P_{\mathsf d}(y)&=\sum_{r,\alpha}\Tr(F_yM_{\mathsf d r\alpha}\rho M_{\mathsf d r\alpha}^\dagger)
 =\Tr(\rho F_y),\\
 P(y\mid\text{detector on})&=P(y\mid\text{detector off})
 \quad\text{before causal contact}.
 \end{aligned}
 \label{eq:record-early-no-signal}
\end{equation}
For continuous readouts this equality holds for every measurable set of readings, so it concerns the complete distribution, not merely its mean. Correlations conditioned on a transmitted right record may differ. The sum evaluates probabilities over runs and does not construct the metric by averaging branches. In relativistic QFT the rigorous formulation uses commuting local algebras and local probe couplings; it does not require a factorization of sharp spatial regions into two Fock spaces \cite[Sec.~3.3]{FewsterVerch2020}. Appendix~\ref{app:source-details} contains the corresponding weak-field support argument.

\paragraph{An exact extension on a prescribed strongly curved spacetime.}
The covariant response can be constructed without assuming small curvature. Fix a smooth, time-oriented globally hyperbolic metric $g$, with no timelike boundary, and smooth resolved data. Global hyperbolicity supplies Cauchy surfaces and a well-posed causal propagation problem; it does not require weak curvature or exclude an event horizon. Singular endpoints and Cauchy horizons are outside this statement.

One globally covariant way to generate a causal scalar response is to introduce a propagating internal signal $\psi_R$, a local memory $q_{\rm rec}$, and a relaxation law. The memory ensures that a passing signal can leave a persistent response, as a completed detector record should:
\begin{equation}
 \begin{aligned}
 \Box_g\psi_R&=j_R,\qquad
 u^\mu\nabla_\mu q_{\rm rec}=\frac{\psi_R^2}{\tau_D},\qquad
 u^\mu\nabla_\mu\chi_R=\frac{q_{\rm rec}}{\tau_D}(1-\chi_R),\\
 \chi_R(\tau)&=1-\exp\!\left[-\frac1{\tau_D^2}
        \int_{\tau_i}^{\tau}(\tau-\tau')\psi_R(\tau')^2d\tau'\right].
 \end{aligned}
 \label{eq:detector-covariant-response-variable}
\end{equation}
Here $j_R$ is a smooth compact record source supported in $\mathcal O_R$; it represents formation of either a click or a completed no-click record, not the scalar field's charge current. The variables $\psi_R$ and $q_{\rm rec}$ are dimensionless, so $j_R$ has dimensions of inverse length squared. The prescribed timelike unit vector $u$ is the physical observer congruence, $\tau$ its proper time, and $\tau_i$ its intersection with an initial section preceding the record. Choose retarded zero wave data and initially $q_{\rm rec}=\chi_R=0$. The integral is along an integral curve of $u$, and $\tau_D$ is a fixed positive relaxation time. Integrating the memory equation and then the relaxation equation gives the final expression in Eq.~\eqref{eq:detector-covariant-response-variable}.

The signal propagates causally; the memory and the metric response begin only after it arrives. The square makes $q_{\rm rec}\geq0$ and $0\leq\chi_R<1$, even if the wave changes sign. If a nonzero finite pulse crosses an observer history, the memory retains a positive value after the pulse. The response then approaches one exponentially along that history, provided it has an indefinitely long future proper time. If no signal ever arrives, both memory and response remain zero. This is a covariant model of causal arrival followed by persistent local relaxation, rather than an instantaneous distant switch. It is a different concrete response model from Eq.~\eqref{eq:detector-response-function}, with the same causal-support property. Its internal variables specify a candidate constitutive dynamics; their calibration, energy interpretation and implementation in a coupled theory remain additional work, not consequences of the POVM. Even $\chi_R\to1$ does not remove the conservation flux discussed above.

For this fixed-background statement, the reference and conditional tensors are compared on the same supplied metric, and the reference total source is assumed conserved with respect to its connection. Use the scalar response to form the tensor $Q^{(r)}_{\mu\nu}=\chi_R\delta T^{(r)}_{\mu\nu}$ and define
\begin{equation}
 \begin{aligned}
 (\mathcal L_gb)_{\mu\nu}
 &=\nabla_\mu b_\nu+\nabla_\nu b_\mu-g_{\mu\nu}\nabla^\alpha b_\alpha,\\
 (P_gb)_\nu&:=\Box_gb_\nu+R_\nu{}^\lambda b_\lambda
 =-\nabla^\mu Q^{(r)}_{\mu\nu},\\
 \delta C^{(r)}&=(\chi_R-1)\delta T^{(r)}+\mathcal L_gb^{(r)},
 \qquad \delta\Theta^{(r)}=Q^{(r)}+\mathcal L_gb^{(r)}.
 \end{aligned}
 \label{eq:detector-covariant-compensation}
\end{equation}
The tensor $R_\nu{}^\lambda$ is the classical Ricci tensor with one raised index; $\Box_g=\nabla^\alpha\nabla_\alpha$ here acts on a covector. With the paper's curvature convention, commuting the derivatives gives the exact identity
\begin{equation}
 \nabla^\mu(\mathcal L_gb)_{\mu\nu}
 =\Box_gb_\nu+R_\nu{}^\lambda b_\lambda.
 \label{eq:detector-covariant-divergence}
\end{equation}
The Ricci term is required for classical conservation. It is not a curvature vertex in the tangent quantum action, and therefore does not change the Minkowski renormalization prescription.

\emph{Strong-curvature support statement on a fixed background.} For the above data, the retarded solution of Eq.~\eqref{eq:detector-covariant-compensation} exists uniquely in the past-compact class and satisfies
\begin{equation}
 \nabla^\mu\delta\Theta^{(r)}_{\mu\nu}=0,
 \qquad
 \supp\delta\Theta^{(r)}\subseteq J_g^+(\mathcal O_R).
 \label{eq:detector-strong-support}
\end{equation}
Past compactness means that the source support intersects the causal past of every event in a compact set. It follows here from the compact record region and causal support. The operator $P_g$ has the metric wave operator as its principal part; the Ricci term contains no second derivatives of $b$. It is therefore normally hyperbolic on covectors and has a unique retarded inverse with causal support \cite{BaerGinouxPfaeffle2007,Baer2015GreenHyperbolic}. Applying that inverse to $-\nabla Q$, and using Eq.~\eqref{eq:detector-covariant-divergence}, proves both claims. Curvature can produce tails inside the light cone, but not propagation outside it. No expansion in the magnitude of curvature is used. This establishes an exact covariant conservation and support result for the candidate response on a prescribed strongly curved geometry.

\paragraph{What a nonlinear strong-field metric theorem requires.}
The preceding theorem does not yet compare two self-consistently evolving metrics. Their covariant derivatives, retarded inverses and causal cones differ. Even if each total source is conserved in its own metric, their difference need not be conserved under either fixed connection. Consequently the fixed-background proof cannot simply replace $\eta$ by $g$ in the linearized Einstein equation and declare the full problem solved.

A precise sufficient-condition statement is available. Suppose a completion of the matter, detector, reference source and response equations admits a local well-posed hyperbolic evolution after gauge fixing, propagates the Einstein constraints, and has no characteristics outside the metric light cone. Suppose the two control settings have identical constraint-compatible initial data outside the controlled region and all record-dependent influences on the total source enter through causal response variables. Then their developments agree, up to an initial-data-preserving isometry, on the common domain of dependence not reached by the intervention. A local gravitational probe there has the same geometry and, with the local-instrument assumptions above, the same unconditioned readout distribution. The proof is local uniqueness and domain of dependence for the coupled Einstein evolution, followed by the quantum probability identity in Eq.~\eqref{eq:record-early-no-signal}. No weak-curvature hypothesis enters this implication; it applies while the smooth developments exist. The vacuum existence and geometric-uniqueness theorem and its underlying Cauchy framework are standard, whereas an added matter/response system must satisfy its own hyperbolicity and constraint hypotheses \cite{ChoquetBruhatGeroch1969,Sbierski2016Cauchy}.

These hypotheses have not been established for the full metric-dependent, spatially extended quantum reconstruction of this paper. In particular, conservation alone does not prove local well-posedness, causal characteristics or stability, and the nonlocal dependence of the comparison and transport must be accounted for in that analysis. The results obtained here are therefore an explicit weak-field no-signalling construction, an exact strong-curvature support theorem on a prescribed metric, and a sufficient-condition theorem for a nonlinear completion. Determining the physical response parameters, its later metric and response energy, and whether the full coupled theory meets the nonlinear hypotheses remains future work. Alternative~1 is retained with precisely this status as a candidate completion.

\section{Gravitational decoupling of the reference vacuum}
\label{sec:vacuum-gravity}

The positive energy prescription has an immediate consequence for the
gravitational role of the quantum vacuum. The reference Minkowski vacuum
contributes exactly zero to the selected energy at every regular cosmological
epoch. This statement concerns the energy that sources the classical metric,
rather than a decrease of vacuum fluctuations. It follows from the fixed
vacuum normalization and the coupling rule already specified in the
construction. Changing the geometry
changes which part of a nonzero energy measure gravitates, but cannot turn a
zero measure into a source.

\subsection{The vacuum has a zero positive energy measure}

At a spacetime event $x$, let $|0_x\rangle$ be the normalized reference vacuum
of the tangent Minkowski theory and let
$\rho_{0,x}=|0_x\rangle\langle0_x|$ be its density operator. The Hamiltonian
$\widehat H_R$ is the complete renormalized tangent Hamiltonian in the
specified energy frame. Its positivity and vacuum normalization were fixed
in Eq.~\eqref{eq:positive-H}. The spectral calculus for a positive
self-adjoint operator consequently gives
\begin{equation}
 \widehat H_R^{1/2}|0_x\rangle=0,
 \qquad
 \widehat H_R^{1/2}\rho_{0,x}\widehat H_R^{1/2}=0.
 \label{eq:vacuum-energy-kernel}
\end{equation}
The second equality is a positive quadratic-form identity; it does not
require multiplication of unbounded operators on arbitrary vectors.

Consider any mode label $A$ and any measurable tangent region
$B\subseteq\Sigma_A$. The bounded positive effect $\Effect_A(B)$ is the
spatially restricted localization effect in
Eq.~\eqref{eq:metric-energy-measure}. Substitution of
Eq.~\eqref{eq:vacuum-energy-kernel} into that definition proves
\begin{equation}
 \begin{split}
 \mu_{A,0}(B)
 &:=\operatorname{Tr}\!\left[
 \widehat H_R^{1/2}\rho_{0,x}\widehat H_R^{1/2}\Effect_A(B)\right]\\
 &=\left\|\Effect_A(B)^{1/2}\widehat H_R^{1/2}|0_x\rangle\right\|^2
 =0.
 \end{split}
 \label{eq:vacuum-zero-measure}
\end{equation}
Here the additional subscript $0$ indicates evaluation in the reference
vacuum. No commutation of $\Effect_A(B)$ with $\widehat H_R$ is assumed.
Thus every part of the vacuum's assigned energy measure vanishes, for any
admitted localization widths, shapes or number of outcomes. In particular,
localization cannot uncover positive vacuum energy hidden by cancellation
between different cells. The energy measure is positive from the outset.

Let $\Gamma_A(X;g)$ denote the geometrical response already used in
Eq.~\eqref{eq:source-energy-fraction}; it can be the sharp metric-comparison
indicator or its prescribed smooth realization, with
$0\leq\Gamma_A\leq1$. Equation~\eqref{eq:vacuum-zero-measure} then implies
\begin{equation}
 \mathcal E^{\rm grav}_{A,0}[g]
 =\int_{\Sigma_A}\Gamma_A(X;g)\,d\mu_{A,0}(X)=0,
 \qquad
 \sum_A\mathcal E^{\rm grav}_{A,0}[g]=0.
 \label{eq:vacuum-zero-gravitating-energy}
\end{equation}
For a continuous POVM, the last sum is the corresponding outcome integral.
The result holds for every admissible classical metric, independently of
its curvature and of the universal metric tolerance $\zeta$. Since the
vacuum's total assigned energy is zero, its gravitating fraction $f_A$ is
undefined and is not needed. Its gravitating contribution is simply zero.

For a free complex scalar, the Hamiltonian in
Eq.~\eqref{eq:complex-scalar-positive-hamiltonian} contains the positive
particle and antiparticle occupation energies. Both annihilation operators
annihilate $|0_x\rangle$, which explicitly verifies the hypothesis above.
The proof itself does not require a free field or a particle decomposition.
For the interacting complex scalar with the already specified potential
$V_R(|\Phi|^2)=m_R^2|\Phi|^2+\lambda_R|\Phi|^4/2$, it applies to the full
interacting Hamiltonian and its true reference vacuum whenever the
positivity, domain and vacuum conditions of the tangent theory hold.
There is no requirement to set $\lambda_R=0$. The free Fock vacuum must not
be substituted for an interacting ground state without justification.
The normalization is fixed for the theory; it is not reset to assign zero
energy to an evolving condensate, a false vacuum or a thermal state. So in the proposed theory vacuum does not gravitate.

\subsection{The classical source and the meaning of ``at every epoch''}

The source prescription states that a zero-energy mode has zero
gravitational contribution. To make this statement precise for the complete
tensor, the reconstruction in Eq.~\eqref{eq:source-reconstruction} must
respect
\begin{equation}
 \mathfrak R_{A,g}[\mathfrak D_{A,0},\Gamma_A]_{\mu\nu}=0.
 \label{eq:vacuum-reconstruction-condition}
\end{equation}
Here $\mathfrak D_{A,0}$ denotes the admitted mode data evaluated in the
reference vacuum. Equation~\eqref{eq:vacuum-reconstruction-condition}
explicitly expresses the zero-contribution convention; it is not inferred
from a vanishing integral of energy alone, which would not determine all
pressure and shear components. Vacuum stress correlations can remain
nonzero, but the present construction does not promote their variance to
an additional mean classical source.

It follows that $T^{{\rm grav},0}_{\mu\nu}=0$, where the
superscript $0$ denotes the intrinsic reference-vacuum contribution.
This contribution has no gain/loss term in
Eq.~\eqref{eq:source-balance}. We impose the corresponding
zero-response convention $C^0_{\mu\nu}=0$, with zero initial
response data and no incoming or previously generated response.
This is an additional specification of the reference-vacuum
sector: conservation alone would also permit an independently
specified conserved classical tensor. The zero-energy case is
treated directly by this convention, without assigning a
gravitating fraction to a vacuum mode.

This convention is compatible with both response constructions
in Subsection~\ref{sec:measurement}. For a history that remains
in the reference-vacuum sector and has no source change,
$\delta T^{(r)}_{\mu\nu}=0$ throughout the relevant history, and
therefore $Q^{(r)}_{\mu\nu}=0$. With zero initial data, the retarded
solution is $b^{(r)}_\nu=0$ in both the weak-field equation
\eqref{eq:record-candidate-response} and the prescribed
curved-background equation
\eqref{eq:detector-covariant-compensation}. Their response
increments consequently vanish. The Ricci term in the curved
equation does not generate a response from zero data. This
establishes compatibility of the zero-response convention with
the stated constructions; the fully coupled nonlinear evolution
retains the conditional status explained in
Subsection~\ref{sec:measurement}.

The condition $C^0_{\mu\nu}=0$ must not be interpreted as saying
that the total response vanishes whenever the scalar field is
found in its vacuum. After absorption by a detector, for example,
the field can be in the vacuum while the detector carries the
absorbed energy and a response to the preceding redistribution
is still present. Such a response belongs to the interaction
history, not to the intrinsic reference-vacuum contribution.
Similarly, $Q^{(r)}_{\mu\nu}=0$ at one event does not by itself
imply a vanishing response there. An independently added
conserved tensor remains separate classical input, rather than
a contribution derived from the vacuum energy measure.

In a homogeneous and isotropic FRW geometry let $u^\mu$ be the cosmic-fluid
unit velocity, and let $a(t)$ be the dimensionless scale factor normalized
by $a(t_0)=1$ today. Define the reference-vacuum source density and pressure
by contraction of its total source tensor. The preceding result gives
\begin{equation}
 \begin{aligned}
 \varepsilon^{\rm grav}_{\rm vac}(t)
 &:=u^\mu u^\nu\bigl(T^{{\rm grav},0}_{\mu\nu}+C^0_{\mu\nu}\bigr)=0,\\
 p^{\rm grav}_{\rm vac}(t)
 &:=\frac13(g^{\mu\nu}+u^\mu u^\nu)
       \bigl(T^{{\rm grav},0}_{\mu\nu}+C^0_{\mu\nu}\bigr)=0.
 \end{aligned}
 \label{eq:vacuum-FRW-zero}
\end{equation}
The density is therefore zero for every regular epoch with $a>0$, and its
time derivative is zero. A formal limit $a\to0^+$ of this quantity is also
zero; this does not assert that the construction is defined at a singular
Big Bang. Expansion can increase the gravitating fraction of excited or
thermal modes while the reference-vacuum contribution remains identically
zero.

One further distinction is essential. Evaluating the reference vacuum in
each fiber is different from proving that an initially vacuum state stays
in the reference vacuum under the prescribed transport. The latter requires
\begin{equation}
 \mathcal U_A[g;s_2,s_1]\,\rho_{0,x_1}\,
 \mathcal U_A[g;s_2,s_1]^\dagger=\rho_{0,x_2},
 \qquad x_i=z_A(s_i).
 \label{eq:vacuum-transport-preservation}
\end{equation}
Here $\mathcal U_A$ is the full propagator and $x_i$ are the endpoints of
the central history. Unitarity alone does not establish this equality.
If transport produces excitations relative to the endpoint reference
vacuum, their energy is tested by the usual metric rule. Their possible
gravitational contribution is not the subtracted reference-vacuum energy.
Equation~\eqref{eq:vacuum-FRW-zero} therefore excludes a reference-vacuum
offset at every epoch without assuming that every evolving state remains
empty.

The physical content and the logical scope are now explicit. Vacuum
subtraction is fixed in the tangent theory and enters the gravitational
coupling through its positive renormalized Hamiltonian. Conventional curved-spacetime renormalization and its geometrical
stress terms constitute a different source prescription
\cite{HollandsWald2015}. In an ordinary Einstein source, a constant vacuum
density $\varepsilon_0$ would give
$T^{\rm vac}_{\mu\nu}=-\varepsilon_0g_{\mu\nu}$ and would shift the effective
cosmological constant to $\Lambda+8\pi G\varepsilon_0$. A change of energy
origin alone therefore does not remove its gravitational effect. Here the
essential additional input is the specified gravitational coupling to the
vacuum-subtracted positive energy measure, with the zero-source condition
\eqref{eq:vacuum-reconstruction-condition}.

The argument consequently does not determine the independent classical
$\Lambda$ in Eq.~\eqref{eq:source-einstein}, or prove that all possible
vacuum-related effects vanish. Determining the observed cosmological
constant remains distinct from excluding this reference-vacuum
contribution \cite{Weinberg2000CosmologicalConstantProblems}.

\section{Complex-scalar and thermal-fluid examples in expanding and contracting universes}
\label{sec:cosmology}

This section applies the metric-comparison rule to a homogeneous and isotropic universe without a detector, allowing flat, negatively curved and positively curved spatial geometries. It separates three questions: how the initial quantum ensemble and its localization are specified, when the geometrical gravitating fraction increases on a prescribed expanding background, and what additional closure is needed when that fraction also determines the background. This separation is necessary because a geometrical selection rule alone neither fixes the quantum state nor closes Einstein's equations. The final subsection describes the corresponding decrease during contraction, including the recollapse of the positively curved homogeneous example.

We use cosmic proper time $t$, with a dot denoting $\dd/\dd t$, and $H=\dot a/a$. The dimensionless scale factor is normalized to $a(t_0)=1$ today. In the big-bang examples $a(t)\to0$ as $t\to0^+$; the state and its localization are specified at a finite $t_i>0$. The dimensional spatial-curvature parameter $k$ is defined by the spatial scalar curvature $6k/a^2$. To distinguish it from the conventional normalized curvature sign, write $k=\kappa/L_{\rm ref}^2$, where $\kappa\in\{-1,0,+1\}$ and $L_{\rm ref}>0$ is a fixed reference length; for nonzero curvature it is the spatial curvature radius at $a=1$.  Units are $c=\hbar=k_B=1$, and the fluid four-velocity $u$ supplies the observer used in the positive metric-comparison norm. No cosmic size or particle horizon is imposed as an additional localization cutoff.

\subsection{Initial state and spherical localization without a detector}
\label{subsec:thermal-localization}

At an initial time $t_i>0$, a homogeneous, isotropic state with specified mean energy and vanishing mean charges (by charge symmetry) can be selected by maximizing the von Neumann entropy. In a finite-volume regulator this gives
\begin{equation}
 \rho_{Q,i}=Z_i^{-1}\exp(-\widehat H_R/T_i),
 \qquad
 Z_i=\Tr\exp(-\widehat H_R/T_i).
 \label{eq:cosmo-initial-thermal}
\end{equation}
Here $T_i$ is the initial temperature, $\widehat H_R$ is the total renormalized tangent-Minkowski Hamiltonian, and $Z_i$ normalizes the state. The maximum-entropy argument is described in \cite{Jaynes1957}. In infinite space one uses its local thermal, or Kubo--Martin--Schwinger, limit; the infinite-volume thermal state is not a trace-class density operator on a single global Fock space. For a free massless boson the regulator also removes or regulates the zero-frequency mode, for example with suitable boundary conditions or an auxiliary positive mass taken to zero after local observables have been defined; finite periodic volume alone would not make that mode's partition function finite. The normalized energy partition is applied before the local or energy-density limit, without treating the infinite-volume total Hamiltonian as a finite random variable. Equation~\eqref{eq:cosmo-initial-thermal} is a specification of initial data, not a measurement update. These thermal constructions refer to the tangent-Minkowski theory. A compact classical spatial slice does not by itself impose a reflecting boundary on a tangent fiber or fix its localization width. If profiles extend beyond the local comparison domains, a compatible global prescription is still required.

The localization prescription supplies additional initial data. For a species $j$, let $\ell_{j,i}>0$ be the initial one-coordinate Gaussian width. A simple comoving family has $\ell_j(a)=a\ell_{j,0}$, where $\ell_{j,0}=\ell_{j,i}/a_i$ and $a_i=a(t_i)$. In the tangent spatial coordinates $\bm X$ centered at the mode's contact point, its dimensionless response window is $w_j(\bm X)=\exp[-|\bm X|^2/(2\ell_j^2)]$. Translations of this window, with the corresponding normalization measure, form one continuous partition of the field energy. They are not independent copies of the same field energy assigned anew at every contact point. If several widths are included for the same species, their weights must themselves be normalized. Taking infinitely many copies of a complete partition would count the same energy infinitely many times.

For an exactly free field, this prescription can be made explicit on the positive-energy particle and antiparticle sectors. Let $h_j=(\bm P^2+m_{R,j}^2)^{1/2}$ be the one-particle Hamiltonian and $\bm Q$ the spatial localization operator in the selected tangent frame. The positive localized energy operator is
\begin{equation}
 \widehat K_j(w)=\dd\Gamma_j\!\left(b_j(w)\right),
 \qquad
 b_j(w)=h_j^{1/2}w(\bm Q)h_j^{1/2}.
 \label{eq:cosmo-local-energy-operator}
\end{equation}
The expectation value $\mathcal E_j=\langle\widehat K_j(w)\rangle_T$
is the mean energy assigned to the entire localization cell, evaluated
in Eq.~\eqref{eq:cosmo-gaussian-energy-profile}; the metric-comparison
criterion then determines what fraction of this energy gravitates.
Specifying the operator itself is also necessary because its second
moment, which is not determined by the mean profile alone, enters the
energy variance in Eq.~\eqref{eq:cosmo-exact-quantum-variance}.
Here $\dd\Gamma_j$ denotes second quantization, namely the sum of the one-particle operator over excitations of species $j$. For a complex scalar both charge sectors are included. For this thermal example we fix the canonical configuration response $w(\bm Q)$ in the selected observer frame and its energy sandwich. It is one free-field realization of the positive energy partition described in Sec.~\ref{sec:povm}; it is not identical to every other admissible localization construction. In particular, the Klein--Gordon energy-density form with a local sum of time-derivative, spatial-gradient, and mass terms generally defines a different response. The choice in Eq.~\eqref{eq:cosmo-local-energy-operator} is retained throughout this calculation and is not identified with a positive renormalized pointwise stress density. Its associated energy-normalized POVM is defined on the non-vacuum sector in Appendix~\ref{app:povm-fock}.

Homogeneity makes the mean energy measure of a translated Gaussian cell proportional to its response window. Consequently its normalized assigned-energy profile is
\begin{equation}
 \begin{aligned}
 \frac{\dd\mu_j}{\mathcal E_j}
 &=p_{\ell_j}(\bm X)\,\dd^3X,
 &p_{\ell}(\bm X)
 &=\frac{\exp[-|\bm X|^2/(2\ell^2)]}{(2\pi\ell^2)^{3/2}},\\
 \mathcal E_j&=\varepsilon_{Q,j}(2\pi)^{3/2}\ell_j^3.
 \end{aligned}
 \label{eq:cosmo-gaussian-energy-profile}
\end{equation}
The measure $\mu_j$ and total cell share $\mathcal E_j$ are the positive assigned energy, while $\varepsilon_{Q,j}$ is the homogeneous tangent energy density. Thus the Gaussian in this example follows from a specified localization response acting on a homogeneous state. A Gaussian complex field amplitude would not, in general, have a Gaussian energy density: its gradient, kinetic, and mass terms must also be evaluated.

For the explicit relativistic-fluid calculation we take free, massless species, zero chemical potentials, a constant set of populated species, and an adiabatic thermal description with $T=T_0/a$. Then
\begin{equation}
 \varepsilon_Q=c_QT^4,
 \qquad p_Q=\frac{\varepsilon_Q}{3},
 \qquad
 c_Q=\frac{\pi^2}{30}\left(g_b+\frac78g_f\right).
 \label{eq:cosmo-radiation-ensemble}
\end{equation}
Here $g_b$ and $g_f$ count bosonic and fermionic internal degrees of freedom, including antiparticles. A free massless complex scalar has $g_b=2$, $g_f=0$, and hence $\varepsilon_{Q,\Phi}=\pi^2T^4/15$. These equations describe excitations of the tangent theory; its subtracted vacuum adds no energy. A nonzero mass or the Hermitian interaction $\lambda_R|\Phi|^4/2$ changes the equation of state and the correlations. The microscopic construction still uses the same renormalized Minkowski dynamics, but the simple radiation formulas then need the appropriate corrections. The assumed thermal evolution is a useful closure in an adiabatic or equilibrated regime, not a theorem that an arbitrary initial thermal state remains exactly thermal under every allowed transport.

\subsection{The gravitating fraction and an exact scaling result}
\label{subsec:thermal-fraction}

On a homogeneous and isotropic background the metric distortion is radial on the tangent spatial slice. On a regular neighborhood where it increases with radius, the accepted region is a ball. Denote its radius by $r_g(t)$. 
To evaluate the gravitating fraction defined in
Eq.~\eqref{eq:metric-gravitating-energy}, we integrate the normalized
Gaussian energy profile of Eq.~\eqref{eq:cosmo-gaussian-energy-profile}
over the ball of radius $r_g(t)$ selected by the metric criterion.
Writing $z=r/\ell_j$ for the radial distance in units of the Gaussian
width, the enclosed fraction is
$\sqrt{2/\pi}\int_0^{q_j} z^2 e^{-z^2/2}\,\dd z$, where
$q_j=r_g(t)/\ell_j(t)$. We denote this fraction by $\mathcal G(q_j)$;
evaluating the integral gives
\begin{equation}
 f_j(t)=\frac{\mathcal E_j^{\rm grav}}{\mathcal E_j}
 =\mathcal G(q_j),
 \qquad 
 \mathcal G(q)=\operatorname{erf}\!\left(\frac{q}{\sqrt2}\right)
 -\sqrt{\frac2\pi}\,q\exp\!\left(-\frac{q^2}{2}\right).
 \label{eq:cosmo-gaussian-fraction}
\end{equation}
The function $\operatorname{erf}$ is the error function. An infinite Gaussian has a tail outside every bounded normal neighborhood. Accordingly, Eq.~\eqref{eq:cosmo-gaussian-fraction} is an analytic profile formula; its use for an exact local comparison requires a support prescription or an explicit tail error. A compactly supported version is given below. This expression integrates the assigned energy inside the permitted sphere; it does not shift the eigenvalues of $\widehat H_R$. The derivative $\mathcal G'(q)=\sqrt{2/\pi}\,q^2e^{-q^2/2}$ is positive. Therefore
\begin{equation}
 \frac{\dd f_j}{\dd\ln a}
 =\sqrt{\frac2\pi}\,q_j^3e^{-q_j^2/2}
 \left(\frac{\dd\ln r_g}{\dd\ln a}
       -\frac{\dd\ln\ell_j}{\dd\ln a}\right).
 \label{eq:cosmo-general-growth-test}
\end{equation}
The physical condition is explicit: the region of sufficiently small metric distortion must grow faster than the assigned-energy width. A larger universe by itself is not sufficient. This condition is independent of the spatial-curvature sign and applies on an expanding interval; a possible turnaround is treated using proper-time derivatives below.

We now determine how the radius $r_g(t)$ of the gravitating region
changes with time. For this exact example, consider a prescribed,
spatially flat FRW universe with scale factor $a(t)=(t/t_0)^p$,
where $0<p<1$, $t$ is cosmic proper time measured from the big bang,
and $t_0$ is the present time. Let $x_t=(t,\bm 0)$ be the central
event of the mode and $u_t$ the four-velocity of the comoving observer
there. We use the metric mismatch $D_{x_t,u_t}(X)$ defined in
Eq.~\eqref{eq:metric-mismatch}. Its argument $X$ tells us where in
the tangent space to make the comparison. At a physical tangent
distance $r$ from the center, in a unit spatial direction $\bm n$,
we set $X=X_r=(0,r\bm n)$ in the observer's orthonormal frame.
The zero means that the initial displacement is purely spatial.
At that displacement, $D_{x_t,u_t}(X_r)$ compares the full classical
metric, pulled back by Eq.~\eqref{eq:metric-pullback}, with the tangent
Minkowski metric: it measures the largest difference in squared
interval for a variation of unit size in the observer's frame.
Thus, although $X_r$ selects a spatial point of the profile, the test
compares both clock and length intervals.

Isotropy makes the result independent of $\bm n$. The flat power-law
geometry has a further property: if the cosmic age $t$ and the
physical tangent distance $r$ are both multiplied by the same factor,
the dimensionless metric mismatch is unchanged. For example, at twice
the cosmic age one must go twice as far from the center to encounter
the same mismatch.
Since, $r$ is the physical distance from the center of the mode in its
tangent space, and $t$ is the cosmic age at which we examine the mode,
for the spatially flat power-law background $a(t)\propto t^p$, the
curvature length grows in proportion to $ct$, so $r/(ct)$ measures
how far the profile extends relative to the curvature scale at that
epoch. 
The scaling symmetry, derived in
Appendix~\ref{app:cosmology-details}, means that the mismatch depends
only on the ratio $y=r/t$ in units with $c=1$. 
Denote this dependence
by $d_p(y)$, where the subscript specifies the expansion exponent.
The gravitational criterion of Eq.~\eqref{eq:metric-gravitating-energy}
then gives
\begin{equation}
 D_{x_t,u_t}(X_r)=d_p(r/t),
 \qquad
 d_p\!\left(y_\zeta(p)\right)=\frac{\zeta^2}{3},
 \qquad
 r_g(t)=t\,y_\zeta(p).
 \label{eq:cosmo-exact-homothety}
\end{equation}
Here $\zeta$ is the universal metric tolerance, and $y_\zeta(p)$
is the first positive value of $r/t$ at which the mismatch reaches
the threshold $\zeta^2/3$. We restrict the comparison to a regular
neighborhood $0\leq r/t\leq y_{\rm dom}$ on which $d_p$ is strictly
increasing and which contains this threshold crossing. Within that
neighborhood, the energy assigned at $r\leq r_g(t)$ contributes to
gravity. The physical consequence is that this radius grows as $t$,
whereas the width of a comoving localization window grows only as
$a(t)\propto t^p$. Since $p<1$, the gravitational criterion admits
a progressively larger portion of that window's energy profile.
To keep the exact theorem within this domain, replace the infinite Gaussian response by $w_{\ell,C}(r)=e^{-r^2/(2\ell^2)}\chi(r/(C\ell))$. Here $C>0$ is a fixed dimensionless support parameter and $\chi$ is a fixed smooth nonincreasing function, equal to one near zero and zero for arguments at least one. The response still defines a spherical localization prescription. Its normalized cumulative energy is
\begin{equation}
 \begin{aligned}
 \mathcal G_C(q)&=
 \frac{\displaystyle\int_0^{\min(q,C)}
 z^2e^{-z^2/2}\chi(z/C)\,\dd z}
 {\displaystyle\int_0^C z^2e^{-z^2/2}\chi(z/C)\,\dd z},\\
 f_j&=\mathcal G_C(q_j),
 \qquad q_j=\frac{y_\zeta(p)t}{\ell_j(t)}.
 \end{aligned}
 \label{eq:cosmo-compact-profile}
\end{equation}
At $t_i>0$, the profile lies inside the monotone normal domain if $C\ell_{j,i}<t_i y_{\rm dom}$. An initially small fraction also requires $q_{j,i}=t_i y_\zeta(p)/\ell_{j,i}$ small enough that $\mathcal G_C(q_{j,i})\ll1$; the hierarchy $y_\zeta(p)\ll\ell_{j,i}/t_i<y_{\rm dom}/C$ is possible only if $C y_\zeta(p)\ll y_{\rm dom}$. Since $\ell_j/t\propto t^{p-1}$ decreases during expansion, the entire support stays inside that domain. The comparison then covers every point with assigned energy, without a particle-horizon cutoff.

For these profiles the exact fraction satisfies
\begin{equation}
 \frac{\dd f_j}{\dd\ln a}
 =\frac{1-p}{p}q_j\mathcal G_C'(q_j)\geq0,
 \qquad
 f_j=1\quad\hbox{once }q_j\geq C.
 \label{eq:cosmo-exact-prescribed-growth}
\end{equation}
For the sharp criterion the increase is strict whenever the moving boundary sweeps through positive energy; $q_j\geq C$ guarantees saturation. For a fixed smooth nonincreasing $W$, the weight at each scaled radius $z$ is nondecreasing as $\ell_j/t$ falls. Its mean reaches unity when the entire support lies in the region where $W=1$. The result applies to prescribed radiation ($p=1/2$) and dust ($p=2/3$) backgrounds from finite initial data. If the taper removes a fraction $\delta_C$ of the original Gaussian mass, its normalization changes any cumulative fraction by at most $\delta_C$ (Appendix~\ref{app:cosmology-details}).

For a general FRW history the leading local metric comparison gives $r_g\simeq\zeta L_{\rm curv}$, where
\begin{equation}
 L_{\rm curv}^{-2}
 =\max\!\left(\left|\frac{\ddot a}{a}\right|,
              \left|H^2+\frac{k}{a^2}\right|\right).
 \label{eq:cosmo-curvature-approximation}
\end{equation}
This relation is a consequence of the full spacetime metric comparison explained in Sec.~\ref{sec:metric-comparison}; it is not the fundamental replacement for that comparison. It applies for either sign of $k$, provided the higher normal-coordinate terms remain small over the selected region. On the spatially flat power-law background, $y_\zeta(p)=\zeta/[\max\{p(1-p),p^2\}]^{1/2}+O(\zeta^2)$, consistent with Eq.~\eqref{eq:cosmo-exact-homothety}. A closed FRW geometry generally has no such power-law homothety, so its exact threshold radius must be obtained from its own full metric comparison.

For orientation, on a prescribed standard radiation background with zero cosmological constant, Einstein's equations give $H^2+k/a^2\propto a^{-4}$ and $\ddot a/a=-(H^2+k/a^2)$ \cite{Carroll1997GR}. Consequently $L_{\rm curv}\propto a^2$, independently of the sign of $k$. In the local approximation, a comoving thermal width $\ell_j\propto a$ therefore gives $q_j\propto a$ and an increasing fraction during expansion. At a closed-universe turnaround, $H=0$ but $L_{\rm curv}=a/\sqrt{k}$ remains finite. These are statements on a prescribed radiation background; the next homogeneous example separately includes the feedback of $f$ on that background.

On the prescribed flat backgrounds $a(t)\propto t^p$, $0<p<1$, the comparison radius grows faster than each comoving width. For any reporting threshold $0<f_*<1$, a cell that reaches $f_j\geq f_*$ stays above it, and cells of other widths can cross later. This is the sense in which more modes can gravitate appreciably during expansion. The theorem concerns this fixed family, without assuming that its selected energy generates the prescribed background.

\subsection{Quantum uncertainties and thermal fluctuations}

The preceding subsection showed when expansion increases $f_j$, the
fraction of a mode's mean energy that gravitates. To assess how well
this mean characterizes the energy entering the classical source, we
must also calculate the fluctuations of the selected energy. In
Eq.~\eqref{eq:cosmo-exact-quantum-variance} below, choosing $w=w_j$
gives the variance of the energy assigned to the entire mode, whereas
choosing $w=w_jW_g$, with $W_g$ the smooth gravitational weight of
Eq.~\eqref{eq:metric-smooth-energy}, gives the variance of its
gravitating portion. If the standard deviation is comparable to or
larger than the mean gravitating energy, the relative fluctuation is
large. This does not reverse the growth of $f_j$, which is a statement
about mean energies; it limits how well the mean alone characterizes
the mode's contribution to the classical gravitational source.

The energy share is a mean, and its uncertainty must be computed from the state and the actual response. It is not fixed by the Gaussian width alone. In a free gauge-invariant thermal state the exact variance of Eq.~\eqref{eq:cosmo-local-energy-operator} is
\begin{equation}
 (\Delta K_j(w))^2
 =\Tr_1\!\left[n_j b_j(w)(1+\sigma_j n_j)b_j(w)\right],
 \qquad
 n_j=\left(e^{h_j/T}-\sigma_j\right)^{-1}.
 \label{eq:cosmo-exact-quantum-variance}
\end{equation}
Here $\Tr_1$ is the one-particle trace, $n_j$ is the occupation-number operator, and $\sigma_j=+1$ for bosons and $-1$ for fermions. Bose enhancement and Pauli suppression therefore enter the calculation directly. The complex-scalar variance includes both charge sectors. Appendix~\ref{app:cosmology-details} derives this expression by the thermal two-point contractions, gives the covariance of overlapping cells, and distinguishes it from the variance of a differently defined local stress observable. The need to retain quantum effects for small thermal subsystems is also demonstrated in explicit relativistic-gas calculations \cite{DasEtAl2021EnergyFluctuations}.

For a massless gas and a smooth window broad compared with $T^{-1}$, the thermodynamic limit gives
\begin{equation}
 \begin{aligned}
 \langle K(w)\rangle&=c_QT^4\int w\,\dd^3X,
 & (\Delta K(w))^2&\simeq4c_QT^5\int w^2\,\dd^3X,\\
 \frac{(\Delta K(w))^2}{\langle K(w)\rangle^2}
 &\simeq\frac{1}{2\pi^{3/2}c_Q(T\ell)^3}.
 \end{aligned}
 \label{eq:cosmo-large-cell-variance}
\end{equation}
with the last expression specific to $w=e^{-r^2/(2\ell^2)}$. The size condition applies to the actual weight: if the selected portion has radius $r_g\ll\ell$, the large-cell estimate can fail even when $T\ell\gg1$. For the comoving thermal family, $T r_g=(T\ell)q$ becomes small on the formal $q\to0$ branch; there the selected-energy variance requires Eq.~\eqref{eq:cosmo-exact-quantum-variance} and the covariances of cells must be retained when summing them.

A literal discontinuous metric indicator is adequate to define the positive mean energy, but may give an infinite second moment of the corresponding sharply localized energy operator. Finite stress and energy fluctuations are therefore tested with the fixed smooth response prescribed in Sec.~\ref{sec:metric-comparison}. The smoothing is part of the universal resolution prescription, not a species-dependent parameter chosen to enhance gravitation. A smooth monotone radial response changes the function $\mathcal G$ quantitatively while retaining the monotonicity statements proved in Appendix~\ref{app:cosmology-details}.

\subsection{A self-consistent homogeneous example for all spatial-curvature signs}

We now let the gravitating energy affect the background. 
To estimate the gravitating radius, we expand the classical metric
about the center of the mode in Riemann normal coordinates and retain
the first curvature correction, which is quadratic in the displacement
[Eq.~\eqref{eq:metric-local-expansion}; see also
\cite[Sec.~8.2]{PoissonPoundVega2011}]. For an FRW background, the
resulting metric mismatch at physical tangent distance $r$ is
$D\simeq r^2/(3L_{\rm curv}^2)$ by Eq.~\eqref{eq:metric-frw-leading}.
It reaches the tolerance $\zeta^2/3$ of
Eq.~\eqref{eq:metric-gravitating-energy} at
$r_g\simeq\zeta L_{\rm curv}$. A local prediction requires control of higher-order terms at the crossing and a monotone comparison domain covering the energy profile, or a bound on the energy outside that domain. If an unresolved tail contains a fraction $\delta_{\rm tail}$, it can change the selected fraction by up to $\delta_{\rm tail}$; relative accuracy at small $f$ requires $\delta_{\rm tail}\ll f$.

Take a common comoving width $\ell=a\ell_0$, a radiation ensemble with $\varepsilon_Q=\varepsilon_{Q0}a^{-4}$, vanishing cosmological constant for this example, and a homogeneous source reconstruction calibrated so that the selected energy per physical cosmic volume is $\varepsilon_g=f\varepsilon_Q$, with isotropic pressure $p_g=f\varepsilon_Q/3$. This calibration is part of the reconstruction: tangent integration with $\dd^3X$ is not identified by coordinate equality with integration over a cosmic slice. Here $\varepsilon_{Q0}=c_QT_0^4$. We first solve the quadratic model with the sharp criterion and untapered Gaussian, $f=\mathcal G(q)$. When the gravitational response is resolved, the same smooth window must be used for its mean and its variance; Appendix~\ref{app:cosmology-details} gives the corresponding mean $\mathcal G_g(q)$.

Changes in the fraction transfer energy into or out of the gravitating source. The response tensor $C_{\mu\nu}$ must supply the balance described in Sec.~\ref{sec:source}. One explicit homogeneous closure is
\begin{equation}
 \varepsilon_C=0,
 \qquad
 p_C=-\frac{\varepsilon_Q}{3}\frac{\dd f}{\dd\ln a},
 \qquad
 \dot\varepsilon_g+3H(\varepsilon_g+p_g+p_C)=0.
 \label{eq:cosmo-response-closure}
\end{equation}
Here $\varepsilon_C$ and $p_C$ are the energy density and pressure of $C_{\mu\nu}$ in the fluid frame. We choose zero response density, so the pressure balances the change of gravitating energy. A response that carries energy density can also satisfy the Bianchi identity and can give a different expansion rate.

In this radiation example, reporting the bosonic and fermionic contributions separately or together gives the same selected energy density $\varepsilon_g$ and hence the same pressure $p_g=\varepsilon_g/3$. The history of the total fraction $f$ is unchanged, so Eq.~\eqref{eq:cosmo-response-closure} gives the same response pressure $p_C$ and the same cosmological evolution.

Define $D_0=8\pi G\varepsilon_{Q0}/3$, where $G$ is Newton's constant, and $B=H^2+k/a^2$. We construct the branch of Eq.~\eqref{eq:cosmo-curvature-approximation} on which $|\ddot a/a|\leq B$, so $L_{\rm curv}^{-2}=B$. Einstein's energy equation and $q=r_g/(a\ell_0)$ then give
\begin{equation}
 \begin{aligned}
 B&=D_0a^{-4}f, & f&=\mathcal G(q),\\
 a&=\alpha_\ell q\sqrt{\mathcal G(q)},
 &\alpha_\ell&=\frac{\ell_0\sqrt{D_0}}{\zeta}.
 \end{aligned}
 \label{eq:cosmo-implicit-solution}
\end{equation}
The dimensionless parameter $\alpha_\ell$ is fixed by the initial width, initial energy density, and universal tolerance. It is not a time-dependent efficiency fitted to enforce the result. Since $q\sqrt{\mathcal G(q)}$ is strictly increasing from zero to infinity, this equation defines a unique $q(a)>0$ on this branch. The dynamically accessible values must also satisfy Einstein's expansion equation,
\begin{equation}
 H^2=\frac{D_0f(a)}{a^4}-\frac{k}{a^2}\geq0.
 \label{eq:cosmo-expansion-domain}
\end{equation}
This distinction matters for positive spatial curvature, where an expanding solution need not reach arbitrarily large $a$.

To check both monotonicity and the curvature used in the calculation, define $\gamma_f(q)=q\mathcal G'(q)/\mathcal G(q)$. Direct integration gives $0<\gamma_f(q)<3$, and differentiation yields
\begin{equation}
 \frac{\dd\ln q}{\dd\ln a}=\frac{2}{2+\gamma_f},
 \qquad
 \frac{\dd\ln f}{\dd\ln a}=\frac{2\gamma_f}{2+\gamma_f}>0,
 \qquad
 \frac{\ddot a}{a}=-\frac{2B}{2+\gamma_f}.
 \label{eq:cosmo-self-consistent-growth}
\end{equation}
Thus $|\ddot a/a|\leq B$, verifying the curvature branch used in Eq.~\eqref{eq:cosmo-implicit-solution}. For every curvature sign, $\dot f=Hf\,2\gamma_f/(2+\gamma_f)>0$ during expansion on this branch. The full quadratic maximum can also admit local solutions on which $|\ddot a/a|>B$ and $f$ decreases during expansion (Appendix~\ref{app:cosmology-details}). The result is an existence statement for the growing branch of this closure.

For the sharp Gaussian model, Eq.~\eqref{eq:cosmo-implicit-solution} gives $f\simeq0.00233$, $0.03594$, $0.4615$ and $0.9738$ at $a/\alpha_\ell=0.01$, $0.1$, $1$ and $3$, respectively. Thus the gravitating fraction rises from approximately $0.23\%$ to $97.38\%$ across these values. A smooth response changes the numerical fractions. For a closed solution, only values reached before its maximum scale factor are physically accessible.

Writing $c_3=\sqrt{2/\pi}/3$, the small-$q$ behavior $\mathcal G(q)=c_3q^3+O(q^5)$ gives
\begin{equation}
 f\sim c_3^{2/5}\left(\frac{a}{\alpha_\ell}\right)^{6/5}
 \quad(a\rightarrow0),
 \qquad
 f\longrightarrow1
 \quad(a\rightarrow\infty,\ k\leq0).
 \label{eq:cosmo-early-late}
\end{equation}
For a flat background the corresponding formal early behavior is $a\propto t^{5/7}$, whereas the late behavior approaches radiation expansion $a\propto t^{1/2}$. In an open background the sufficiently late expansion becomes curvature dominated; the quadratic model still gives $B\to0$ and the formal limit $f\to1$. Extending this limit to the full metric rule requires control of the comparison domain and higher-order terms over the relevant energy profile; small $B$ alone does not validate an arbitrarily large quadratic estimate of $r_g$. The small-$a$ fraction in Eq.~\eqref{eq:cosmo-early-late} also describes the formal early branch with $k>0$; its large-$a$ limit is not a future limit of the recollapsing closed example.

For $k>0$, inserting Eq.~\eqref{eq:cosmo-implicit-solution} into Eq.~\eqref{eq:cosmo-expansion-domain} gives $\dot a^2=\zeta^2/(\ell_0^2q^2)-k$. Since $a(q)$ is strictly increasing, there is exactly one positive maximum scale factor. Denoting quantities at this turnaround by a subscript ``max'', one obtains
\begin{equation}
 q_{\max}=\frac{\zeta}{\ell_0\sqrt{k}},
 \qquad
 f_{\max}=\mathcal G(q_{\max}),
 \qquad
 a_{\max}=\sqrt{\frac{D_0f_{\max}}{k}}.
 \label{eq:cosmo-closed-turnaround}
\end{equation}
In this closed radiation model with $k>0$ and $\Lambda=0$, the
universe momentarily stops expanding at $a=a_{\max}$. At that
instant, the gravitating fraction also stops increasing:
$\dot f=(\dd f/\dd a)\dot a=0$. Its slope $\dd f/\dd a$ remains
finite and positive, however, as follows from
Eq.~\eqref{eq:cosmo-self-consistent-growth}. The response pressure
in Eq.~\eqref{eq:cosmo-response-closure} can therefore be evaluated
directly as $p_C=-(\varepsilon_Q/3)\,a\,\dd f/\dd a$. It remains
finite and negative at maximum expansion, even though the rate of
change of the gravitating fraction vanishes there.

The fraction increases throughout the expanding branch, but near-complete gravitation before recollapse requires $q_{\max}\gg1$; it is not guaranteed by expansion alone. The acceleration in Eq.~\eqref{eq:cosmo-self-consistent-growth} is strictly negative at $a_{\max}$, so the solution continues into contraction. At that event the comparison radius is $r_g=\zeta a_{\max}/\sqrt{k}$, namely $\zeta$ times the spatial curvature radius. The formulas therefore retain their local approximation and controlled-profile qualifications; a comparison radius of order the curvature radius requires evaluation of the full metric. Recollapse here follows from the specified radiation and response closure with $\Lambda=0$, not from positive spatial curvature alone.


The same growing branch exists for a normalized distribution of comoving widths with fixed energy weights (Appendix~\ref{app:cosmology-details}); Eq.~\eqref{eq:cosmo-closed-turnaround} is the single-width result. For a thermal massless complex scalar take $g_b=2$, $g_f=0$. The total selected energy is the sum of the positive species contributions.

At $a(t_0)=1$, an illustrative present-epoch test for one sharp Gaussian cell is $f_j(t_0)\geq1-\delta$ if and only if $r_g(t_0)/\ell_j(t_0)\geq\mathcal G^{-1}(1-\delta)$, for $0<\delta<1$. At leading curvature order this requires $\zeta\geq [\ell_j(t_0)/L_{\rm curv}(t_0)]\,\mathcal G^{-1}(1-\delta)$. For illustration, $99\%$ of the energy lies within $3.37$ widths; this illustration is not an experimental bound. Applying the condition to measured sources requires their actual profiles, controlled tails and an experimental bound on the absolute deficit of the source in question, as in Eq.~\eqref{eq:today-mass-calibration}. A bound on a difference between materials cannot be used as that $\delta$. A resolved mask has its own cumulative function. The normalization of $a$ fixes none of these data; in the closed model an expanding present also requires $D_0f(1)>k$.

The prescribed geometry of Subsection~\ref{subsec:thermal-fraction} and the growing solution of the homogeneous feedback model therefore give two distinct expansion results. On the latter branch the selected radiation contributes to the geometry, reaches its largest fraction at the closed-model turnaround, and loses fraction during recollapse.

\subsection{A localized complex-scalar packet is a different problem}

The thermal example fixes a family of Gaussian response windows. It does not assert that every freely propagated scalar packet keeps that profile. A useful check comes from a massive complex scalar in a prescribed spatially flat radiation background, where $m_R\gg H_i$, $m_R\ell_i\gg1$, and the packet has negligible mean peculiar momentum. Here $H_i=H(t_i)$ and $\ell_i$ is its initial one-coordinate position width. In the leading nonrelativistic, adiabatic one-particle regime, an initially unchirped minimum-uncertainty Gaussian obeys
\begin{equation}
 \ell_{\rm pkt}^2(a)
 =\left(\frac a{a_i}\right)^2\ell_i^2
 \left[1+\gamma_i^2\ln^2\!\left(\frac a{a_i}\right)\right],
 \qquad
 \gamma_i=\frac{1}{2m_RH_i\ell_i^2}.
 \label{eq:cosmo-scalar-spreading}
\end{equation}
Here $\ell_{\rm pkt}(a)$ is the physical position width of the evolving
scalar packet along one spatial direction. It is a property of the
quantum state and need not equal the prescribed width $\ell_j(a)$ of
the Gaussian POVM window introduced in Subsection~\ref{subsec:thermal-localization}.
The parameter $\gamma_i$ compares dispersive spreading with expansion. This follows from the redshifting kinetic generator $\bm p_{\rm com}^2/(2m_Ra^2)$, with $\bm p_{\rm com}$ the comoving momentum.

Equation~\eqref{eq:cosmo-scalar-spreading} follows by specializing
the free-particle propagation and position-moment formulas of
Al-Hashimi and Wiese~\cite{AlHashimiWiese2009}
to an initially unchirped minimum-uncertainty Gaussian in the
nonrelativistic regime of a spatially flat, radiation-dominated universe.
In this regime the assigned-energy profile is Gaussian to leading order in kinetic energy divided by $m_R$; it is not exactly Gaussian at finite relativistic momentum.

On the monotone normal neighborhood of this spatially flat radiation background the threshold radius is proportional to $t\propto a^2$. Put $v=\ln(a/a_i)$. The radius-to-width ratio therefore satisfies
\begin{equation}
 \frac{\dd\ln(r_g/\ell_{\rm pkt})}{\dd\ln a}
 =1-\frac{\gamma_i^2v}{1+\gamma_i^2v^2}
 \geq1-\frac{\gamma_i}{2}
 \qquad(v\geq0).
 \label{eq:cosmo-scalar-growth-condition}
\end{equation}
For this unchirped Gaussian, $\gamma_i\leq2$ gives a nondecreasing leading fraction throughout the interval of validity. If $\gamma_i>2$, the derivative is negative near $v=1/\gamma_i$ whenever that epoch is within the nonrelativistic regime and the profile is covered or its tail controlled. Initial position--momentum correlation changes this test. Appendix~\ref{app:cosmology-details} gives the covariance and the momentum and energy uncertainties; the initial width alone cannot determine the result.

We now follow the same freely propagating Gaussian packet for a given
scale factor $a(t)$, without assuming radiation domination. The packet
spreads because its momentum components have different peculiar
velocities. In the nonrelativistic regime these velocities decrease as
$1/a(t)$ as the momenta redshift. Converting a physical displacement
to a comoving displacement supplies a second factor of $1/a(t)$, so
the accumulated comoving spreading depends on
$\int_{t_i}^{t}\dd t'/a(t')^2$. The resulting physical width of the
packet is
\begin{equation}
 \begin{aligned}
 \ell_{\rm pkt}^2(t)
 &=\left(\frac{a(t)}{a_i}\right)^2\ell_i^2
   \left[1+\vartheta_i(t)^2\right],\\
 \vartheta_i(t)
 &=\frac{a_i^2}{2m_R\ell_i^2}
   \int_{t_i}^{t}\frac{\dd t'}{a(t')^2}.
 \end{aligned}
 \label{eq:cosmo-general-scalar-spreading}
\end{equation}
Here $\vartheta_i$ is the dimensionless accumulated dispersive spreading, initially zero, and $t'$ is the integration time. Equation~\eqref{eq:cosmo-scalar-spreading} is recovered in the flat radiation case, where $\vartheta_i=\gamma_i\ln(a/a_i)$. For $k\neq0$ this local formula additionally requires $\ell_{\rm pkt}\ll a/\sqrt{|k|}$, with $a/\sqrt{|k|}$ the spatial curvature radius. The nonrelativistic and adiabatic approximations must remain valid, including $m_RL_{\rm curv}\gg1$; small $H$ alone does not ensure this near a turnaround. It is not a global Gaussian solution on a compact curved slice.

On a prescribed standard radiation background with $\Lambda=0$, the local comparison gives $r_g\propto a^2$ for either curvature sign, as discussed after Eq.~\eqref{eq:cosmo-curvature-approximation}. Combining it with Eq.~\eqref{eq:cosmo-general-scalar-spreading} yields the proper-time growth test
\begin{equation}
 \frac{\dd}{\dd t}\ln\!\left(\frac{r_g}{\ell_{\rm pkt}}\right)
 =H-\frac{\vartheta_i\dot\vartheta_i}{1+\vartheta_i^2},
 \qquad
 \dot\vartheta_i=\frac{a_i^2}{2m_R\ell_i^2a^2}.
 \label{eq:cosmo-general-scalar-growth}
\end{equation}
Equation~\eqref{eq:cosmo-general-scalar-growth} applies to all three
spatial-curvature signs on the prescribed radiation background with
$\Lambda=0$. In the flat ($k=0$) and open ($k<0$) cases, the universe
continues to expand and there is no turnaround. The gravitating
fraction of the Gaussian packet increases whenever the comparison
radius $r_g$ grows faster than the packet width $\ell_{\rm pkt}$,
that is, whenever the right-hand side of
Eq.~\eqref{eq:cosmo-general-scalar-growth} is positive. Expansion
alone does not guarantee this, because the packet also spreads.

Consider now the closed case ($k>0$). The universe reaches a finite
maximum scale factor, where $H=0$. As it approaches this point from
the expanding side, the growth of $r_g\propto a^2$ slows to zero,
whereas a packet that has already begun to spread continues to broaden.
Its gravitating fraction can therefore begin to decrease even before
the universe starts to contract. \textit{This conclusion concerns a freely
evolving scalar packet; the prescribed thermal windows of the preceding
subsection have widths proportional to $a$ and do not undergo this
additional dispersive spreading.}

There is no universal monotonic theorem for all scalar states. For example, an outward-moving, spherically symmetric massless packet can develop an energy shell whose physical radius grows like $a\eta$, where $\eta$ is conformal time. In the spatially flat radiation example this is proportional to $t$, just as $r_g$ is. Depending on the fixed tolerance and initial packet, much of the energy can remain outside the central region. The general prediction is obtained from the transported positive energy measure and the exact metric window; the thermal and sufficiently slowly dispersing examples above identify conditions under which increased gravitation is actually established.

\subsection{Contraction and a decreasing gravitating fraction}
\label{sec:collapse}

The same rule governs increasing and decreasing contributions; it introduces no independent switch for collapse. At each contact event the transported quantum data determine the positive energy profile, and the classical metric determines the tolerated central region. For a radial energy profile of fixed shape in units of its width, a decreasing ratio of accepted radius to width gives a nonincreasing fraction, strictly decreasing wherever the shrinking boundary passes through positive energy [Eq.~\eqref{eq:cosmo-general-growth-test}]. A changing shape requires the full energy integral. The tangent quantum state and its energy are not removed when its geometrical weight decreases. A decreasing fraction must also be distinguished from a decreasing absolute source density: compression can increase the latter even while a smaller fraction gravitates.

For a homogeneous contracting continuation with the same instantaneous thermal and comoving-width closure, the function $f(a)$ in Eq.~\eqref{eq:cosmo-implicit-solution} remains increasing as a function of $a$. Since $\dot a<0$ during contraction, $\dot f<0$. In particular, the closed solution in Eq.~\eqref{eq:cosmo-closed-turnaround} increases its gravitating fraction up to $f_{\max}$, has $\dot f=0$ at maximum expansion, and then decreases its fraction as it recollapses. The pressure specified by Eq.~\eqref{eq:cosmo-response-closure} remains finite through this turnaround. This continuation assumes that the same thermal state prescription and comoving localization family remain applicable; it is not a statement that arbitrary quantum states retrace their expansion history. A specified spatially flat contracting power-law background gives the corresponding exact metric-scaling example by replacing $t$ with the proper time remaining to the homogeneous singularity, for as long as the quantum profile stays inside the regular comparison domain. These are controlled statements about the stated homogeneous families, rather than an assumption that every collapsing object follows their equation of state.

Energy balance remains essential. For an isotropic radiation ensemble, the selected source alone obeys
\begin{equation}
 \dot\varepsilon_g+3H(\varepsilon_g+p_g)
 =\varepsilon_Q\dot f,
 \qquad
 \nabla^\mu C_{\mu\nu}
 =-\nabla^\mu T^{\rm grav}_{\mu\nu}.
 \label{eq:cosmo-collapse-balance}
\end{equation}
The right-hand side of the first equation is negative during the contraction just described. It describes a loss relative to the compressional evolution of the selected radiation, not necessarily $\dot\varepsilon_g<0$. For example, the formal small-$a$ solution still has $\varepsilon_g\propto a^{-14/5}$. The second equation is the covariant balance condition: the response tensor accounts for the change in energy--momentum that gravitates. A time-dependent multiplier of a conserved matter tensor, with no corresponding response, would not satisfy Einstein's equations.

In an inhomogeneous collapse the full metric window is generally directional, the packet can focus as well as spread, and pressure and shear affect the source. One must therefore evolve the state moments, the metric comparison, and the stress balance together. Neither the existence of a trapped surface nor the growth of a single scalar curvature invariant automatically sets the fraction to zero. Conversely, a decreasing fraction alone does not prove that a black-hole interior is regular: the response stress and the remaining gravitating energy can still become large. The construction supplies a specific criterion to investigate such solutions, while their existence, stability, and regularity remain dynamical questions for future research consideration.

\section{Summary of the construction and open questions}
\label{sec:algorithm}

\subsection{The proposed coupling, step by step}

The quantum state and the geometry supply different inputs at each step. The following sequence specifies how they couple
\begin{enumerate}
\item \emph{Set the initial data.} On a classical spacelike slice specify its spatial metric and its normal rate of change, subject to the Einstein constraints for the total source in Eq.~\eqref{eq:source-einstein}. Fix the observers and clocks, the tangent-Minkowski matter theory and state with the moments in Eq.~\eqref{eq:main-state-moments}, and the physical localization instrument and widths. For a complex scalar these inputs include its renormalized mass and interaction, and either a prepared packet or a specified thermal ensemble.
\item \emph{Transport the field.} Follow a free massive packet's central geodesic and frame by Eq.~\eqref{eq:main-reference-trajectory-clock}; for massless profiles use the specified observer's spacelike sections. Evolve the canonical profile and separate its geometric change from the Minkowski reference evolution by Eqs.~\eqref{eq:P-flat-limit}--\eqref{eq:main-residual-exponential}. The algebra map and its unitary implementation in Eqs.~\eqref{eq:main-algebra-lift}--\eqref{eq:main-quantum-transport} give the quantum evolution in Eq.~\eqref{eq:main-full-ordered-propagator}, with ordinary tangent interactions included once.
\item \emph{Localize and assign energy.} Apply one complete instrument to the unconditioned state at the endpoint. Equation~\eqref{eq:main-localization} gives the probabilities and conditional states; Eq.~\eqref{eq:main-energy-partition} partitions the original state's energy. Equation~\eqref{eq:metric-energy-measure} locates each share within its profile. These shares, rather than the conditional mean energies, are the quantities to be added.
\item \emph{Determine each mode's extent.} Use the position and momentum moments in Eqs.~\eqref{eq:main-position-moments}--\eqref{eq:ellipsoid-new} and \eqref{eq:main-intrinsic-covariance}--\eqref{eq:main-covariance-uncertainty} to follow its width, spreading and correlations. Equation~\eqref{eq:worldtube-map-new} places the successive profile contours in spacetime, and Eq.~\eqref{eq:main-resolved-boundary-flux} samples the flux through their moving boundary. The gravitational test uses the energy profile, including its tails.
\item \emph{Determine the energy that gravitates.} At each contact point pull the classical metric back by Eq.~\eqref{eq:metric-pullback} and compute its difference from the tangent metric with Eq.~\eqref{eq:metric-mismatch}. Integrate the assigned energy within the universal tolerance by Eq.~\eqref{eq:metric-gravitating-energy}. At finite resolution use Eq.~\eqref{eq:metric-smooth-energy} and evaluate the fluctuations of that same resolved quantity.
\item \emph{Construct one gravitational source.} Transfer the selected energies, stresses and fluxes to the classical slice with Eq.~\eqref{eq:source-reconstruction}, using its proper volume and a common energy frame. Reconstruct each contribution at the prescribed physical resolution, including each interaction contribution once. If several labels are reported together, add their already reconstructed tensors as in Eq.~\eqref{eq:source-einstein}; carry the same total source into the Einstein evolution in the next step. For example, the energy gained by an absorbing detector is transferred from the field in the joint budget in Eq.~\eqref{eq:detector-gravitating-budgets}.
\item \emph{Evolve the metric and matter together.} Solve Einstein's equation \eqref{eq:source-einstein} with the specified response $C_{\mu\nu}$ satisfying the balance law \eqref{eq:source-balance}. Evaluate the transport and the energy fractions on that same evolving geometry. In the homogeneous radiation example, Eqs.~\eqref{eq:cosmo-response-closure}--\eqref{eq:cosmo-self-consistent-growth} carry out this last step on the expanding solution studied here.
\end{enumerate}

\subsection{Results and their scope}
\label{subsec:results-scope}

The positive energy partition and its geometrically weighted refinement are exact operator-form statements on their stated domains. The full metric difference is also exact wherever the comparison map is defined. Its normal-coordinate expansion explains, rather than assumes, the appearance of a curvature radius. The fixed-background FRW scaling result establishes increasing gravitating energy fraction for the stated profile family and comparison domain. The feedback example constructs an expanding solution on the curvature branch verified by Eq.~\eqref{eq:cosmo-self-consistent-growth}. For an individual scalar packet, Eq.~\eqref{eq:cosmo-general-scalar-growth} compares the growth of the accepted radius with the actual spreading of the packet.

The case of ``more modes gravitate'' is made quantitative by the energy fraction. An extended mode can already have a nonzero contribution at an early epoch and a larger fraction later; a Gaussian has no finite radius containing its entire energy. For a chosen operational threshold $0<f_*<1$, the number or outcome measure of cells with $f_A\geq f_*$ is nondecreasing if each member's fraction is nondecreasing and the outcome labels and their counting or integration measure are held fixed. The threshold $f_*$ is a reporting convention, not an additional gravitational constant. The unambiguous source is always the sum of the actual weighted energies and reconstructed tensors. A count of modes above threshold refers to the original physical localization family. Combining its labels can change the reported count, but cannot change the total gravitating energy, stress or curvature.

A comparison of gravitational and inertial mass can test the universal tolerance for ordinary matter today. For a weakly bound, slowly moving body $B$ at rest, let $M_{i,B}$ be its inertial mass, $M_{g,B}$ its gravitational mass as inferred from the field it produces, and $F_B$ the fraction of its energy entering the gravitational source. Write $E_B^{\rm cl}$ for the energy of any separately specified classical matter in $B$, measured in the same rest frame as the quantum shares and counted only once. In units $c=1$, if pressure, binding and the response make negligible corrections to the measured mass, a comparison that allows a fractional deficit $\delta_B$ requires
\begin{equation}
 \frac{M_{g,B}}{M_{i,B}}\simeq F_B(\zeta)
 :=\frac{E_B^{\rm cl}+\sum_A\mathcal E_A^{\rm grav}(\zeta)}{E_B^{\rm cl}+\sum_A\mathcal E_A},
 \qquad 1-F_B(\zeta)\leq\delta_B .
 \label{eq:today-mass-calibration}
\end{equation}
The sums run over the quantum shares of $B$. A wholly classical body has $F_B=1$ for every $\zeta$ and by itself does not constrain the tolerance. For finite quantum profiles, a mass comparison constrains $\zeta$ only through a predicted dependence of the measured quantity on $\zeta$; the profiles must be fixed independently. If stresses, binding or $C_{\mu\nu}$ matter at that precision, the limit applies to the mass inferred from the full solution of Eq.~\eqref{eq:source-einstein}. MICROSCOPE compared the free-fall accelerations of titanium and platinum in the same terrestrial field \cite{Touboul2022MICROSCOPE}. With universal free fall, this differential measurement does not determine the fraction of either body's energy that produces gravity. Lunar laser ranging tests a different quantity: within the adopted lunar model, the reported bound of $3.9\times10^{-14}$ is on the difference between the active-to-passive gravitational mass ratios of aluminum and iron \cite{Singh2023ActivePassive}. Here \emph{active} refers to the field a body produces and \emph{passive} to its response to an applied field; these are the terms of the experimental analysis. Neither experiment directly measures the absolute deficit $1-M_{g,B}/M_{i,B}$ in Eq.~\eqref{eq:today-mass-calibration}.

\subsection{Further construction and physical tests}

The explicit scalar transport and thermal energy distributions give starting points for future work. We have to study the algebra map, unitary implementation and short-distance correlations for interacting fields, and determine how endpoint instruments and resolved stress data can be reconstructed on one common geometry. Further applications will test the local-equilibrium approximation and specify how to continue the metric comparison for profiles extending beyond one normal neighborhood.

For a detector coupled to separated packets, Eq.~\eqref{eq:record-early-no-signal} gives identical early readout distributions under the local-instrument assumptions, and Eq.~\eqref{eq:detector-strong-support} gives causal support for the change in the total source on a prescribed curved metric. Future work  need to identify the physical response function, calculate the later metric and response energy, and study stability and causal evolution when the metric is solved for dynamically. It should also test whether one response law can describe both the detector and the cosmological pressure in Eq.~\eqref{eq:cosmo-response-closure}. Measurements of relative phases, source masses and free fall probe complementary parts of the construction; cosmological perturbations and source fluctuations offer further tests.

One possible laboratory test is to compare the gravitational fields
produced by two preparations of the same atoms with equal inertial mass
but different measured packet widths. If the metric criterion assigns
different gravitating fractions to the two preparations, it predicts a
difference in their fields beyond that expected from the ordinary change
in mass distribution.

\section{Conclusions}

The construction distinguishes how quantum matter propagates through
classical spacetime from how its energy contributes to the gravitational
field. Along the reference trajectory of a freely propagating massive
packet, a component of fixed local energy accumulates the usual phase
with respect to the proper time defined in
Eq.~\eqref{eq:main-reference-trajectory-clock}. In the geometrical-optics
limit, Eq.~\eqref{eq:main-observed-packet-frequency} gives the frequency
measured by each observer; comparing the frequencies at emission and
reception yields the standard redshift.
For a free massless complex scalar in a spatially flat radiation universe we exhibit a calibration with unitary field transport. The localized energy shares are positive and sum to the inertial energy. Applying the full metric comparison to each share determines how much enters the common Einstein source, whose total stress satisfies the Bianchi balance. The vacuum-subtracted tangent Hamiltonian and the zero-source condition \eqref{eq:vacuum-reconstruction-condition} give exactly zero gravitational contribution from the reference vacuum at every epoch, independently of the classical cosmological constant.

In the decelerating expanding FRW examples, the region accepted by the metric grows faster than the prescribed comoving thermal profiles. A progressively larger fraction of the energy of bosons and fermions therefore gravitates; Eq.~\eqref{eq:cosmo-exact-quantum-variance} also quantifies the fluctuations of that energy at a specified resolution.
The same increase occurs in the homogeneous radiation solution of
Eq.~\eqref{eq:cosmo-self-consistent-growth}, where the Friedmann equation
determines the expansion rate from the gravitating energy density.
For a massive complex-scalar packet the competition is between the widening accepted region and quantum spreading, with the growth condition given by Eq.~\eqref{eq:cosmo-general-scalar-growth}. 

An independently classical body contributes its full stress tensor; the local metric rule gives $F_B=1$ in Eq.~\eqref{eq:classical-full-fraction}. For matter described by quantum profiles, the same universal tolerance must be consistent with measurements comparing gravitational and inertial mass, under the conditions of Eq.~\eqref{eq:today-mass-calibration}.

\appendix

\section{Tangent fields, normalization and finite observables}
\label{app:bundle-foundations}

The physical purpose of these conditions is to ensure that the energy and stress used in the gravitational construction are actual finite observables. The Hilbert bundle and its observable representations are defined in Sec.~\ref{sec:geometry}; they do not require a particle-number decomposition. The following free complex-scalar normalization supplies a concrete example, while interactions remain part of the fixed renormalized Minkowski theory \cite{Haag1996,Collins1984}:
\begin{align}
 \widehat\Phi_{R,x}(X)&=\int d\mu_R(\bm k)
 \left[a_x(\bm k)e^{-i\omega^R_{\bm k}X^0+i\bm k\cdot\bm X}
 +b_x^\dagger(\bm k)e^{i\omega^R_{\bm k}X^0-i\bm k\cdot\bm X}\right],
 \label{eq:flatfield}\\
 d\mu_R(\bm k)&=\frac{d^3\bm k}{(2\pi)^3\,2\omega^R_{\bm k}},
 \qquad \omega^R_{\bm k}=\sqrt{|\bm k|^2+m_R^2},
 \nonumber\\
 [a_x(\bm k),a_x^\dagger(\bm q)]
 &=[b_x(\bm k),b_x^\dagger(\bm q)]
 =(2\pi)^3\,2\omega^R_{\bm k}\delta^{(3)}(\bm k-\bm q).
 \label{eq:complex-scalar-energy-commutators}
\end{align}
Here $a_x,b_x$ annihilate particles and antiparticles; independent mixed commutators vanish. No rescaling of these invariantly normalized operators is made. Their vacuum-subtracted energy is
\begin{align}
 \widehat H_R^{\rm M}=\int d\mu_R(\bm k)\,\omega^R_{\bm k}
 (a_x^\dagger a_x+b_x^\dagger b_x),
 \label{eq:complex-scalar-positive-hamiltonian}\\
 \widehat H_R\geq0,
 \qquad \widehat H_R|0\rangle=0.
 \label{eq:positive-H}
\end{align}
The positivity and vacuum convention are also assumptions on the admitted interacting theory; its Hamiltonian is not generally the displayed free sum. A normalized packet has $a_x^\dagger(\varphi_A)=\int d\mu_R\,\varphi_Aa_x^\dagger$ and $\int d\mu_R|\varphi_A|^2=1$. Equation~\eqref{eq:main-scalar-state-profile} then gives $\Phi_A=\int d\mu_R\,\varphi_A e^{-i\omega^R X^0+i\bm k\cdot\bm X}$ and $\langle\widehat H_R^n\rangle=\int d\mu_R(\omega^R)^n|\varphi_A|^2$. This makes the physical meaning of finite energy moments explicit.

For the interacting complex-scalar example the tangent Hamiltonian is instead
\begin{equation}
 \widehat H_{\Phi,R}^{\rm M}=\int d^3X\,
 \left[\widehat\Pi_R^\dagger\widehat\Pi_R
 +\bm\nabla\widehat\Phi_R^\dagger\!\cdot\!\bm\nabla\widehat\Phi_R
 +m_R^2\widehat\Phi_R^\dagger\widehat\Phi_R
 +\tfrac{\lambda_R}{2}(\widehat\Phi_R^\dagger\widehat\Phi_R)^2\right]_R^{\rm M}.
 \label{eq:scalar-Minkowski-Hamiltonian}
\end{equation}
Here $\widehat\Pi_R=\partial_{X^0}\widehat\Phi_R^\dagger$ is the canonical momentum, and the brackets mean the fixed Minkowski composite renormalization with its vacuum convention. Couplings to other fields are included once in the total Hamiltonian. This quantum generator is distinct from the c-number profile action in Appendix~\ref{app:profile-foundations}.

The state domain $\mathfrak D_{\rm adm}(A)$ is specified in Eq.~\eqref{eq:main-state-moments}. In addition, resolved stress means and covariances must depend continuously on their smooth sampling functions. On each compact test support $K$, require some finite integer $N$ and constant $C_{\rho,K}$ such that
\begin{equation}
 \bigl[\Tr(\rho\widehat T_R[f]^2)\bigr]^{1/2}
 \leq C_{\rho,K}p_{K,N}(f),\qquad
 p_{K,N}(f)=\max_{\hat a,\hat b}\sum_{|\alpha|\leq N}
 \sup_{X\in K}|\partial^\alpha f^{\hat a\hat b}(X)|.
 \label{eq:stress-test-continuity}
\end{equation}
Here $\alpha$ is a derivative multi-index and $p_{K,N}$ is a smooth-test-function seminorm in fixed coordinate units. Cauchy--Schwarz also controls the mean and bilinear covariance. These are state/domain hypotheses, not consequences of trace normalization. Gaussian scalar momentum amplitudes illustrate finite energy moments; sharply localized states can fail the stress conditions. Infinite homogeneous equilibrium requires a local algebraic state or a controlled finite-volume limit, rather than an infinite-volume density operator with finite total energy.

The tangent Ward identity is Eq.~\eqref{eq:main-stress-Ward}. Multiplication by a smooth localization or gravitational weight $w$ instead gives $\partial^{\hat a}(w\widehat T^R_{\hat a\hat b})=(\partial^{\hat a}w)\widehat T^R_{\hat a\hat b}$. This is why the selected source needs the response of Sec.~\ref{sec:source}. Independently adjustable curvature vertices are excluded from tangent loops; admitting them would change the renormalization problem \cite{BunchParker1979,Toms1982,HollandsWald2003RG}.

\section{Canonical transport and the complex-scalar generator}
\label{app:profile-foundations}\label{app:scalar-generators}

This appendix checks that geometrical deformation preserves the field's canonical data and that Minkowski evolution is subtracted in the same coordinates. These checks prevent double counting ordinary quantum phases or confusing Hubble dilution with loss of probability. The full complex data and their real symplectic pairing are Eqs.~\eqref{eq:main-complex-data}--\eqref{eq:main-complex-symplectic}. For linear transport, or the tangent map of a nonlinear Hamiltonian flow, canonicity means
\begin{equation}
 \Omega_{\mathbb C}(\mathscr T_A^{\rm min}\delta_1Z,
                    \mathscr T_A^{\rm min}\delta_2Z)
 =\Omega_{\mathbb C}(\delta_1Z,\delta_2Z).
 \label{eq:scalar-symplectic-preservation}
\end{equation}
The variations obey the complex conjugacy relations. Finite Cauchy data have finite $\int_{\Sigma}(|n^\mu\nabla_\mu\Phi|^2+|D\Phi|_h^2+m_R^2|\Phi|^2)d\Sigma$, with $n$ the section normal, $h$ its spatial metric and $D$ its intrinsic derivative. Support may be compact or have tails whose required moments converge; ``finite'' does not mean a finite-dimensional field space. The profile problem, boundary data and endpoint identifications must be well posed before a propagator exists \cite{BaerGinouxPfaeffle2007}.

For the explicit spatially flat FRW example use cosmic time, dimensionless $a(t_0)=1$, and a fixed comoving spatial chart. With $V_R(u)=m_R^2u+\lambda_Ru^2/2$, the complex profile action and momenta are
\begin{equation}
 S^g_{A,\rm prof}=\int dt\,d^3x\,a^3
 (|\dot\Phi_A|^2-a^{-2}|\bm\nabla\Phi_A|^2-V_R(|\Phi_A|^2)),
 \qquad P_{\Phi,A}=a^3\dot\Phi_A^*,\quad
 P_{\Phi^*,A}=a^3\dot\Phi_A.
 \label{eq:FLRW-scalar-profile-action}
\end{equation}
The symbol $S$ in Eq.~\eqref{eq:FLRW-scalar-profile-action} denotes the profile action. Its Hamiltonian density is $a^{-3}P_\Phi P_{\Phi^*}+a|\bm\nabla\Phi|^2+a^3V_R$. For nonzero $\lambda_R$ this is an effective mean-profile closure: exact interacting matrix elements contain the renormalized cubic operator, rather than the product of three one-point profiles. A single such equation does not transport all interacting correlations.

For a free Fourier amplitude, retain both complex canonical pairs in $Z_{\bm k}=(\phi_{\bm k},\phi_{\bm k}^*,p_{\phi,\bm k},p_{\phi^*,\bm k})^{\mathsf T}$, with $p_{\phi}=a^3\dot\phi^*$ and $p_{\phi^*}=a^3\dot\phi$. The wave vector $\bm k$ is comoving. The generators of $\dot Z=\mathscr L Z$ in this one chart are
\begin{equation}
 \mathscr L_g=
 \begin{pmatrix}0&0&0&a^{-3}\\0&0&a^{-3}&0\\
 0&-\Omega_{a,\bm k}^2&0&0\\-\Omega_{a,\bm k}^2&0&0&0\end{pmatrix},
 \qquad
 \mathscr L_{\rm M}=
 \begin{pmatrix}0&0&0&1\\0&0&1&0\\
 0&-\omega^2&0&0\\-\omega^2&0&0&0\end{pmatrix},
 \label{eq:FLRW-curved-canonical-generator}
\end{equation}
where $\Omega_{a,\bm k}^2=a^3m_R^2+a|\bm k|^2$ and $\omega^2=m_R^2+|\bm k|^2$. They act on profile data, whereas the operators in Eqs.~\eqref{eq:scalar-finite-mode-K}--\eqref{eq:scalar-finite-mode-total} give the finite quantum example. Subtracting the two matrices gives off-diagonal coefficients $d=a^{-3}-1$ and $e=m_R^2(1-a^3)+|\bm k|^2(1-a)$. In the reordered complex pairs $(\phi,p_{\phi^*})$ and $(\phi^*,p_\phi)$, each difference block is $\left(\begin{smallmatrix}0&d\\e&0\end{smallmatrix}\right)$. With $c_\omega=\cos[\omega(t-t_1)]$ and $s_\omega=\sin[\omega(t-t_1)]$, conjugation by the reference oscillator gives each interaction-picture block
\begin{equation}
 \mathscr G_I^{(2)}=
 \begin{pmatrix}
 -c_\omega s_\omega(d\omega+e/\omega)&d c_\omega^2-e s_\omega^2/\omega^2\\
 e c_\omega^2-d\omega^2s_\omega^2&c_\omega s_\omega(d\omega+e/\omega)
 \end{pmatrix}.
 \label{eq:scalar-interaction-picture-geometric-generator}
\end{equation}
The full generator contains both identical complex blocks. This explicit multiplication verifies the conjugated difference in Eq.~\eqref{eq:main-residual-generator}; using the unrotated difference in its ordered exponential is generally incorrect. For an interacting profile the cubic Fourier term couples distinct momenta, $[|\Phi|^2\Phi]_{\bm k}=\int d^3q\,d^3r\,\phi_{\bm q}\phi_{\bm r}^*\phi_{\bm k-\bm q+\bm r}/(2\pi)^6$. The displayed linear matrices are therefore free-field examples, not an interacting closure theorem.

A common time-dependent coordinate change $Y=\mathsf C(t)Z$ transforms either generator as $\widetilde{\mathscr L}=\mathsf C\mathscr L\mathsf C^{-1}+\dot{\mathsf C}\mathsf C^{-1}$, so $\Delta\widetilde{\mathscr L}=\mathsf C\Delta\mathscr L\mathsf C^{-1}$. In velocity coordinates the same map multiplies the momenta by $a^{-3}$ and reorders them. Both transformed generators then contain the chart term $-3H$, which cancels in their difference. The reference entries are rescaled reference momenta, not separately chosen flat-space velocities. Endpoint factors likewise transform the residual by $\widetilde{\mathfrak P}_I=\mathsf C(t_1)\mathfrak P_I\mathsf C(t_1)^{-1}$. This explains why a coordinate change cannot repair a failed physical comparison.

The complete map has the usual composition law. Its residual in Eq.~\eqref{eq:P-flat-limit}, anchored to the initial data space, instead satisfies
\begin{equation}
 \mathfrak P_I(s_3,s_1)=
 [\mathscr T_{\rm M}(s_2,s_1)]^{-1}\mathfrak P_I(s_3,s_2)
 \mathscr T_{\rm M}(s_2,s_1)\mathfrak P_I(s_2,s_1).
 \label{eq:profile-residual-cocycle}
\end{equation}
The common mode label is suppressed. In the inertial flat limit the residual is the identity while ordinary Minkowski evolution remains. Tensor, spinor and gauge profiles require their representation connections and physical constraints; curvature terms generated by commuting these derivatives are fixed geometrical identities, not additional adjustable quantum vertices \cite{BaerGinouxPfaeffle2007,HollandsWald2015}.

\section{Quantum implementation and short-distance consistency}
\label{app:unitary-foundations}\label{app:transport-tests}
\label{subsec:endpoint-short-distance}

The aim is to ensure that a transported scalar profile represents a usable quantum state, with the same algebra and finite source observables. Canonical preservation alone concerns classical initial data. The dual test action and algebra map are Eqs.~\eqref{eq:main-dual-transport}--\eqref{eq:main-algebra-lift}. For a free scalar, spacetime tests are equivalence classes modulo $(\Box_\eta-m_R^2)C_0^\infty$; the causal Green operator relates these classes to solution data \cite{BaerGinouxPfaeffle2007}. Thus dual transport is not unrestricted pointwise multiplication of a spacetime test function by a Cauchy-data matrix. For interacting fields the actual operator products, renormalized composites and Ward identities must also be preserved \cite{Haag1996,HollandsWald2001}.

An algebra isomorphism need not have a unitary implementer in a specified infinite-dimensional representation. Require a strongly continuous, differentiable lift on an admitted common domain and fix its phase convention along the path. In an irreducible representation the algebra determines the unitary only up to a scalar phase; relative path phases cannot simply be discarded \cite{AgulloAshtekar2015,CombescureRobert2006}. For the total generator $\widehat G_A=\widehat K_{A,R}^{\rm M}-i\mathcal A_{{\rm Sch},A}^{\rm geo}$, assume self-adjointness, a stable common dense domain and sufficient time regularity for nonautonomous evolution. A common-domain family continuously differentiable in its graph norm, together with the required stability bounds, is one standard sufficient setting \cite{Kato1995}. Once forward and inverse evolutions exist,
\begin{equation}
 \frac{d}{ds}(\mathcal U_A^\dagger\mathcal U_A)
 =i\mathcal U_A^\dagger(\widehat G_A^\dagger-\widehat G_A)\mathcal U_A=0,
 \qquad
 \mathcal U_A(s_3,s_2)\mathcal U_A(s_2,s_1)=\mathcal U_A(s_3,s_1).
 \label{eq:conditional-unitarity-proof}
\end{equation}
This is a conditional proof of unitarity, not a proof of the hypotheses from a formal exponential. The full state domain of Eq.~\eqref{eq:main-state-moments} must also be preserved by transport and by each instrument used for conditional data.

Physical endpoint calibration is distinct from a change of coordinates. If $I_s$ maps raw canonical data to the specified tangent data, consistency requires
\begin{equation}
 \mathscr T_g^{\rm tan}(s_2,s_1)
 =I_{s_2}\mathscr T_g^{\rm raw}(s_2,s_1)I_{s_1}^{-1},\qquad
 \mathscr L_{\rm M}^{\rm raw}
 =I_s^{-1}\mathscr L_{\rm M}^{\rm tan}I_s-I_s^{-1}\dot I_s.
 \label{eq:endpoint-identifications}
\end{equation}
Use the same endpoint prescription for the classical and reference maps. Consider a free massless complex scalar in a spatially flat radiation universe with a periodic comoving box; omit the homogeneous zero mode, independently of the ultraviolet limit. In conformal time $d\eta=dt/a$, the scale factor has $a''=0$, and $\chi=a\Phi$ satisfies $\chi''-\nabla^2\chi=0$; primes denote $d/d\eta$. For one Fourier amplitude $\phi_{\bm k}$ and $k=|\bm k|>0$, write $\chi_{\bm k}=a\phi_{\bm k}$. Its unscaled momentum block contains $p_{\phi^*,\bm k}=a^2\phi_{\bm k}'=a\chi_{\bm k}'-a'\chi_{\bm k}$, with the conjugate relation for $p_{\phi,\bm k}$.

Take the comparison from $t_0$ (today, $a(t_0)=1$) to an epoch $t_*$ with $a_*=a(t_*)\ne1$; the implementability test is the same in either direction. In the scaled variables $(\sqrt{k}\chi_{\bm k},\chi_{\bm k}'/\sqrt{k})$ evolution is an oscillator rotation $R_{k\Delta\eta}$ through conformal-time separation $\Delta\eta$. Here $R_\theta$ has rows $(\cos\theta,\sin\theta)$ and $(-\sin\theta,\cos\theta)$. In the fixed unscaled Minkowski comparison of Eqs.~\eqref{eq:scalar-finite-mode-K}--\eqref{eq:scalar-finite-mode-total}, let $\mathsf M_{\bm k}$ propagate $(\sqrt{k}\phi_{\bm k},p_{\phi^*,\bm k}/\sqrt{k})$, with the conjugate block understood. Since $\phi_{\bm k}=\chi_{\bm k}/a$ and $p_{\phi^*,\bm k}=a\chi'_{\bm k}-a'\chi_{\bm k}$, the endpoint rescaling gives, as $k\to\infty$,
\begin{equation}
 \mathsf M_{\bm k}=\operatorname{diag}(a_*^{-1},a_*)R_{k\Delta\eta}+O(k^{-1}),
 \qquad |\beta_{\bm k}|=\tfrac12|a_*-a_*^{-1}|+O(k^{-1}).
 \label{eq:unscaled-radiation-test}
\end{equation}
Here $\beta_{\bm k}$ is the coefficient of negative frequency relative to the fixed Minkowski oscillators. A reference rotation changes only its phase. Its nonzero limiting magnitude makes $\sum_{\bm k}|\beta_{\bm k}|^2$ diverge, so the unscaled comparison has no unitary on that Fock space when the cutoff is removed \cite{Shale1962,Castello2012Fock}.

For the alternative physical calibration, use the complex conjugate canonical pairs $(\chi,(\chi^*)')$ and $(\chi^*,\chi')$ at each endpoint, including the momentum shift $\chi'=a^{-1}P_{\Phi^*}+a'\Phi$. Evolution is now a rotation with $\beta_{\bm k}=0$ for every nonzero mode. For parallel reference rest planes translated through cosmic time $\Delta t=t_*-t_0$, the residual is $R_{k(\Delta\eta-\Delta t)}$ with unitary lift $\exp[-i\widehat H_R^{\rm M}(\Delta\eta-\Delta t)]$. It preserves the reference vacuum, finite energy moments and the smooth tangent two-point difference in Eq.~\eqref{eq:main-tangent-smoothness} by Minkowski time translation. This gives a full-field example without a high-frequency cutoff. The two calculations identify different measured variables with the endpoint canonical data; merely changing coordinates on a fixed comparison cannot change implementability. Relating the scaled variables to physical momenta $k/a$, lengths and calorimetric energy is a separate calibration.

The proposed smooth tangent difference in Eq.~\eqref{eq:main-tangent-smoothness} must not be confused with subtracting a flat kernel from a curved-spacetime bisolution. The conventional four-dimensional scalar Hadamard singular model is
\begin{equation}
 H_g(x,y)=\frac{1}{8\pi^2}
 \left[\frac{U_g(x,y)}{\sigma_+(x,y)}
 +V_g(x,y)\ln(\mu^2\sigma_+(x,y))\right],
 \quad
 V_g(x,x)=\frac12\left[m_R^2+\left(\xi-\frac16\right)R(x)\right].
 \label{eq:curved-Hadamard-parametrix}
\end{equation}
Here $\sigma$ is half the squared geodesic interval, $\sigma_+$ its time-oriented distributional prescription, $\mu$ an inverse-length scale, and $U_g,V_g$ smooth geometrically fixed biscalars. The first is the square root of the Van Vleck determinant. The repeated argument means coincidence of the two spacetime points. A Hadamard two-point function differs smoothly from this local singular model \cite{Radzikowski1996,SahlmannVerch2001,DecaniniFolacci2008}. Even at the prescribed minimal value $\xi=0$, $V_g(x,x)=m_R^2/2-R/12$ contains curvature. It is not the scalar potential $V_R$. Consequently, flat subtraction of a genuinely curved bisolution generally leaves singular curvature terms. Equation~\eqref{eq:main-tangent-smoothness} instead compares two distributions in one calibrated tangent algebra. Outside the explicit free sector above, preservation and nonemptiness must be checked for the proposed calibration; higher interacting correlations need additional control \cite{HollandsWald2015,HollandsWald2003RG}.

Finally, a differentiable unitary frame change $\mathscr W(s)$ transforms the total evolution by $\mathcal U'_A(s_2,s_1)=\mathscr W(s_2)\mathcal U_A(s_2,s_1)\mathscr W(s_1)^{-1}$. If $\widehat K'_A=\mathscr W\widehat K_A\mathscr W^{-1}$, the connection acquires $\mathcal A'_A=\mathscr W\mathcal A_A\mathscr W^{-1}-\dot{\mathscr W}\mathscr W^{-1}$. This is a frame change, not a new propagation law. When cells describe one endpoint ensemble, its effects must form one complete POVM and its conditional states and probabilities must come from one unconditioned state and compatible instrument. Unrelated cellwise conjugations guarantee neither property.

\section{Positive energy forms and field localization}
\label{app:localization}\label{app:povm-fock}

The purpose is exact, nonnegative energy accounting even when localization does not commute with the Hamiltonian. POVMs and instruments in Sec.~\ref{sec:povm} are standard measurement-theory objects, but their physical detector-free choice is additional input \cite{DaviesLewis1970,Busch2016QuantumMeasurement,Luders2006}. A change of basis does not change a fixed POVM; choosing different spatial effects can change the energy assignment.

For $\rho=\sum_n r_n|\psi_n\rangle\langle\psi_n|$ with finite mean energy, every eigenvector with $r_n>0$ lies in $D(\widehat H_R^{1/2})$. The quadratic forms of Eq.~\eqref{eq:main-energy-form} therefore extend to the same mixed-state assignment:
\begin{equation}
 \mathcal E_A[\rho]=\sum_n r_nq_A[\psi_n]
 =\Tr(B_\rho B_\rho^\dagger\Effect_A),\qquad
 B_\rho=\widehat H_R^{1/2}\rho^{1/2},\qquad
 \sum_A\mathcal E_A=\|B_\rho\|_{\rm HS}^2=\Tr(\rho\widehat H_R).
 \label{eq:energy-share-state-new}
\end{equation}
The Hilbert--Schmidt norm is $\|B\|_{\rm HS}^2=\Tr(B^\dagger B)$; its finiteness makes $B_\rho B_\rho^\dagger$ positive trace class. Thus only legitimate trace cyclicity with bounded effects is needed. The result is ensemble-independent and applies to countable partitions by monotone convergence. An operator representative of $q_A$ requires closability; its specified closure gives a unique nonnegative self-adjoint operator by the representation theorem for closed forms \cite[Chap.~VI]{Kato1995}. The mean-energy form itself is primary.

The conditional state does not automatically contain only its cell. For two equal-energy particles $|1_L,1_R\rangle$, an energy-weighted pair of cell effects can each act as $\I/2$ on that state. Both conditional states then retain both particles, although their additive energy shares are correct. This is why the measure in Eq.~\eqref{eq:metric-energy-measure}, rather than an assumed normalized conditional spatial density, enters the gravitational rule. A genuine absorbing detector has the different instrument described in Sec.~\ref{sec:measurement}.

Only a short free-field extension is needed here. Let $h_i$ be the positive one-particle Hamiltonian of a species or charge sector $i$, and $E_i^{(1)}(B)$ its normalized spatial POVM. On the bosonic or fermionic Fock space define the dominated regional energy forms
\begin{equation}
 K(B)=\sum_i d\Gamma_i\!\left(h_i^{1/2}E_i^{(1)}(B)h_i^{1/2}\right),
 \quad H_0=\sum_i d\Gamma_i(h_i),\qquad
 F^E(B)=H_0^{-1/2}K(B)H_0^{-1/2}.
 \label{eq:app-Fock-energy-POVM}
\end{equation}
The operation $d\Gamma_i$ sums a one-particle operator over the constituents in that sector. The inequality $0\leq K(B)\leq H_0$ makes the last form a bounded effect on $(\ker H_0)^\perp$, including when the inverse has no mass gap; its total is the identity there. A separate zero-energy outcome covers the kernel. A complex scalar includes both particle and antiparticle sectors. Permutation-symmetric configuration observables and field-theoretic localization constructions are treated in Refs.~\cite{DurrGoldsteinTumulkaZanghi2005,Moretti2023,Moretti2026QFT,CeleriKiossesTerno2016}; they have distinct physical localization properties. The canonical position response chosen in Eq.~\eqref{eq:cosmo-local-energy-operator} is not asserted to satisfy all local-algebra properties of those alternatives.

Interaction energy is not generally a sum of one-particle operators. The interacting construction uses the full $\widehat H_R$ and its specified normalized POVM directly in Eq.~\eqref{eq:energy-share-state-new}. Local quantum-probe instruments provide concrete measurement precedents \cite{FewsterVerch2020,MandryschNavascues2025}, not a unique cosmological localization observable. Neither the free extension nor the square-root instrument proves a general interacting localization theorem.

\section{Uncertainty, spreading and resolved boundaries}
\label{app:worldtube-details}

The physical role of this appendix is to distinguish the extent of a quantum packet from a sharp material boundary, and intrinsic fluctuations from measurement resolution. The covariance contour in Sec.~\ref{sec:worldtube} summarizes the state; it is not an additional gravitational switch. For position first-moment operators $X^i=\int X^i\POVM(dX)$ and second moments $\mathsf M_2^{ij}=\int X^iX^j\POVM(dX)$, the actual outcome covariance is
\begin{equation}
 C_X=C_X^{\rm op}+N_X^{\rm res},\qquad
 (N_X^{\rm res})^{ij}=\Tr\rho_A
 \left(\mathsf M_2^{ij}-\tfrac12\{X^i,X^j\}\right).
 \label{eq:POVM-intrinsic-outcome-noise}
\end{equation}
The intrinsic covariance uses the operator first moments. Positivity of the POVM implies positive semidefinite $N_X^{\rm res}$ in every real direction; apparatus noise is added only if independent and not already included \cite{Busch2016QuantumMeasurement}. The same distinction holds without a detector.

For a complex linear combination $B=\sum_I z_I\Delta R_I$ of the six centered position and momentum observables, positivity of $\langle B^\dagger B\rangle$ proves Eq.~\eqref{eq:main-covariance-uncertainty}, since $\langle\Delta R_I\Delta R_J\rangle=C^{\rm op}_{IJ}+i\Omega_{IJ}/2$. Here $z_I$ are arbitrary complex coefficients and $\boldsymbol\Omega$ is the commutator matrix defined in the main text. This is the Robertson--Schr\"odinger inequality \cite{Robertson1929,Schrodinger1930}. If exact centered Heisenberg evolution is $\Delta\bm R(s)=\mathcal M\Delta\bm R(s_1)$, substitution proves Eq.~\eqref{eq:main-covariance-evolution}. For nonlinear dynamics it is only a linearization and higher moments may be needed \cite{Weedbrook2012}. A massive, slowly moving complex-scalar packet has $X(t)=X(0)+tP(0)/m_R$, explaining the momentum-spread and position--momentum terms in Eq.~\eqref{eq:Minkowski-Gaussian-spreading}.

For the characteristic function in Eq.~\eqref{eq:main-time-characteristic}, finite energy variance gives $|\mathcal C_A^{\rm coh}(\delta\varsigma)|^2=1-(\Delta E_A^{\rm loc})^2\delta\varsigma^2+o(\delta\varsigma^2)$. Its first threshold crossing measures phase dispersion, not necessarily irreversible decoherence. Likewise, the flux in Eq.~\eqref{eq:main-resolved-boundary-flux} must be smeared in space and time: the stress is an operator-valued distribution. A moving contour samples both transport through it and its own motion, with local energy flux $T^{i0}s_i-v_bT^{00}$, where $s_i$ is its outward unit normal and $v_b$ its normal velocity. The contour does not remove the energy in its tails.

The scalar short-wavelength limit explains the reference geodesic without making a packet rigid. Writing $\Phi_A=\mathcal C_Ae^{i\theta_A}$ gives, at leading orders, $k_\mu=\partial_\mu\theta_A$, $g^{\mu\nu}k_\mu k_\nu+m_R^2=0$ and $\nabla_\mu(|\mathcal C_A|^2k^\mu)=0$. Thus $k^\nu\nabla_\nu k_\mu=0$ while the amplitude can focus or dilute \cite{HollandsWald2015}. In flat FRW a free Fourier profile rescaled as $v_{\bm k}=a^{3/2}\phi_{\bm k}$ obeys
\begin{equation}
 \ddot v_{\bm k}+\varpi_{\bm k}^2v_{\bm k}=0,
 \qquad \varpi_{\bm k}^2=m_R^2+|\bm k|^2/a^2
 -\tfrac32\dot H-\tfrac94H^2.
 \label{eq:FRW-adiabatic-frequency-check}
\end{equation}
This profile frequency is not a new quantum curvature coupling. An oscillatory adiabatic approximation requires $\varpi^2>0$ and $|\dot\varpi|/\varpi^2\ll1$, with higher derivative bounds as appropriate. Near a singular epoch these assumptions can fail.

\section{Metric distortion and the domain of the energy rule}
\label{app:metric-comparison}

The physical question is how neighboring clock and ruler intervals depart from their tangent predictions. This appendix justifies the norm and the local curvature estimate, without introducing a separate mode metric. The symmetric tensor $\Delta=\widetilde g-\eta$ defines a self-adjoint endomorphism $q^{-1}\Delta$ in the positive observer inner product $q$. Its real eigenvalues $d_j$ give $D=\max_j|d_j|$ by the Rayleigh quotient. This proves Eq.~\eqref{eq:metric-mismatch} and avoids division by a Lorentzian null interval. Simultaneous transformation of the metric, observer and exponential map preserves this number; changing the physical observer also changes the energy frame.

For the convention $[\nabla_\mu,\nabla_\nu]V^\alpha=R^\alpha{}_{\beta\mu\nu}V^\beta$, a variation of the radial geodesic $\gamma(\lambda)=\exp_x(\lambda X)$ yields a Jacobi field $J_V$:
\begin{equation}
 \frac{D^2J_V^\alpha}{d\lambda^2}
 +R^\alpha{}_{\beta\mu\nu}\dot\gamma^\beta J_V^\mu\dot\gamma^\nu=0,
 \quad J_V(0)=0,\quad \frac{DJ_V}{d\lambda}(0)=V,
 \quad J_V(1)=(d\exp_x)_XV.
 \label{eq:metric-jacobi}
\end{equation}
Here $\lambda\in[0,1]$ is affine and $D/d\lambda$ the classical derivative along the geodesic. The pullback is $\widetilde g_X(U,V)=g(J_U(1),J_V(1))$. Expanding this relation gives the normal-coordinate series \cite{PoissonPoundVega2011,Nesterov1999}
\begin{equation}
 \widetilde g_{ab}(X)=\eta_{ab}-\tfrac13R_{acbd}X^cX^d
 -\tfrac16R_{acbd;e}X^cX^dX^e
 +O\!\left[(\|\nabla^2R\|_q+\|R\|_q^2)|X|_q^4\right].
 \label{eq:metric-rnc-cubic}
\end{equation}
All coefficients are at $x$; the semicolon is a covariant derivative and the remainder uses local positive frame bounds. Thus rapidly varying curvature can invalidate a quadratic estimate even if $|R|r^2$ is small. Integrating the geodesic and Jacobi equations instead evaluates the full comparison.

In FRW put $A=\ddot a/a$, $B=H^2+k/a^2$ and $X=(0,\bm X)$, with $r^2=|\bm X|^2$. The curvature components are $R_{0i0j}=-A\delta_{ij}$ and $R_{ijkl}=B(\delta_{ik}\delta_{jl}-\delta_{il}\delta_{jk})$. The leading metric differences are $\Delta_{00}=Ar^2/3$, $\Delta_{0i}=0$, and $\Delta_{ij}=-B(r^2\delta_{ij}-X_iX_j)/3$. Their eigenvalues are $Ar^2/3,0,-Br^2/3,-Br^2/3$, proving Eq.~\eqref{eq:metric-frw-leading}. The Ricci scalar alone would miss radiation curvature. This is a point-normal comparison, not a Fermi chart along a whole history; the latter has a different quadratic clock coefficient \cite{ManasseMisner1963}.

Parallel transport is an isometry, so replacing $d\exp_x$ by parallel transport would erase the mismatch. Comparing only the radial geodesic interval also gives equality, $\sigma(x,\exp_xX)=\eta_x(X,X)/2$. The full endpoint variations in Eq.~\eqref{eq:metric-jacobi} are essential. Normal domains can end at conjugate points, nonunique geodesics or incompleteness. Extending a profile beyond them needs a specified continuation and overlap prescription, not a rule declaring the omitted energy nongravitating. Within a domain the exact sublevel set can have several components; a central ball requires the monotonic radial assumptions stated in the cosmological example.

Finally the spatial weight $0\leq\Gamma_A\leq1$ refines, rather than duplicates, each positive effect: $\Effect_A^{\rm grav}=\int w_A\Gamma_A\POVM(dX)$ and $\Effect_A^{\rm rest}=\Effect_A-\Effect_A^{\rm grav}$. Hence both are positive and $\sum_A(\mathcal E_A^{\rm grav}+\mathcal E_A^{\rm rest})=\Tr(\rho\widehat H_R)$. This includes both charges of a complex scalar. It is an energy-accounting identity, not positivity of its unrestricted renormalized pointwise stress.

\section{Source reconstruction and the detector-response test}
\label{app:source-details}

This appendix explains how selected tangent energy becomes a tensor source, and why conservation and absence of an early detector signal are different requirements. For the spatial reconstruction map $\iota_{A,s}$ of Sec.~\ref{sec:source}, define its proper-volume Jacobian by $\iota_{A,s}^*d\Sigma_g=j_{A,s}(X)d^3X$. If $d\mu_A=e_A(X)d^3X$, conservation of the selected observer energy requires
\begin{equation}
 \varepsilon_A^{\rm grav}(\iota_{A,s}X)=\frac{\Gamma_A(X)e_A(X)}{j_{A,s}(X)},
 \qquad
 T^{{\rm grav},A}_{\mu\nu}
 =\varepsilon_A^{\rm grav}n_\mu n_\nu+2n_{(\mu}q^A_{\nu)}
 +p_A^{\rm grav}h_{\mu\nu}+\pi^A_{\mu\nu}.
 \label{eq:source-density-jacobian}
\end{equation}
Here $h_{\mu\nu}=g_{\mu\nu}+n_\mu n_\nu$, $q^A$ is spatial energy flux, $p_A^{\rm grav}$ pressure, and $\pi^A$ spatial symmetric trace-free stress; these are not the norm $q_x$, probability $p_A$, or a representation map. The actual resolved stress data and reconstruction fix them. An isotropic massless scalar ensemble has zero flux and shear and pressure $\varepsilon^{\rm grav}/3$. A localized packet generally does not. Overall normalization cannot correct a wrongly shaped density or determine missing pressure. This is a source-reconstruction requirement, not another criterion for gravitation. The explicit homogeneous closure is already Eq.~\eqref{eq:cosmo-response-closure}.

For the candidate in Sec.~\ref{sec:measurement}, all derivatives below are flat-background derivatives and all perturbations are first order. The identity $\partial^\mu(\mathcal Lb)_{\mu\nu}=\Box b_\nu$ proves conservation of $\delta\Theta^{(r)}=\chi_R\delta T^{(r)}+\mathcal Lb^{(r)}$ using Eq.~\eqref{eq:record-candidate-response}. Assume smooth resolved, past-compact sources, zero response perturbation before the experiment, and the retarded solution. A wave equation has finite propagation speed \cite{BaerGinouxPfaeffle2007}, so the support of both $b^{(r)}$ and $\delta\Theta^{(r)}$ lies within $J^+(\mathcal O_R)$. In Lorenz gauge the trace-reversed metric perturbation satisfies
\begin{equation}
 \Box\delta\overline h^{(r)}_{\mu\nu}=-16\pi G\delta\Theta^{(r)}_{\mu\nu},
 \qquad
 \partial^\mu\delta\overline h^{(r)}_{\mu\nu}=0,
 \qquad
 \delta\overline h_{\mu\nu}=\delta h_{\mu\nu}
 -\tfrac12\eta_{\mu\nu}\eta^{\alpha\beta}\delta h_{\alpha\beta}.
 \label{eq:record-retarded-metric}
\end{equation}
Common initial constraints and the conserved source make the gauge condition consistent. Retarded evolution implies $\delta h=0$ outside this causal future, hence no change of locally measured linearized curvature there \cite{Carroll1997GR}. The instantaneous matter contribution can nevertheless change: outside the future, the algebraic term in $\delta C$ cancels it. Merely imposing $\partial^\mu\delta C_{\mu\nu}=-\partial^\mu\delta T_{\mu\nu}$ would not establish this support property. The factor $\chi_R$ and its compensating term are the extra assumption.

To pass from an unchanged early metric to readout statistics, require that the reference histories for switching the detector on and off themselves agree in the causal past of the left probe. Also require ordinary local quantum instruments there, no additional direct coupling of the probe to the auxiliary response, and the same early probe preparation. This includes any field, detector and probe entanglement in the initial joint state $\rho$. Let $d$ denote the right control setting, $M_{dr\alpha}$ its local instrument operators with $\sum_{r,\alpha}M_{dr\alpha}^\dagger M_{dr\alpha}=\I$, and $F_y$ the left readout effect evolved with the common early metric. Local commutativity gives $[F_y,M_{dr\alpha}]=0$. Therefore
\begin{equation}
 \begin{aligned}
 P_d(y)&=\sum_{r,\alpha}\Tr
 (F_yM_{dr\alpha}\rho M_{dr\alpha}^\dagger)\\
 &=\Tr\left[\rho F_y\sum_{r,\alpha}
 M_{dr\alpha}^\dagger M_{dr\alpha}\right]
 =\Tr(\rho F_y).
 \end{aligned}
 \label{eq:record-readout-proof}
\end{equation}
The tensor-product notation of the packet model is only a convenient finite-mode representation; in local QFT the appropriate statement concerns commuting local algebras and localized instruments \cite{FewsterVerch2020}. Locality must be established for the actual detector and probe instruments; normalization of a spatial energy POVM alone does not imply it. The equality is for unpostselected distributions; $P(y\mid r,d)$ can change. The sum computes probabilities over runs and does not average metrics to construct an individual source.

This result assumes full acceptance of the relevant energy profiles, prescribed reference histories and a fixed weak-field comparison. It is not a theorem for the nonlinear metric-dependent transport and reconstruction. The additional response can carry either sign of energy relative to the reference. Even if $\chi_R$ tends to one, the retarded $\mathcal Lb$ term can persist, so recovery of a desired later branch metric requires a further calculation. Subsection~\ref{sec:measurement} gives the covariant response on a prescribed background. The physical calibration, later metric and response energy, and stability of a fully coupled completion remain open. In a reference-vacuum sector with zero reconstructed source and no record-induced change, the candidate adds zero response when its homogeneous data are zero.

\section{Cosmological scaling and quantum fluctuations}
\label{app:cosmology-details}

The aim is to identify what actually proves increased gravitational participation, and where quantum spreading can prevent it. The exact prescribed-background result and the approximate homogeneous feedback solution have different assumptions. Their definitions and physical interpretation remain in Sec.~\ref{sec:cosmology}; only the supporting steps are collected here.

For $a(t)=(t/t_0)^p$, $0<p<1$, the homothety $\varphi_\lambda(t,\bm x)=(\lambda t,\lambda^{1-p}\bm x)$ obeys $\varphi_\lambda^*g=\lambda^2g$. Constant metric rescaling leaves the Levi--Civita connection unchanged, so $\varphi_\lambda\exp_xX=\exp_{\varphi_\lambda(x)}(d\varphi_\lambda)_xX$. In orthonormal comoving components the differential multiplies lengths by $\lambda$. Both forms in the mismatch quotient scale equally, giving
\begin{equation}
 D_{(\lambda t,\bm0)}(\lambda X^{\hat a})=D_{(t,\bm0)}(X^{\hat a}),
 \qquad d_p(y)=\tfrac13\max\{p(1-p),p^2\}y^2+O(y^3).
 \label{eq:cosmo-distortion-homothety}
\end{equation}
Here $y=r/t$ in units with $c=1$. The scaling identity shows that,
at twice the cosmic age, the same metric mismatch is reached at twice
the physical tangent distance. This is why the function $d_p$ defined in
Eq.~\eqref{eq:cosmo-exact-homothety} depends on the ratio $r/t$:
the ratio follows from the geometry and imposes no light-travel-time
or horizon restriction on the mode. On its stated monotone normal domain the threshold radius is proportional to $t$, whereas the prescribed comoving width grows as $t^p$. The compact support-to-time ratio decreases, so initial domain coverage is preserved. Positivity of the energy profile then proves Eq.~\eqref{eq:cosmo-exact-prescribed-growth}. For the smooth weight, $W(3d_p(z\ell/t)/\zeta^2)$ is nondecreasing at each fixed $z$, and integration preserves this ordering. For the tail estimate let $P$ be the Gaussian probability, $P_C$ its normalized taper, and $\delta_C=\int(1-\chi) \,\dd P$, with $\chi$ evaluated at $r/(C\ell)$. For $\delta_C>0$ the removed mass defines a probability $P_{\rm rem}$ and $P=(1-\delta_C)P_C+\delta_C P_{\rm rem}$. The two means under $P$ and $P_C$ of any weight between zero and one therefore differ by at most $\delta_C$. The zero-loss case is immediate.

Quantum and thermal fluctuations concern the specific energy observable, not just the classical profile. In a finite orthonormal one-particle basis write $K(b)=\sum_{mn}b_{mn}a_m^\dagger a_n$, with $b$ self-adjoint. For a gauge-invariant quasifree thermal state, Wick contraction gives
\begin{equation}
 \langle K(b)\rangle=\Tr_1(nb),\qquad
 \langle K(b)K(c)\rangle-\langle K(b)\rangle\langle K(c)\rangle
 =\Tr_1[nb(1+\sigma n)c].
 \label{eq:cosmo-wick-covariance}
\end{equation}
Here $c$ is a second one-particle operator, $n=(e^{h/T}-\sigma)^{-1}$, and $\sigma=+1$ for bosons and $-1$ for fermions. This follows directly from the CCR or CAR and thermal pair contractions; the real part is the symmetrized covariance. Setting $b=c=h^{1/2}w(\bm Q)h^{1/2}$ gives Eq.~\eqref{eq:cosmo-exact-quantum-variance}. Independent species add; overlapping cells generally have nonzero cross covariance. Standard quasifree thermal constructions and relativistic energy-fluctuation calculations are discussed in Refs.~\cite{BratteliRobinson1997,DasEtAl2021EnergyFluctuations}.

In a translation-invariant scalar sector, let $\widetilde w(\bm k)=\int e^{-i\bm k\cdot\bm X}w(\bm X)d^3X$, $\omega_{\bm p}=\sqrt{|\bm p|^2+m_R^2}$ and $n_{\bm p}=(e^{\omega_{\bm p}/T}-\sigma)^{-1}$. The same trace becomes
\begin{equation}
 \operatorname{Var}K(w)=\int\frac{d^3p\,d^3p'}{(2\pi)^6}
 \omega_{\bm p}\omega_{\bm p'}n_{\bm p}(1+\sigma n_{\bm p'})
 |\widetilde w(\bm p-\bm p')|^2.
 \label{eq:cosmo-variance-integral}
\end{equation}
Each independent complex-scalar charge sector contributes once. Spin traces are needed for nontrivial fermionic responses. This trace representation does not rescale the field operators of Appendix~\ref{app:bundle-foundations}. For a broad smooth window, $\bm p'\simeq\bm p$ and $\partial_Tn=\omega n(1+\sigma n)/T^2$ give $\operatorname{Var}K\simeq T^2(\partial_T\varepsilon_Q)\int w^2d^3X$, proving Eq.~\eqref{eq:cosmo-large-cell-variance}. A sharp spherical step has a Fourier tail of order $|\bm k|^{-2}$ and can cause a logarithmic boundary divergence in the second moment. Smooth finite responses avoid that particular problem. The result is for the specified positive energy assignment; a differently renormalized local stress observable has its own fluctuations.

For the feedback example, $\mathcal G(q)=\sqrt{2/\pi}\int_0^qz^2e^{-z^2/2}dz$ implies $0<q\mathcal G'/\mathcal G<3$. At quadratic order the universal smooth response is $\chi_g(v)=W(v^2)$, $v=r/r_g$. It replaces the sharp result by $\mathcal G_g(q)=\sqrt{2/\pi}\,q^3\int_0^1v^2\chi_g(v)e^{-q^2v^2/2}dv$, with the same logarithmic slope bounds: increasing $q$ increases the resolved response at fixed profile radius, while the residual integral multiplying $q^3$ decreases. For normalized radiation-energy weights $\nu(d\ell_0)$ over strictly positive comoving widths, set
\begin{equation}
 F(z)=\int\mathcal G\!\left(\frac{\zeta z}{\ell_0\sqrt{D_0}}\right)\nu(d\ell_0),
 \qquad z=\frac a{\sqrt f},\qquad f=F(z),\qquad a=z\sqrt{F(z)}.
 \label{eq:cosmo-mixture-solution}
\end{equation}
The constant $D_0$ is defined in Eq.~\eqref{eq:cosmo-implicit-solution}. These weights only express the ratio of additive species energies. The logarithmic slope $\gamma_F=zF'/F$ lies between zero and three; differentiation gives $d\ln f/d\ln a=2\gamma_F/(2+\gamma_F)>0$. With the specified pressure closure, $\ddot a/a=-2B/(2+\gamma_F)$, which verifies the curvature branch used in the solution. The same proof holds with $\mathcal G_g$. Dominated convergence gives the early and late limits; the common early exponent $6/5$ further requires a finite nonzero inverse-cube width moment. Species thresholds or changing weights need extra terms and are outside this example.

To see the scope of this solution, consider one width on an expanding interval and set $s=\dd\ln f/\dd\ln a$. The acceleration equation with Eq.~\eqref{eq:cosmo-response-closure} gives $(\ddot a/a)/B=s/2-1$. Before selecting a curvature branch the radius equation reads
\begin{equation}
 \frac{a^2}{\alpha_\ell^2 q^2\mathcal G(q)}
 =\max\{1,|s/2-1|\}.
 \label{eq:feedback-full-branch-test}
\end{equation}
The solution in Sec.~\ref{sec:cosmology} sets this ratio to one and verifies the required inequality. If the ratio is $\Xi>1$, the two alternatives give $s=2(1\pm\Xi)$. Since $s=\gamma_f\,\dd\ln q/\dd\ln a$ and $\gamma_f>0$, each sign defines a local differential equation for $q(a)$ in that open region. For example, flat expanding data with $q=1$ and $a=\alpha_\ell\sqrt{3\mathcal G(1)/2}$ give $\Xi=3/2$; the minus sign yields $s=-1$. The pressure closure alone therefore does not select the growing branch.

For the individual massive scalar packet, in the stated nonrelativistic adiabatic radiation regime let $X_i,P_i$ be one initial physical canonical pair and $v=\ln(a/a_i)$. Integrating the kinetic spreading gives
\begin{equation}
 X(a)=\frac a{a_i}\left(X_i+\frac{v}{m_RH_i}P_i\right),
 \quad P(a)=\frac{a_i}{a}P_i,\qquad
 \operatorname{Cov}(X,P)=\frac{\gamma_i v}{2}
 \label{eq:cosmo-packet-linear-map}
\end{equation}
for the unchirped minimum-uncertainty Gaussian of Sec.~\ref{sec:cosmology}. Its variances are $\operatorname{Var}X=\ell_{\rm pkt}^2$ and $\operatorname{Var}P=(a_i/a)^2/(4\ell_i^2)$, so their determinant remains $1/4$. In three dimensions its leading energy moments are
\begin{equation}
 \overline E=m_R+\frac{3}{8m_R\ell_i^2}\left(\frac{a_i}{a}\right)^2,
 \qquad (\Delta E)^2=\frac{3}{32m_R^2\ell_i^4}
 \left(\frac{a_i}{a}\right)^4.
 \label{eq:cosmo-packet-energy-moments}
\end{equation}
Relativistic and higher adiabatic corrections are omitted \cite{AlHashimiWiese2009}. Initial phase matters: $\psi_c(X)\propto\exp[-X^2/(4\ell_i^2)+icX^2/(2\ell_i^2)]$, with real dimensionless $c$, has $\operatorname{Cov}(X_i,P_i)=c$ and $\operatorname{Var}P_i=(1+4c^2)/(4\ell_i^2)$. Thus the general width follows by inserting these moments into the first map in Eq.~\eqref{eq:cosmo-packet-linear-map}; it is not fixed by $\ell_i$ alone. This is the physical reason the thermal-window theorem does not assert monotonic gravitational participation for every freely spreading scalar state.

\bibliographystyle{unsrtnat}
\bibliography{18_2tangent_metric_comparison_gravity_references}
\end{document}